\PassOptionsToPackage{table}{xcolor}
\ifdefined\mainsubmission
  \PassOptionsToPackage{draft}{hyperref}
\fi
\documentclass[sigconf,nonacm]{acmart}
\ifdefined\pdfsuppressptexinfo
\fi
\ifdefined\mainsubmission
\fi
\usepackage{tikz}
\usepackage{subcaption}
\usetikzlibrary{fadings, arrows.meta, matrix, positioning, fit, calc, backgrounds}
\DeclareRobustCommand{\circnum}[1]{\tikz[baseline=(cn.base)]{%
  \node[circle, draw, line width=0.4pt, inner sep=1.15pt,
        font=\scriptsize] (cn) {#1};}}

\usepackage{array}
\usepackage{bbm}

\usepackage{float}
\usepackage{multirow}
\usepackage{enumitem}
\usepackage[linesnumbered,ruled,vlined]{algorithm2e}
\def\nocomment{1}  % comment this line to re-enable author comments and figure placeholders
\ifx\nocomment\undefined
\newcommand{\laxman}[1]{{\color{brown}{\bf Laxman:} #1}}
\newcommand{\kuba}[1]{{\color{violet}{\bf Kuba:} #1}}
\newcommand{\guy}[1]{{\color{olive}{\bf Guy:} #1}}
\newcommand{\lars}[1]{{\color{blue}{\bf Lars:} #1}}
\newcommand{\tobias}[1]{{\color{purple}{\bf Tobias:} #1}}
\newcommand{\richard}[1]{{\color{magenta}{\bf Richard:} #1}}
\newcommand{\raj}[1]{{\color{cyan}{\bf Raj:} #1}}
\newcommand{\figplaceholder}[1]{\textcolor{purple}{\textbf{[Figure placeholder:} #1\textbf{]}}}
\else
\newcommand{\laxman}[1]{{}}
\newcommand{\kuba}[1]{{}}
\newcommand{\guy}[1]{{}}
\newcommand{\lars}[1]{{}}
\newcommand{\tobias}[1]{{}}
\newcommand{\richard}[1]{{}}
\newcommand{\raj}[1]{{}}
\newcommand{\figplaceholder}[1]{{}}
\fi

\newcommand{\hide}[1]{}

\newcommand{\algoname}[1]{\textsf{#1}}

\newcommand{\diskann}{\algoname{DiskANN}}

\newcommand{\pynn}{\algoname{PyNNDescent}}

\newcommand{\mP}{\mathcal{P}}

\newcommand{\mX}{\mathcal{X}}

\newcommand{\Cmax}{\ensuremath{C_{max}}}
\newcommand{\Cmin}{\ensuremath{C_{min}}}
\newcommand{\Psamp}{\mathcal{P}_{samp}}
\newcommand{\lMax}{\ensuremath{\ell_{max}}}

\newcommand{\sfac}{\sigma}  % per-vector TQ scaling factor (orig.\ norm / quantized norm)

\providecommand{\abs}[1]{\lvert#1\rvert}
\providecommand{\norm}[1]{\lVert#1\rVert}

\newcommand{\emp}[1]{\emph{\textbf{#1}}} % highlight
\newcommand{\myparagraph}[1]{\noindent\emp{#1.} \,}

\SetCommentSty{mycommfont}

\usepackage{cleveref}
\crefname{section}{Sec.}{Sec.}
\crefname{theorem}{Thm.}{Thm.}
\crefname{lemma}{Lem.}{Lem.}
\crefname{corollary}{Col.}{Col.}
\crefname{table}{Tab.}{Tab.}
\crefname{algorithm}{Alg.}{Alg.}
\crefname{figure}{Fig.}{Fig.}
\crefname{fact}{Fact}{Fact}
\Crefname{table}{Tab.}{Tab.}
\crefname{problem}{Problem}{Problem}

\newcommand{\ourmethod}{PiPNN II\xspace}
\newcommand{\ouroldmethod}{PiPNN\xspace}
\newcommand{\ourgpumethod}{GPiPNN\xspace}
\newcommand{\tq}{TurboQuant\xspace}

\newtheorem{defn}{Definition}

\newcommand{\HashPrune}{\textsc{HashPrune}\xspace}
\newcommand{\RobustPrune}{\textsc{RobustPrune}\xspace}
\newcommand{\DirectEdgePrune}{\textsc{DirectEdgePrune}\xspace}
\long\def\hide#1{}

\DeclareMathOperator*{\argmax}{argmax}
\DeclareMathOperator*{\argmin}{argmin}

\newcommand{\mstdeg}{\ensuremath{m}\xspace}
\newcommand{\fanout}{\ensuremath{F}\xspace}

\begin{document}

\title[Turbocharging PiPNN for Proximity and $k$-NN Graph Building]{Turbocharging PiPNN
for Proximity and $k$-NN Graph Building}
\hypersetup{
  pdftitle={Turbocharging PiPNN for Proximity and k-NN Graph Building}
}
%Fast Graph-Based Approximate Nearest-Neighbor Search \\ by Partitioning and Pruning}

%%%
%%% The "author" command and its associated commands are used to define the authors and their affiliations.
\author{Tobias Rubel}
\affiliation{%
  \institution{UMD}
  \city{College~Park}
  \country{USA}
}
\email{trubel@umd.edu}

\author{Richard Wen}
\affiliation{%
  \institution{UMD}
  \city{College~Park}
  \country{USA}
}
\email{rwen1@umd.edu}

% \author{Kishen Gowda}
% \affiliation{%
%   \institution{UMD}
%   \city{College~Park}
%   \state{Maryland}
%   \country{USA}
% }
% \email{kishen19@umd.edu}

\author{Guy Blelloch}
\affiliation{%
  \institution{CMU and Google Research}
  \city{Pittsburgh}
  \country{USA}
}
\email{blelloch@cs.cmu.edu}

\author{Laxman Dhulipala}
\affiliation{%
  \institution{UMD and Google Research}
  \city{College~Park}
  \country{USA}
}
\email{laxman@umd.edu}

\author{Lars Gottesb\"uren}
\affiliation{%
  \institution{Google Research}
  \city{Zürich}
  \country{Switzerland}
}
\email{gottesbueren@google.com}

\author{Jakub {\L}{\k a}cki}
\affiliation{%
  \institution{Google Research}
  \city{New~York~City}
  \country{USA}
}
\email{jlacki@google.com}

\author{Vahab Mirrokni}
\affiliation{%
  \institution{Google Research}
  \city{New~York~City}
  \country{USA}
}
\email{mirrokni@google.com}

% Set author metadata explicitly as well as the visible author blocks.
\hypersetup{pdfauthor={Tobias Rubel, Richard Wen, Guy Blelloch,
  Laxman Dhulipala, Lars Gottesbüren, Jakub Łącki, Vahab Mirrokni}}

\begin{abstract}
Graph-based Approximate Nearest Neighbor Search (ANNS)  indexes such as HNSW and Vamana deliver state-of-the-art query performance at the cost of large indexing times.
The recently introduced \ouroldmethod{} method drastically accelerated index construction by way of a partitioning-based algorithm which leverages efficient dense matrix multiplication and top-k aggregation kernels.

In this paper, we unlock the full potential of \ouroldmethod{}'s hardware-friendly algorithm design.
We introduce \ourmethod{}, which optimizes multiple steps of the \ouroldmethod{} recipe, giving up to $8\times$ speedups for graph-based ANNS index construction.
In addition, we present \ourgpumethod{}, a GPU implementation of \ouroldmethod{}, which builds comparable ANNS indices up to $43.7\times$ faster than Jasper and up to $50\times$ faster than CAGRA.
We also extend \ourmethod{} and \ourgpumethod{} to compute high-quality $k$-nearest-neighbor ($k$-NN) graphs directly, which unlocks massive speedups over methods used in many existing systems for clustering, anomaly detection, and more. 
For 10-NN graph building at $90\%$ recall, we obtain up to a $13\times$ speedup on CPUs and $6.6\times$ on GPUs as compared to the prior best methods. \ourmethod{} can build a billion-scale 10-NN graph in under 16 minutes.

Common to both \ourmethod{} and \ourgpumethod{} is the use of \emph{quantization} to reduce memory traffic and leverage fast low-precision matrix operations.
Top-$K$ emerges as a new bottleneck, which we optimize by specializing the code to the specific $K$, as well as fusing with the matrix kernel to avoid materializing the distance matrix to memory.
For CPUs, we demonstrate further performance improvements by leveraging our matrix kernels in the RobustPrune procedure originally proposed in Vamana and used in PiPNN.
On GPUs, we flatten recursive partitioning into segmented batches, use one fused top-$K$ abstraction for partitioning and leaf scoring, and prune with already-witnessed direct edges rather than a new candidate-to-candidate scoring pass.

\end{abstract}

\maketitle

%%% VLDB submission block (disabled for arXiv) %%%
%%% VLDB block start %%%
% \vldbtopmatter
%%% VLDB block end %%%

%\input{00_paper_outline}

\section{Introduction}

\begin{figure}[t]
    \centering
    \includegraphics[width=\linewidth]{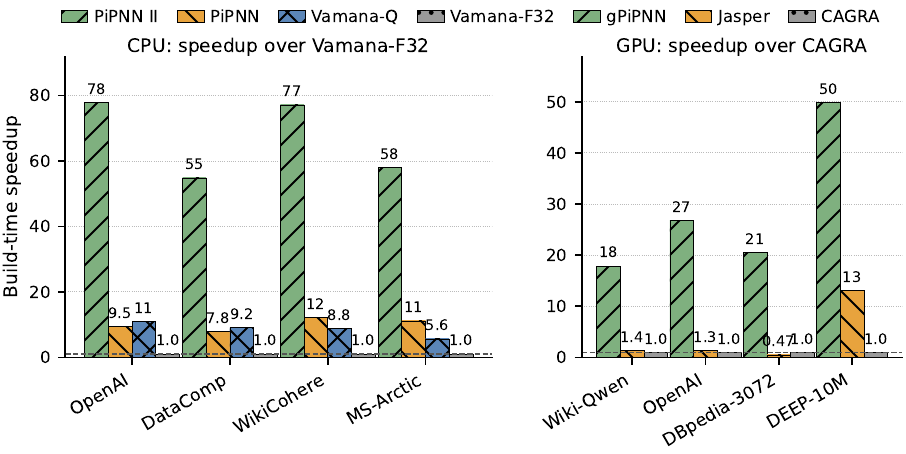}
    \caption{Index-build speedup of \ourmethod{} and \ourgpumethod{} over well-optimized baselines. \emph{Left:} CPU speedup  of \ourmethod{} over unquantized Vamana. \emph{Right:} GPU speedup of \ourgpumethod{} over CAGRA.}
    \label{fig:teaser}
\end{figure}

Vector embeddings power modern search, recommendation, entity resolution, and retrieval-augmented generation systems~\cite{LiLiSuhara:2020,ZeakisPapadakisSkoutas:2023,LewisPerezPiktus:2020,gao2023retrieval,mitra2018introduction,roy2022systematic,XiongXLTLBAO21}.
%
%These systems solve nearest-neighbor search problems over large collections of vectors, and extensively rely on \emph{approximate nearest-neighbor search (ANNS)} as exact search over large corpora is prohibitively expensive.
These systems rely on \emph{approximate nearest-neighbor search (ANNS)} because exact search is prohibitively expensive and small errors are tolerable.
Graph-based indexes such as HNSW, Vamana (DiskANN), and NSG have become the standard for high-recall, low-latency ANNS~\cite{malkov2018efficient,jayaram2019diskann,nsg}.
Unfortunately, their excellent query performance has come at the cost of slow index construction: building a high-quality search graph (also termed ``proximity graph'') over a large dataset has historically required many hours or even days of compute on powerful multi-core machines~\cite{manohar2024parlayann}.

Fast index construction is not merely a convenience:  build time is often a major factor in the total cost of the system.
Indexes must be rebuilt as data drifts, parameters need to be tuned across datasets, and ANNS appears as an inner loop in fundamental tasks including accelerated $k$-means~\cite{spalding2025scalable}, hierarchical clustering~\cite{yu2025pecann, dhulipala2023terahac}, minimum spanning trees, and single-linkage clustering~\cite{andoni2023sub, beretta2025approximating, azarmehr2025massively, bateni2017affinity}.
Small modifications to graph-index construction also yield high-quality $k$-NN graphs~\cite{carey2022stars}, a primitive that we directly target in this paper.
An indexing method that takes hours or days is simply unusable in many modern workflows.
%
%The root cause is the \emph{search bottleneck}: traditional graph-based indexing methods insert points one at a time or in small batches, each performing a beam search on the partial index, so a large fraction of build time is spent on irregular pointer-chasing rather than dense numerical work~\cite{rubel2026pipnn,jayaram2019diskann,malkov2018efficient}.

The recently introduced Pick-in-Partitions Nearest Neighbors (\ouroldmethod{}) method significantly speeds up high-quality index construction by replacing build-time graph search with regular partition-local scoring work~\cite{rubel2026pipnn}.
%
%changed this build-time tradeoff by eliminating graph search from construction altogether~\cite{rubel2026pipnn}.
%
It recursively partitions the input vectors into small overlapping collections (termed ``leaves''), generates candidate edges through batched all-pairs distance comparisons and top-$K$ aggregation inside each leaf, and streams those candidates through an online history-independent pruning procedure (\HashPrune{}) to retain sparsity.
The bulk of \ouroldmethod{}'s construction is comprised of all-to-all comparisons over small, cache-resident sets of vectors; an operation far more amenable to modern hardware than the irregular pointer-chasing of incremental graph construction~\cite{malkov2018efficient,jayaram2019diskann,nsg}.
The payoff of this approach is that the existing version of \ouroldmethod{} is up to $11.6\times$ faster than the ParlayANN implementation of Vamana on CPU and produces indexes with equivalent QPS-recall performance~\cite{rubel2026pipnn}.

Despite these significant improvements in build times, the first version of \ouroldmethod fails to leverage modern hardware to its full potential.
For instance, it keeps vectors in full-precision float32 form to perform GEMM (General Matrix Multiply) and materializes the full distance matrices to memory to perform per-row top-$k$ reduction.
In this paper we significantly optimize the entire approach from first principles.
We start by leveraging \emph{quantization} to replace the full-precision vectors with compact integer codes that are cheaper to compare~\cite{avq_2020, soar_2023, douze2024faiss, johnson2019billion, manohar2024parlayann, krishnaswamy2024diskann}.
Quantizing the vectors not only speeds up memory accesses, but it also enables low-precision multiply-accumulate instructions.
Decoding cost may seem concerning at first glance, but is easily amortized over many distance computations.
%Quantization is an especially good fit for \ouroldmethod{}: its
%In particular, \ouroldmethod{}'s
%batched all-pairs structure scores each vector against many others, so a quantized code can be decoded once and reused across the entire batch, amortizing the cost of decoding.
%
%Thus quantization can help not only improve the memory-pressure inherent to all irregular graph-based ANNS indexing methods, but also benefits from hardware accelerated bulk top-k and amortized decoding which is difficult to leverage in frameworks that have less regular memory access patterns.
%
%Realizing this potential, however, is more involved than simply decoding and then running a quantized GEMM.
%
Lastly, we fuse the GEMM kernel with top-k reduction in register, to avoid materializing an $M \times N$ distance matrix to memory. 
While fused distance computation and in-register $k$-selection are established in GPU similarity search~\cite{johnson2019billion,douze2024faiss}, our contribution is a family of compressed-domain kernels specialized to \ouroldmethod{}'s leader-assignment and leaf-processing shapes and instantiated across CPUs and GPUs.
%Materializing and re-reading an $M \times N$ score matrix would squander the bandwidth savings, so \ouroldmethod{}'s inner loop must fuse pairwise distance computation with per-query top-$k$ reduction into a single in-register pass.

The aforementioned opportunity and challenge carries over to hardware acceleration using GPUs.
Compared to existing GPU ANNS systems~\cite{ootomo2024cagra, mccoy2026jasper} which struggle to leverage the full potential of the hardware due to relying on either beam search or iterative refinement on graphs, \ouroldmethod{}'s batched many-to-many structure is an ideal fit for tensor cores. 
However, implementing \ouroldmethod{} on GPUs presents non-trivial challenges: \ouroldmethod{}'s recursive partitioning produces irregularly-sized subproblems whose control flow is not handled well by the GPU scheduler and must be flattened into uniform passes; the vectors needed for each local comparison are scattered in device memory and must be gathered into dense batches; and the \HashPrune{} reservoirs for each point must be maintained without the use of locks.

%Thus, existing GPU ANNS systems therefore either work around this so-called ``search bottleneck''~\cite{rubel2026pipnn} (CAGRA reorganizes around nn-descent and rank-based pruning~\cite{ootomo2024cagra}) or simply try to optimize around it (Jasper ports ParlayANN's Vamana implementation approach to GPU via CUDA batch-parallel insertion~\cite{mccoy2026jasper,manohar2024parlayann}).

\emp{In this paper we realize the full potential of \ouroldmethod{} by carefully restructuring \ouroldmethod{}'s construction pipeline around fused quantized batched top-$K$ kernels}.
%Crucially our new kernels fold candidate generation and per-query top-$K$ selection into a single in-register pass over compressed-domain integer dot products, never materializing the $M \times N$ score matrix.
%
We instantiate this design on both shared memory CPU systems (\emp{\ourmethod{}}\footnote{Pronounced ``Pippin the second''.}) as well as GPUs (\emp{\ourgpumethod{}}), thus cashing in on the promise for hardware-accelerated ANNS builds within the \ouroldmethod{} framework. 
Our results show that our GEMM kernels take full advantage of the hardware: the CPU score-and-select kernels run near instruction throughput roofline, whereas the corresponding GPU kernels are memory bound for high-dimensional vectors typical of modern embeddings.

%For \ourmethod{} on CPU, we present  presents new VNNI-optimized fused quantized batched top-$K$ kernels for all of the kernel shapes dominating \ouroldmethod{}'s runtime, namely top-$1$ and top-$K$ assignment of points to leaders during partitioning and a symmetric top-$2$ pass inside each leaf.
%
%On GPUs, \ourgpumethod{} \emph{flattens} \ouroldmethod{}'s recursion into sorted segment-membership arrays, streams leaf batches through the same segmented top-$K$ primitive used for partitioning, accumulates candidates with a deterministic emit-sort-merge pipeline, and applies a novel \DirectEdgePrune{} that avoids candidate-to-candidate scoring during the final prune by leveraging cached distances in the \HashPrune{} structure.
On GPUs, \ouroldmethod{} \emph{flattens} the recursion used in \ouroldmethod{}'s partitioning routines by refactoring them as radix sorts over leader-assignment keys. In addition, it processes multiple collections of points in a single kernel using a segmented top-$K$ primitive, which ameliorates workload imbalances. Finally, it avoids a costly final prune, by way of a novel \DirectEdgePrune{} over the cached distances in the \HashPrune{} structure.

Our extensive experimental study establishes \ourmethod{} and \ourgpumethod{} as the new state of the art for proximity-graph index construction on CPUs and GPUs, delivering large speedups over the strongest prior methods while preserving high-recall query performance.
On CPUs, \ourmethod{} builds indexes up to $\mathbf{8\times}$ \emp{faster than \ouroldmethod{}} and substantially outperforms all prior graph-based ANNS builders on the datasets we evaluate.
On GPUs, \ourgpumethod{} achieves speedups of up to $\mathbf{43.7\times}$ \emp{over Jasper} and $\mathbf{50\times}$ \emp{over CAGRA} across a diverse benchmark of text, image, and text-to-image datasets, while matching or exceeding their query throughput at high recall.
On CPU, we show that our new methods enable the construction of billion-point graph indexes on a single machine in under 10 minutes, as opposed to hours for prior best methods.
%We construct a high-quality billion-point graph index in under ten minutes on a single GPU machine.

Beyond proximity-graph construction, we show that the same methods produce high-quality \emph{$k$-NN graphs} (a connection between partitioning-based construction and $k$-NN graphs also briefly noted in prior work~\cite{gottesburen2024unleashing}), making \ourmethod{} and \ourgpumethod{} state-of-the-art for $k$-NN graph construction as well. At recall$10$@$10 \geq 0.9$, \ourmethod{} is up to $\mathbf{13\times}$ faster than the best prior CPU $k$-NN builder (Vamana-Q), and \ourgpumethod{} is up to $\mathbf{6.6\times}$ faster than the best prior GPU $k$-NN builder (\textsc{cuvs NN-Descent}).

\noindent In summary, the main contributions of this paper are:
% SIGMOD submission: use the acmart list geometry and spacing defaults.
% \begin{enumerate}[label=(\arabic*),topsep=0pt,itemsep=0pt,parsep=0pt,leftmargin=15pt]
\begin{enumerate}
    \item  Highly optimized \emph{quantized GEMM fused with top-$K$ kernels}.

    \item
    \ourmethod{}, a shared-memory CPU implementation leveraging quantization and our new fused kernels to achieve state-of-the-art index construction and $k$-NN graph building performance. 
    
    \item \ourgpumethod{}, a GPU implementation of the same method which redesigns the \ouroldmethod{} framework for excellent performance on modern accelerators. %\ourgpumethod{} replaces recursive partition objects with flat segmented batches and avoids a new candidate-to-candidate scoring pass during final pruning.
    
    %whose four scoring kernels build on one templated WGMMA + TMA skeleton, whose recursive partitioning is flattened into sorted routing-key arrays processed by radix sorts, and whose final prune avoids candidate-to-candidate distance recomputation. \ourgpumethod{} attains up to $22.9\times$ and $140\times$ speedups over Jasper and CAGRA respectively.

    \item An extensive experimental evaluation spanning twelve datasets across text, image, and text-to-image modalities at scales up to one billion points (one hundred million on GPU), including direct $k$-NN graph construction results, that establishes \ourmethod{} and \ourgpumethod{} as state of the art for fast high-quality ANNS index construction on modern CPUs and GPUs.
\end{enumerate}

%\lars{Still some duplication here}

% \guy{Is quantization also an argument of why quantization is important, or is the memory bottleneck somewhere else?} \tobias{I added a little bit about this above, please see if it suffices. We had discussed not belaboring the ``search bottleneck'' in particular in this paper, but memory bottleneck is important to drive home IMO.}

\section{Preliminaries and Prior Work}
\label{sec:preliminaries}

%\subsection{Problem Statement}

\myparagraph{Approximate Nearest Neighbor Search (ANNS)}
Let $\mX \subseteq \mathbb{R}^d$ be a collection of $n$ vectors and let
$d(\cdot,\cdot)$ denote a dissimilarity measure on $\mathbb{R}^d$
(e.g.\ Euclidean distance or the negative of an
inner product).

%\kuba{In the following definition, should we say "a set" instead of "the set" and explain that it is not uniquely defined?}
\begin{defn}\label{defn:knns}
  ($k$-nearest neighbor search)
  Given a query $q \in \mathbb{R}^d$, the $k$-nearest neighbor set of $q$ in
  $\mX$ is the set $\mathcal{K} \subseteq \mX$ of size $k \le \lvert \mX \rvert$ containing the
  $k$ points of smallest dissimilarity to $q$:
  \[
    \forall p \in \mathcal{K},\ \forall p' \in \mX \setminus \mathcal{K},
    \quad d(q,p) \le d(q,p').
  \]
\end{defn}

We assume $\mathcal{K}$ is unique. In ANNS, the goal is to return a set $\mathcal{K}'$ that approximates $\mathcal{K}$ as efficiently as possible.

%\kuba{The following definition is not precise in the case when the k-NN set is not unique. Also, do we ever use $k' @ k$ recall for $k' \neq k$? If not, we could simplify the formula, and (a) define $\mathcal{K}$ as the union of \emph{all} k-NN sets (b) divide by k. Finally, we should assume that $k$ is at most the index size.} 
\begin{defn}
  ($k@k'$ recall)
  Let $\mathcal{K}$ be the exact $k$-nearest neighbor set of query $q$ and let
  $\mathcal{K}' \subseteq \mX$ be a returned set of size $k'$. The
  $k@k'$ recall of $\mathcal{K}'$ is $\frac{|\mathcal{K} \cap \mathcal{K}'|}{|\mathcal{K}|}$.
\end{defn}

We focus on $10@10$ recall and use \emph{recall} to denote this quantity; the recall of a query set $Q$ is the mean recall over all queries in $Q$.

\myparagraph{Graph-based ANNS}
Graph-based ANNS methods build a sparse \emph{proximity graph} over the points, with outgoing edges to a small set of nearby neighbors.
The graph need not exactly equal the $k$-NN graph; it must only be navigable enough that a local search can locate the nearest neighbors. HNSW and Vamana are canonical examples, both constructed via greedy graph traversal \cite{malkov2018efficient,jayaram2019diskann}. A long line of work has refined proximity-graph constructions \cite{malkov2014approximate, dong2011efficient, fu2016efanna, harwood2016fanng, nsg, fu2021high, munoz2019hierarchical, mcinnes2020pynndescent, iwasaki2018optimization, yang2024revisiting, zhao2023towards, voruganti2025mirage}, developed parallel or disk-resident variants \cite{krishnaswamy2024diskann, singh2021freshdiskann, manohar2024parlayann}, accelerated query-time search \cite{chen2023finger, gao2023high, gou2025symphonyqg}, and ported graph indexes to GPUs \cite{ootomo2024cagra, khan2024bang, mccoy2026jasper}.

\myparagraph{Beam search}
The standard graph-query procedure is beam search. It maintains a beam $B$ of the $L$ best points seen so far; at each step it expands the best unvisited point in $B$, inserting its neighbors into $B$ and truncating back to size $L$. The search terminates when every point in $B$ has been expanded, and the final beam is returned or used as candidates for reranking (see code in Appendix~\ref{app:beam-search}, Algorithm~\ref{alg:beam_search}).

\myparagraph{The ``search bottleneck''}
Prior to the introduction of \ouroldmethod, the highest-quality graph indexes were built by \emph{searching during construction}: HNSW and Vamana insert points incrementally, traversing the partial index to find candidate neighbors for each point, then pruning for sparsity \cite{jayaram2019diskann,malkov2018efficient}. This produces excellent graphs but creates a \emph{search bottleneck}: beam search is rife with random memory accesses and cache misses, so traversal dominates construction time. 
\ouroldmethod's partitioning approach breaks altogether from prior solutions which used distance-comparison-skipping heuristics and cheap graph-initialization.
It builds indexes up to $11.6\times$ faster than the efficient ParlayANN implementation of Vamana \cite{manohar2024parlayann}.

\subsection{Quantization}

Quantization is a lossy compression technique that maps high-precision vectors to a smaller, discrete set of representative points.
%The primary goal of quantization is to reduce the space required to store each point.
%It is a staple in production ANNS systems where some degree of error is permissible and computations are memory-bound.
%\kuba{Isn't that an overstatement? Up until recently it was not known how to combine graph-based indices with quantization in an efficient way}
%
Broadly speaking, three types of quantization methods are widely-used in the ANNS literature: scalar uniform quantization, which independently maps each vector coordinate to an $X$-bit integer; product quantization, which partitions a vector into subvectors and performs lookups in learned codebooks \cite{jegou2010product, ge2013optimized}; and rotation-based quantizers, which apply a random rotation and leverage distributional properties of the rotated vectors, including RaBitQ \cite{gao2024rabitq, gao2025rabitq, gao2025extendedrabitq}, DRIVE/EDEN \cite{vargaftik2021drive, vargaftik2022eden, benbasat2026eden}, and TurboQuant \cite{zandieh2025turboquant}.
In this paper, we consider two quantization methods: \emph{RSQ8} (rotated scalar quantization with 8 bits per coordinate) and \emph{TQ4}, a 4-bit variant of TurboQuant, which we outline next.

\myparagraph{TurboQuant}
TurboQuant is a scalar quantizer with non-uniform codebook.
It first applies a random rotation and then quantizes each post-rotation coordinate independently with a scalar Lloyd--Max codebook trained against the post-rotation marginal \cite{zandieh2025turboquant}. This achieves near-optimal distortion across bit widths and dimensions, but the resulting inner-product estimator is biased because the MSE-optimal codebook is not matched to the inner-product objective. The original TurboQuant addresses this with a two-stage estimator that applies a 1-bit JL quantizer to the residual to recover an unbiased dot product.

\myparagraph{TurboQuant variant}
We adopt TurboQuant's rotation-plus-scalar-quantizer structure but replace the QJL residual with an extra codebook bit and a per-vector rescaling factor $\sfac$.
We first normalize the vector before quantization, so to get the correct dot product, $\sfac$ is the ratio of the original norm to the quantized norm.
The estimator multiplies the dot product from the decoded vectors by the two scaling factors $\sfac_q$ and $\sfac_d$ (for the query and database vectors, respectively). 
The rotation is done with a Fast Walsh--Hadamard Transform composed with a small random orthogonal matrix, so encoding runs in $O(d \log d)$ and is data-oblivious.
Post-rotation quantization is done by independently mapping each coordinate to its nearest codebook entry. 
With TQ-4 we use an int8 version of the codebook scaled to $[-127, 127]$.
For further details, we refer to the original paper.
For the remainder of this paper, the important part is that we can quickly decode vectors from 4-bit to int8 with a 16-entry codebook, in-register using the \emph{vpshufb} shuffle instruction.

A goal of this paper is to develop fast kernels in which the distance computation relies on an integer inner product plus a small per-vector correction. 
Related quantizers such as RaBitQ, DRIVE, EDEN, \cite{vargaftik2021drive, vargaftik2022eden, benbasat2026eden} all fit this template and could be used instead.
We use the TurboQuant variant because it has fast encoding time, and after decoding, maps cleanly onto integer dot-product instructions on CPUs (VNNI) and GPUs (WGMMA); any quantization fitting the template plugs into the same kernels.

\section{\ouroldmethod{}}
\label{sec:pipnn}

\ouroldmethod{} (\emph{Pick-in-Partitions Nearest Neighbors}, KDD'26) is a recent graph construction algorithm designed to circumvent the build-side search bottleneck of most existing graph-based construction methods \cite{rubel2026pipnn}. Instead of performing expensive global beam searches to find neighbors, \ouroldmethod{} localizes the search by partitioning the dataset and generating graph edges within each partition.
\Cref{alg:pipNN} gives high-level pseudocode for \ouroldmethod{}~\cite{rubel2026pipnn}, which proceeds in three phases:

\emp{(1) Partition.} We build an \emph{overlapping} partition $\mathcal{B} = \{b_1, \dots, b_t\}$ of $\mX$ via recursive randomized ball carving until each leaf cluster $b_i$ satisfies $\Cmin \le |b_i| \le \Cmax$.
% TODO(laxman): briefly describe the num_clusters / "replicas" knob here
% (running the partitioning $\mathsf{nc}$ times with independent randomness
% and unioning the resulting candidate edges). Referenced in
% Fig.~\ref{fig:cpu-index-quality-main} and Fig.~\ref{fig:additional-qps}
% as "(2 replicas)".

\emp{(2) Pick.} For each leaf, we compute the within-leaf $k$-NN graph for a very small $k = 2$ and bi-direct it. Because points participate in multiple overlapping leaves, they accumulate a diverse set of candidate edges overall, despite the small local $k$.

\emp{(3) Prune.} Candidates from different leaves are streamed into a per-point reservoir of size $\lMax$ using \HashPrune{} to ensure directional diversity, followed by a final \RobustPrune{} pass for sparsification.

\begin{algorithm}
    \SetKwData{X}{$\mathcal{X}$}
    \SetKwData{BP}{$BP$}
    \SetKwData{G}{$G$}
    \SetKwData{B}{$\mathcal{B}$}
    \SetKwData{E}{$new\_edges$}
    \SetKwFunction{FBucket}{Partition}
    \SetKwFunction{FPrune}{Prune}
    \SetKwFunction{FPick}{Pick}
    \SetKwFunction{FaddG}{Add\_Edges}
    \SetKwFunction{FPaddG}{Prune\_And\_Add\_Edges}
    \SetKwFunction{FPruneNode}{PruneNode}
    \caption{\ouroldmethod{}}
    \label{alg:pipNN}
    \KwIn{\X a collection of points with dissimilarity measure, \BP a collection of hyper-parameters}
    \KwOut{\G$=(V,E)$ a graph over $\mathcal{X}$}
    G $\gets (\mathcal{X}, \varnothing)$\;
    $\mathcal{B}$ $\gets$ \FBucket{\X,\ \BP}\;\label{line:partition}
    \ForPar{$b_i \in  \mathcal{B}$}{
        $\mathsf{new\_edges}$ $\gets$ \FPick{$b_i$,\ \BP}\;\label{line:leafprune}
        G.\FPaddG{$\mathsf{new\_edges}$}\;\label{line:addedges}
    }
    
    \ForPar{$x \in$ \X}{
        \FPruneNode{$x$}\;\label{line:finalprune}
    }
    
    \Return \G\;
\end{algorithm}

Accelerating \ouroldmethod{} is the primary goal of this paper.
In total there are four pieces that dominate the runtime: \texttt{AssignLeaders} during Partition, \texttt{ProcessLeafTop2} during Pick, \HashPrune{}, and \texttt{RobustPrune}.
The main cost in the first two is many-to-many top-$k$ computations over small collections of vectors.

\myparagraph{The \texttt{AssignLeaders} Kernel}
\ouroldmethod{} partitions via \emph{Randomized Ball Carving (RBC)}: given a subproblem $\mP \subseteq \mX$, RBC samples a subset of $\Psamp \cdot |\mP|$ points as \emph{leaders}. Each point in $\mP$ is assigned to its $\fanout$ nearest leaders. This multi-assignment creates the overlap between subproblems. To control the growth of subproblems, $\fanout$ decays geometrically per level (typically from $10$ down to $3$, and finally $1$). RBC recurses on any subproblem exceeding size $\Cmax$.
The dominant cost of partitioning is the distance comparisons to score each point against all leaders and retain the top $\fanout$ leaders (\texttt{AssignLeaders} kernel).
%This per-level assignment---each point scoring itself against all leaders and retaining the top $\fanout$---is the dominant cost of partitioning.
%
%$\fanout \in \{1, 3, 10\}$ in practice, we specialize it to those three $K$ values .

\myparagraph{The \texttt{ProcessLeafTop2} Kernel}
Within each leaf $b_i$, we compute all $|b_i|^2$ pairwise distances and extract the top-$2$ neighbors per point.
Each point participates in multiple leaves (typically $30$), so a small per-leaf $k = 2$ is sufficient to deliver a diverse set of candidate neighbors~\cite{rubel2026pipnn}.
Because each pairwise comparison can symmetrically update the top-$2$ of \emph{both} endpoints, bi-directing these edges yields on average $4$ candidates per point per leaf.
\texttt{ProcessLeafTop2} is typically the bottleneck in \ouroldmethod{}.

\myparagraph{\HashPrune{}}
\HashPrune{} is an online procedure that pre-filters the set of candidate edges generated for each index point. %, in order to reduce space usage and speed up the following \RobustPrune{} step.
For each point $p$, the goal is to keep neighbors that are spread out in different directions.
To achieve this, for each point $p$ we maintain a reservoir $M_p$ of capacity $\lMax$ indexed by an $m$-bit hash.
For a candidate $c$, we define its hash $h_p(c) = \bigoplus_{i=1}^{\mstdeg} \mathbbm{1}[\mathcal{H}_i \cdot (c-p) \ge 0]$ for $m$ random hyperplanes $\mathcal{H}_i$,
where $\bigoplus$ denotes the concatenation of bits to form an $m$-bit signature.
Mathematically, this projects the direction of the residual vector $c-p$ onto $m$ hyperplanes, partitioning the space around $p$ into $2^m$ angular cones (sectors).
Two candidates $c, c'$ fall into the same cone (collide) with probability $(1 - \theta/\pi)^{\mstdeg}$, where $\theta$ is the angle between $c-p$ and $c'-p$.
The reservoir $M_p$ retains only the single nearest candidate that falls into each cone.
Crucially, the final state of the reservoir depends only on the set of candidates and not their arrival order. The PiPNN graph is deterministic, despite streaming candidates in parallel.
Pseudocode is in \Cref{app:hashprune}.
% The per-pair distance evaluations $\lVert p, c\rVert$ inside \HashPrune{} are batched in compressed form by our \texttt{Pruning} kernel, so the dot-product portion of \HashPrune{} runs at the same VNNI/WGMMA throughput as \texttt{AssignLeaders} and \texttt{ProcessLeafTop2}.

\myparagraph{\RobustPrune{}}
\RobustPrune{} selects a sparse out-neighbor set from a candidate list via a greedy occlusion rule~\cite{jayaram2019diskann}: it repeatedly picks the nearest remaining candidate $p^*$, adds it as a neighbor, then discards any candidate $p'$ satisfying $\alpha \cdot d(p^*, p') \le d(p, p')$---meaning $p^*$ already ``occludes'' $p'$ from $p$.
The parameter $\alpha \ge 1$ controls pruning aggressiveness; larger $\alpha$ retains more edges and yields denser, more navigable graphs.
\ouroldmethod{} applies a \RobustPrune{} pass over each reservoir to impose the \diskann{} $\alpha$-rule.
Because each reservoir is already small ($\le \lMax$ entries), this pass is relatively cheap compared to graph-based builders that invoke \RobustPrune{} on much larger candidate lists~\cite{rubel2026pipnn}.
%After optimizing \texttt{AssignLeaders} and \texttt{ProcessLeafTop2} however, we will see that \texttt{RobustPrune} becomes a significant bottleneck, which we address.

\myparagraph{Building $k$-NN Graphs with \ouroldmethod{}}
A by-product of the partitioning phase is that each point already accumulates a pool of nearby candidates within its leaves, which we can use to produce a $k$-NN graph directly.
Many prior graph-based approaches to building $k$-NN graphs take a \emph{build-then-search} approach: they first construct a full navigable index over $\mX$ and then issue a beam-search query for each base point, doubling the work and making the query pass dominant.
In this paper, we study using \ouroldmethod{} to retain the top-$k$ neighbors per point directly from its leaves and skip the search pass entirely.
%
%Our earlier work on graph partitioning for ANNS~\cite{gottesburen2024unleashing} sketched this idea in passing but never benchmarked or carefully studied it.
We show in \Cref{sec:experiments} that this strategy matches the recall of build-and-search at a fraction of the cost, and surpasses \textsc{NN-Descent} which is the most popular method for direct $k$-NN construction.

\section{Quantized and Fused Top-$K$ CPU Kernels}
\label{sec:cpu-kernels}

The primary bottleneck of \ouroldmethod{} is computing distances between query points and data points during top-$k$ assignments (either leaders during partitioning, or leafmates during leaf processing). 
Naively, this could be done by materializing a large distance matrix, which is costly due to writing and subsequently re-reading to post-process the distances, e.g., reducing to the top-$K$ neighbors.

To eliminate this bottleneck, \ourmethod{} \emph{fuses} the distance computation with the downstream reduction.
Registers with computed distances for a subset of the neighbors in the distance matrix are immediately consumed instead of being written out to an explicit distance matrix.
In this section we describe the matrix-multiplication structure (\Cref{sec:cpu-kernels:overview}), then the reduction kernels for partitioning (\Cref{sec:cpu-kernels:topk}), and finally leaf processing (\Cref{sec:cpu-kernels:leaf}).

\subsection{Vectorized Quantized MatMul Structure}
\label{sec:cpu-kernels:overview}

\myparagraph{Overview}
In this subsection we describe our core matrix multiplication primitive.
Before diving into the technical details, we give a high level overview.
Our goal is to enable \ourmethod{} to work directly over the quantized data.
The inputs to the kernel are TQ coordinates, which we decode on-the-fly with an \texttt{i8} codebook.
The loop structure of the matrix multiplication iterates over queries in the outer loop, database points in the middle loop and reduces on dimensionality in the inner loop to obtain the dot product between query and database point. 

For efficiency, we use SIMD vector instructions to compute partial dot products.
%
%SIMD (vector) registers can hold multiple arithmetic (e.g., \texttt{f32}, \texttt{i32}, \texttt{u8} or \texttt{i8}) values, and perform bulk operations on them in a single instruction.
%
A SIMD vector register consists of multiple \emph{lanes}, where each lane corresponds to one data item, e.g., a 512-bit AVX register holds 16 \texttt{i32} lanes which can be processed in parallel with a single instruction.
%
%
%A SIMD register holds and processes multiple numbers. floats, int32, or bytes. A register contains multiple lanes, e.g., a 512-bit register holds 16 floats, or 64 bytes. We call each item held a lane.
%
The performance driver of our code is a dot-product instruction from VNNI (Vector Neural Network Instructions) which requires us to make some adaptations to the data layout. 
We give a detailed description of our approach next.

\myparagraph{The Dot Product Instruction}
We leverage the $u8 \times i8$ \emph{quad} dot product instruction \texttt{vpdpbusd} (AVX-512 VNNI). 
For concreteness, the instruction computes:
\[
  \forall i \in [16] : \mathit{acc}[i] \mathrel{+}= \sum_{j=0}^{3} a[4i+j] \cdot b[4i+j]
\]
where $a[k] \in \text{u8}$, $b[k] \in \text{i8}$, $\mathit{acc}[i] \in \text{i32}$, and $i$ iterates over the 16 lanes in a 512-bit register.
Given two 64-byte registers of \texttt{u8} and \texttt{i8}, it computes 16 parallel 4-dimensional dot products in a single instruction.
We use Highway's \texttt{SumOfMulQuadAccumulate} function, which uses \texttt{vpdpbusd} when available and emulates it via widen + pairwise-add instructions (\texttt{vpmaddubsw}) on earlier targets (AVX2) \cite{google_highway}.
We note that unfortunately there is no equivalent $\text{i8} \times \text{i8}$ instruction.

\myparagraph{Multiplying Signed Bytes}
Because the \texttt{vpdpbusd} instruction only supports multiplying unsigned bytes (\texttt{u8}) by signed bytes (\texttt{i8}), we cannot directly multiply the signed codes for both the query and data points. % $\tilde a \in [-128, 127]$.
We resolve this by shifting database codes by $+128$ to make them unsigned, multiplying unsigned codes by signed codes, then subtracting the error from the accumulated inner products. 

Specifically, let $\tilde{a}_i$ be the original signed database byte, $b_i$ be the signed query byte, and $a_i = \tilde{a}_i + 128 \in [0, 255]$ be the shifted unsigned database byte.
We recover the true signed inner product $\sum \tilde{a}_i \cdot b_i$ by evaluating $\sum a_i \cdot b_i - 128 \sum b_i$.

\begin{figure}[t]
  \centering
  \begin{tikzpicture}[
      font=\small,
      x=1mm, y=1mm,
      >={Latex[length=1.5mm,width=1.1mm]},
      box/.style       ={draw=black!70, rounded corners=2pt,
      inner sep=2.2pt, align=center},
      cell/.style      ={draw=black!55, minimum width=3.6mm, minimum height=3.4mm,
      inner sep=0, anchor=center, font=\tiny},
      dbcell/.style    ={cell, fill=teal!9},
      qcell/.style     ={cell, fill=blue!7},
      acccell/.style   ={cell, fill=orange!12, minimum width=4.6mm, minimum height=3.8mm},
      acchot/.style    ={cell, fill=orange!30, draw=orange!60!black,
      minimum width=4.6mm, minimum height=3.8mm},
      region/.style    ={draw=black!40, dashed, rounded corners=3pt, inner sep=2.5pt},
      reglabel/.style  ={font=\sffamily\bfseries\scriptsize, anchor=west,
      fill=white, inner sep=1pt},
      lab/.style       ={font=\scriptsize, align=left},
      arr/.style       ={->, line width=0.5pt, draw=black!35},
    ]

    % ============================================================
    % (1) DATABASE  row-major codes  -> pre-transposed u8 panel tile
    %     distinct points interleaved, one 4-byte group per lane
    % ============================================================
    \matrix[matrix of nodes, nodes={dbcell}, column sep=-0.4pt,
    ampersand replacement=\&] (db0) at (14,66)
    { $31$ \& $\text{-}18$ \& $\text{-}6$ \& $44$\\ };
    \node[lab, anchor=east, font=\tiny] at (db0.west) {$x_0$};
    \node[font=\tiny] at (14,61.5) {$\vdots$};
    \matrix[matrix of nodes, nodes={dbcell}, column sep=-0.4pt,
    ampersand replacement=\&] (db15) at (14,57)
    { $\text{-}44$ \& $31$ \& $18$ \& $\text{-}18$\\ };
    \node[lab, anchor=east, font=\tiny] at (db15.west) {$x_{15}$};
    \begin{scope}[on background layer]
      \node[fit=(db0-1-1)(db0-1-4), draw=teal!55!black, line width=0.5pt,
      inner sep=0pt, rounded corners=1pt] (db0hot){};
      \node[fit=(db15-1-1)(db15-1-4), draw=teal!55!black, line width=0.5pt,
      inner sep=0pt, rounded corners=1pt] (db15hot){};
    \end{scope}

    \matrix[matrix of nodes, nodes={dbcell}, column sep=-0.4pt,
    ampersand replacement=\&] (P) at (20,46)
    { $159$\&$110$\&$122$\&$172$\& $\cdots$ \& $84$\&$159$\&$146$\&$110$\\ };
    \begin{scope}[on background layer]
      \node[fit=(P-1-1)(P-1-4), draw=teal!55!black, fill=teal!28,
      inner sep=0pt, rounded corners=1pt] (Px0){};
      \node[fit=(P-1-6)(P-1-9), draw=teal!55!black, fill=teal!28,
      inner sep=0pt, rounded corners=1pt] (Px15){};
    \end{scope}
    \node[font=\tiny, below=0.2mm of P-1-2.south] {$x_0$};
    \node[font=\tiny, below=0.2mm of P-1-8.south] {$x_{15}$};

    \draw[arr] (db0hot.south)  .. controls +(0,-3) and +(0,3)  .. (Px0.north);
    \draw[arr] (db15hot.south) .. controls +(7,-3) and +(0,3) .. (Px15.north);
    \node[lab, anchor=west, font=\tiny] (ileave) at (22,63)
    {$+128$ bias,\\interleave\\$L$ pts};
    \node[lab, anchor=north, font=\tiny] (panlab) at (20,41.5)
    {one $W=64$ byte load};

    % ============================================================
    % (2) QUERIES  row-major decoded i8 -> broadcast tile
    %     ONE query replicated into every lane (mirror of the panel)
    % ============================================================
    \matrix[matrix of nodes, nodes in empty cells, nodes={qcell},
      row sep=-0.4pt, column sep=-0.4pt,
    ampersand replacement=\&] (Q) at (54,61.5)
    {
      $5$\&$1$\&$\text{-}3$\&$2$\\
      \& \& \& \\
      \& \& \& \\
    };
    \node[font=\tiny, left=0.6mm of Q-1-1] {$q_1$};
    \node[font=\tiny, left=0.6mm of Q-2-1] {$q_2$};
    \node[font=\tiny, left=0.6mm of Q-3-1] (q3lab) {$\vdots$};
    \begin{scope}[on background layer]
      \node[fit=(Q-1-1)(Q-1-4), draw=blue!55!black, line width=0.5pt,
      inner sep=0pt, rounded corners=1pt] (qword){};
    \end{scope}

    \matrix[matrix of nodes, nodes={qcell}, column sep=-0.4pt,
    ampersand replacement=\&] (QT) at (60,46)
    { $5$\&$1$\&$\text{-}3$\&$2$\& $\cdots$ \& $5$\&$1$\&$\text{-}3$\&$2$\\ };
    \begin{scope}[on background layer]
      \node[fit=(QT-1-1)(QT-1-4), draw=blue!55!black, fill=blue!22,
      inner sep=0pt, rounded corners=1pt] (Qln0){};
      \node[fit=(QT-1-6)(QT-1-9), draw=blue!55!black, fill=blue!22,
      inner sep=0pt, rounded corners=1pt] (Qln15){};
    \end{scope}

    \draw[arr] (qword.south) .. controls +(-2,-3) and +(0,3) .. (Qln0.north);
    \draw[arr] (qword.south) .. controls +(9,-3)  and +(0,3) .. (Qln15.north);
    \node[lab, anchor=west, font=\tiny] (bcastop) at (62,63)
    {broadcast\\$q$'s $4$B,\\$M_q{=}6$};
    \node[lab, anchor=north, font=\tiny] (qtilelab) at (60,41.5)
    {$q$ replicated in every lane};

    % ============================================================
    % (3) FUSED VNNI SCORE  -> accumulator + top-K
    % ============================================================
    \node[box, fill=orange!9, minimum width=72.4mm, align=center] (vnni) at (40,33)
    {\texttt{vpdpbusd}: per lane $j$, \;$x_j \!\cdot\! q$ \;
    ($4\times$\texttt{u8} $\cdot$ $4\times$\texttt{i8} $\to$ \texttt{i32})};

    \node[box, draw=black!45, fill=black!3, font=\tiny, align=left,
    inner sep=2.5pt] (num) at (25,19)
    {%
      quad $0$: $a{=}[159,110,122,172]$,\, $b{=}[5,1,\text{-}3,2]$\\[1pt]
      $\textstyle\sum_k a_k b_k = 883$;\quad $\tilde a{=}a{-}128{=}[31,\text{-}18,\text{-}6,44]$\\[1pt]
      $883 = \underbrace{243}_{\sum_k \tilde a_k b_k}
      + \underbrace{640}_{128\sum_k b_k}$
    \;$\to$\; subtract $128\,S_q$};

    \draw[arr] (vnni.south) .. controls +(-10,-2) and +(0,4) .. (num.north);

    % explicit i32 output: one dot product per panel point (lane)
    \node[lab, anchor=south, font=\tiny, align=center] (ahead) at (62,23)
    {\texttt{i32} output $A_j=\langle x_j,\,q\rangle$};
    \matrix[matrix of nodes, nodes={acccell, minimum width=8mm},
    column sep=-0.4pt, ampersand replacement=\&] (A) at (62,19)
    { $243$ \& $\cdots$ \& $\text{-}279$\\ };
    \begin{scope}[on background layer]
      \node[fit=(A-1-1), draw=orange!60!black, fill=orange!30,
      inner sep=0pt, rounded corners=1pt] {};
    \end{scope}
    \node[font=\tiny, below=0.3mm of A-1-1.south] {$x_0$};
    \node[font=\tiny, below=0.3mm of A-1-3.south] {$x_{15}$};
    \node[lab, anchor=north, font=\tiny] (anote) at (62,14)
    {$\to$ rescale, top-$K$};
    \draw[arr] (num.east) -- (A.west);

    % ============================================================
    % region outlines + headers
    % ============================================================
    \begin{scope}[on background layer]
      \node[region, fit={(db0)(db15)(P)(Px0)(Px15)(ileave)(panlab)}] (rDB) {};
      \node[region, fit={(Q)(QT)(Qln0)(Qln15)(bcastop)(qtilelab)
      (rDB.north -| QT)(rDB.south -| QT)}] (rQ) {};
      \node[region, fit=(vnni)(num)(A)(ahead)(anote)] (rV) {};
    \end{scope}
    \node[reglabel] at ([xshift=2mm]rDB.north west)
    {\circnum{1}\, Transpose to \texttt{u8} quads};
    \node[reglabel] at ([xshift=2mm]rQ.north west)
    {\circnum{2}\, Broadcast 4 query bytes};
    \node[reglabel] at ([xshift=2mm]rV.north west)
    {\circnum{3}\, Fused VNNI score $\to$ top-$K$};

  \end{tikzpicture}
  \caption{\textbf{16 quads in a panel of the CPU GEMM kernel} (AVX-512,
    $L=16$). \circnum{1}~The database side is first decoded from TurboQuant to u8 using i8 + 128 codebook, then transposed to a $4x16$ \texttt{u8} tile.
    \circnum{2}~The query side
    stays row-major and decoded to i8; the matching $4$-byte word of one query is
    \emph{broadcast} into all $L$ lanes, then recast to $i8$.
    \circnum{3}~One \texttt{vpdpbusd} multiplies the two tiles lane by
    lane, folding $4\,L$ byte products into $L$
    \texttt{i32} lanes.
    After processing all dimensions, lane 0 recovers the
    true signed dot product $243$ for datapoint 0 from the unsigned-by-signed value $883$ via
    the $128\,S_q$ byte-sum correction (lane $15$ gives $-279$); the epilogue
    then rescales by $\sfac_q$ and $\sfac_d$.}
  \label{fig:cpu-kernel}
\end{figure}
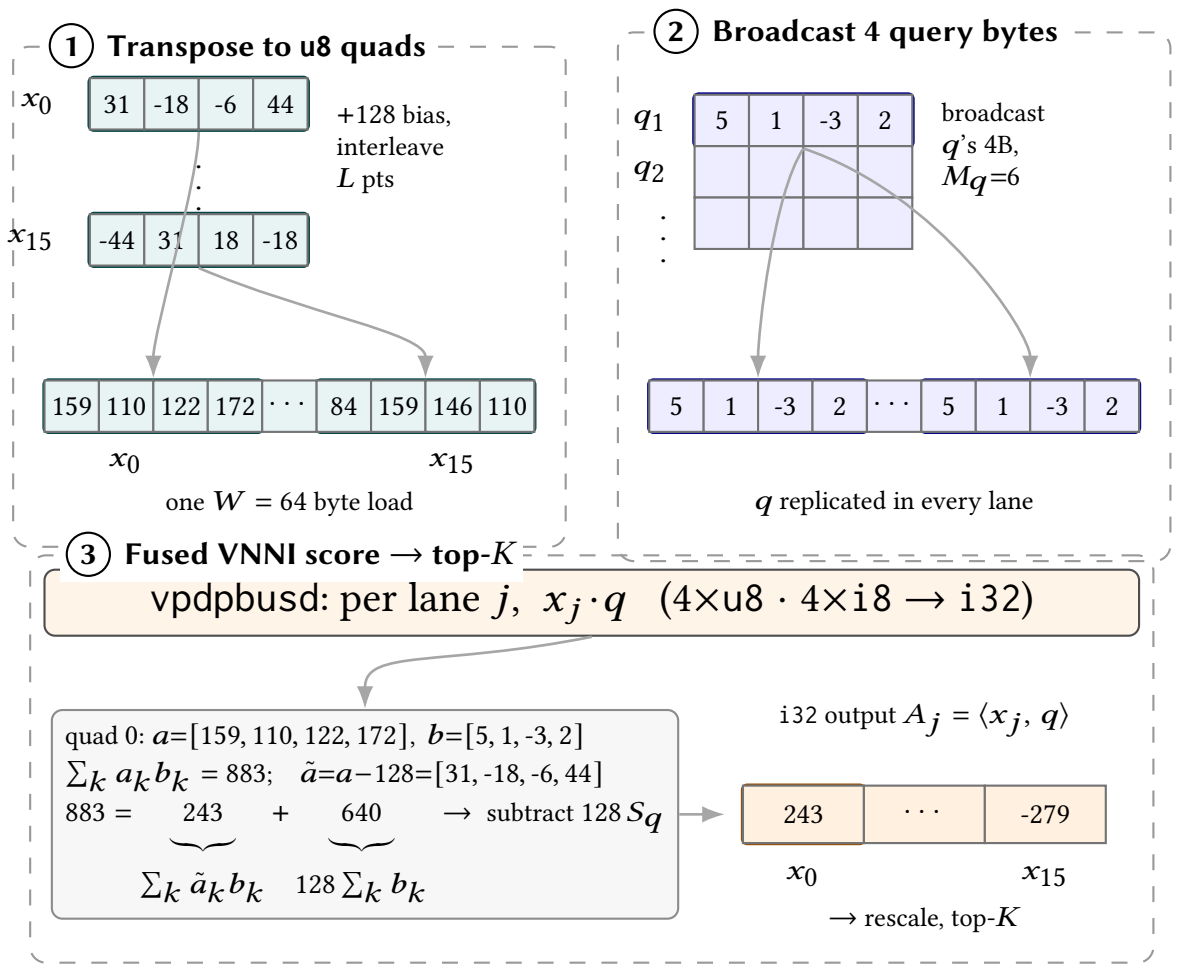

\myparagraph{Data Layout for Vectorization}
The quad instruction computes 16 independent 4-dimensional dot products, one per \texttt{i32} lane, so scoring 16 database points at once requires each lane to hold a different point.
A row-major layout stores each point's coordinates contiguously, which spreads a single point across the lanes rather than 16 distinct points, forcing a costly horizontal reduction per point and one point per instruction.
We instead interleave the database into contiguous blocks called \emph{panels}: a panel groups $L = 16$ database points and a panel tile stores one 4-dimension (quad) coordinate group from each, so 64 consecutive bytes hold the same 4 dimensions across all 16 points (\Cref{fig:cpu-kernel}).
A panel tile is then scored against a broadcasted query quad in a single instruction, yielding 16 partial dot products, one per lane.

\myparagraph{Loop Structure}
We want to perform the inner reduce on the dimensions so that we materialize a finalized dot product after the loop.
Hence, we do not have to spill partial dot products to memory before down-stream consumption in the top-$K$ aggregation.
This motivates our loop structure to process queries in the outer loop, database points in the middle, and dimensions in the inner loop, as shown in Algorithm~\ref{alg:gemm-asymmetric-dot}.
We omit cache tiling as it slightly hurts performance: the inputs are small enough to remain L2-resident.
Note that the pseudocode shows the process for computing negative dot products as the distance.
When we use L2 distance, we precompute the squared norms of the datapoints to compute L2 from dot product.

\myparagraph{Unrolling}
Unrolling over $M_q = 6$ queries at a time amortizes the cost of streaming over the database points repeatedly and exposes instruction-level parallelism; we found $M_q = 6$ to perform best.
Unrolling over multiple panels at a time (we found four to perform best) amortizes the cost of loading and broadcasting the 4 query dimensions: because all four panels share the same broadcast of four query bytes, one broadcast instruction serves four \texttt{vpdpbusd} instructions instead of one.
In the inner loop we thus perform $4 \times M_q = 24$ \texttt{vpdpbusd} instructions per iteration.
For simplicity, unrolling is not shown in~\cref{alg:gemm-asymmetric-dot}.

%\myparagraph{Unrolling}
%%
%Furthermore, query broadcasts dispatch only to port 5 on the CPU, which would bottleneck the compute pipeline.
%%
%To prevent this, we unroll the scoring loop along both axes, processing $M_q = 6$ queries and $N_p = 4$ database panels (64 points) simultaneously. This $6 \times 4$ unrolling tile reuses each database load across 6 queries, and each query broadcast across 4 panels, thus utilizing the 32-register AVX-512 file without register spilling.
%\lars{register file has 188 entries. register file is not the same as physical registers, but used for register renaming by the compiler. we have no idea if register spilling is happening and we should not be claiming so. also 24 is not the number of registers we use here. it should be 6 for query, 4 for DB, 6 for accumulators plus backups for ILP}
%\myparagraph{Memory Bandwidth of Intermediate Distances}
%Finally, writing the intermediate $M \times N$ similarity matrix to memory and reading it back for Top-$K$ selection dominates cache traffic\lars{not validated...}.
%We eliminate this by fusing candidate generation and Top-$K$ selection. As shown in Algorithm~\ref{alg:gemm-asymmetric-dot}, our loop order (Queries outer, Panels middle, Dimensions inner) ensures that finished scores are materialized directly in registers, which we immediately forward to the Top-$K$ aggregation state\lars{state??} without touching memory.\lars{we do touch memort in leaf processing...}

\begin{algorithm}
  \caption{Simplified Fused GEMM Loop}
  \DontPrintSemicolon
  \label{alg:gemm-asymmetric-dot}

  \KwIn{%
    Queries $\mathcal{Q}$, Transposed Panels $\mathcal{P}$, Scaling factors $\sfac_q, \sfac_d$\;
  }
  \KwOut{%
    Fused update to Top-$K$ state\;
  }

  \For{$q = 0$ \KwTo $N_q-1$}{
    $\mathbf{q_{vec}} \gets \mathcal{Q}[q]$;\;
    $\mathbf{q_{bias}} \gets \sum_k 128 \cdot \mathbf{q_{vec}}[k]$;\;
    \For(){$p = 0$ \KwTo $\lceil N_d / L \rceil -1$}{
      $\mathbf{acc} \gets \mathbf{0}_{L}$\;

      \For(\tcp*[f]{Reduce dimensions in 4-byte blocks}){$t = 0$ \KwTo $\lceil d / 4 \rceil - 1$}{
        $\mathbf{a} \gets \mathrm{Load}_{\text{SIMD}}(\mathcal{P}[p][t])$\;
        $\mathbf{q_v} \gets \mathrm{Broadcast}(\mathbf{q_{vec}}[4t \dots 4t+3])$\;
        $\mathbf{acc} \mathrel{+}= \mathrm{SumOfMulQuadAccumulate}(\mathbf{a},\, \mathbf{q_v})$\;
      }

      $\mathbf{acc} \gets \mathbf{acc} - \mathrm{Broadcast}(\mathbf{q_{bias}})$\tcp*[r]{u8 bias correction}
      $\mathbf{acc} \gets \mathrm{ConvertToFloat32}(\mathbf{acc}) \odot (-\boldsymbol{\sfac}_d) \cdot \sfac_q[q]$
      $\mathrm{UpdateTopK}(\mathit{state}[q],\, \mathbf{acc}, \, \text{panel\_indices}(p))$\;
    }
  }
\end{algorithm}

% For RSQ8, we apply an in-place $\texttt{XOR } 0x80$ as we transpose, reinterpreting the raw \texttt{i8} code as the $+128$-shifted \texttt{u8} the instruction wants.
% For TQ4, the centroid table is stored in \texttt{u8} form for the same reason.
% In either case, the byte-sum correction handles the reinterpretation algebraically.

The kernels described next share this GEMM core but differ in the epilogue: \Cref{sec:cpu-kernels:topk} fuses an asymmetric top-$K$ reduction, while \Cref{sec:cpu-kernels:leaf} exploits score symmetry to halve the work.
The same GEMM core is reused for the final \RobustPrune{} pass, evaluated in \Cref{sec:cpu-ablations}.

\subsection{\texttt{AssignLeaders} Kernels}
\label{sec:cpu-kernels:topk}

During each level of partitioning, each point must be assigned to its nearest leader(s)---a batched nearest-neighbor query that the \texttt{AssignLeaders} step performs as an asymmetric GEMM with per-query top-$K$ filtering.
The database side (the leaders) is small, so we can pretranspose the leaders once and amortize the cost over many queries.
Moreover we can simply parallelize over cache-sized chunks of queries and call the single-threaded GEMM kernel.
%As a micro-optimization, when we cluster with dot product (not L2), we can omit the query-scaling factor $\sfac_q$, since we only care about the ranking of the leaders.

We inherit the fanout recipe from \ouroldmethod{}: the number of leaders each point is assigned to is $\fanout = 10$ on the first level, $3$ on the second, and $1$ on deeper levels.
This enables us to specialize the top-$K$ aggregation to these specific values.

\myparagraph{Top-1 ArgMin}
For $K=1$, we use a \emph{vector-of-indices} approach to ArgMin, which operates in a single pass over the scores of a query without spilling to memory.
A score accumulator holds the scores of one query for $L$ database points in its lanes.
We track the per-lane minimum score and database index in registers.
Given the score accumulator for the next panel, we compare lane-wise to obtain a mask and blend (\textrm{IfThenElse}) the new scores and indices into the running tracker.
After all panels, we perform a horizontal reduction across lanes to obtain the actual minimum.

\myparagraph{Displacement sort ($K \in \{3, 10\}$)}
For $K=3$ and $K=10$ we generalize the ArgMin approach to a \emph{displacement sort} (\cref{alg:simd-displacement-sort}).
The key idea is to track the top-$K$ per lane independently, deferring the expensive horizontal merge until all panels are processed; this is faster than vectorized QuickSelect for small $K$.
We maintain a displacement register $\mathbf{d}$, initialized with the incoming scores.
In each of $K$ iterations, a lane-wise comparison of $\mathbf{d}$ with the current top-$i$ entry produces a mask, where $i$ is the current iteration.
Four conditional blends per iteration swap scores and indices between the tracker and the displacement register.
If the incoming score beats the $i$-th best, the displaced entry is carried in $\mathbf{d}$ into iteration $i{+}1$, where it may in turn displace the $(i{+}1)$-th best.
After $K$ iterations, $\mathbf{d}$ holds the element that was pushed out of the top-$K$.
With query unrolling, this still fits in registers for $K=3$, but for $K=10$ the tracker state spills to L1; this is acceptable since the L1 round-trip is negligible relative to the GEMM work and $K=10$ is called for $n$ queries, whereas $K=3$ is called for $10n$ queries.

After the loop over all panels, each of the $K$ tracker registers holds $L$ independent per-lane candidates.
We dump the $K \cdot L$ scores to a stack buffer and resolve to a per-query top-$K$ with scalar insertion sort.
The scalar work is minimal compared to the GEMM and SIMD top-$K$ work.

\begin{algorithm}[h]
  \caption{SIMD Displacement Sort Insertion}
  \DontPrintSemicolon
  \label{alg:simd-displacement-sort}

  \KwIn{%
    Score vector $\mathbf{d} \in \mathbb{R}^L$\;
    Index vector $\mathbf{d}_{\mathrm{idx}} \in \mathbb{N}^L$\;
  }
  \KwData{%
  Tracker score registers $\mathbf{s}_i \in \mathbb{R}^L$ for $i \in \{0, \dots, K-1\}$ (updated in place)\;
  Tracker index registers $\mathbf{i}_i \in \mathbb{N}^L$ for $i \in \{0, \dots, K-1\}$ (updated in place)\;
  }

  \For{$i = 0$ \KwTo $K-1$}{
    $\mathbf{mask} \gets \mathbf{d} < \mathbf{s}_i$ \tcp*[r]{per-lane comparison}
    $\mathbf{s_{new}} \gets \mathrm{IfThenElse}(\mathbf{mask}, \mathbf{d}, \mathbf{s}_i)$\;
    $\mathbf{i_{new}} \gets \mathrm{IfThenElse}(\mathbf{mask}, \mathbf{d}_{\mathrm{idx}}, \mathbf{i}_i)$\;
    $\mathbf{d} \gets \mathrm{IfThenElse}(\mathbf{mask}, \mathbf{s}_i, \mathbf{d})$\;
    $\mathbf{d}_{\mathrm{idx}} \gets \mathrm{IfThenElse}(\mathbf{mask}, \mathbf{i}_i, \mathbf{d}_{\mathrm{idx}})$ \;
    $\mathbf{s}_i \gets \mathbf{s_{new}}$\;
    $\mathbf{i}_i \gets \mathbf{i_{new}}$\;
  }
\end{algorithm}

% \myparagraph{Assign-to-leaders, top-$K$ for $K \in \{3, 10\}$}
% Per-query state is $K$ vectors of $(\texttt{score}, \texttt{idx})$
% maintained in sorted (ascending) order using a branchless SIMD
% insertion network: for each incoming score, we compare against
% positions $0, 1, \ldots, K-1$ in turn and bubble the displaced entry
% down using \texttt{IfThenElse}. The state is $2K \cdot \texttt{Mq}$
% vectors; combined with the $4 \cdot \texttt{Mq}$ accumulators of the
% inner loop and a few broadcasts, this fits in the register file for
% both $K = 3$ and $K = 10$.

% The cost of $K$ shows up only in the final cross-lane reduction. Each
% of the $K$ stored vectors holds $\kpoints$ per-lane candidates that
% must be merged across lanes into a single per-query top-$K$.
% For $K = 3$ on AVX-512 ($\kpoints = 16$) this is a $48$-element
% scalar top-$3$ per query, which compiles to a short straight-line
% sequence; for $K = 10$ it is a $160$-element scalar top-$10$ per
% query, which is where the ``gory scalar'' code lives. The scalar
% work is $O(K \cdot \kpoints)$ per query while the SIMD GEMM is
% $O(N \cdot d / 4)$ per query, so for any leader set of meaningful
% size the SIMD fraction continues to dominate. We confirm this in
% \Cref{sec:experiments}.

\subsection{The Leaf Kernel}
\label{sec:cpu-kernels:leaf}

The leaf processing step computes a self-join: every point in a leaf serves as both query and database entry, and the kernel finds the top-2 nearest neighbors for each point.
Because the similarity relation is symmetric, computing both $d(i, j)$ and $d(j, i)$ is redundant.
To halve the distance computations, we restrict our GEMM to the lower triangle of the score matrix and update the top-2 trackers of both points symmetrically.
Concretely, for a query batch of $M_q$ consecutive points starting at index $q_0$, the kernel only sweeps panels $p < \lceil (q_0 + M_q) / L \rceil$.
For each computed score $s(i, j)$, we scatter updates to the top-2 states of both query $i$ and database point $j$.

Let $\mathbf{s} \in \mathbb{R}^L$ denote the vector of rescaled distances
between query~$q$ and the $L$ database points in panel~$p$, occupying
positions $b, b{+}1, \ldots, b{+}L{-}1$ where $b = p \cdot L$.
We process~$\mathbf{s}$ in three stages.

\myparagraph{Masking}
To enforce the lower-triangular computation and exclude out-of-bounds entries,
we set $\mathbf{s}[\ell] \gets +\infty$ for every lane~$\ell$ satisfying
$b + \ell \geq q$ (eliminating self-distances and pairs already accounted
for when $b + \ell$ served as a query) or $b + \ell \geq n$ (padding
lanes beyond the point set).
After masking, each finite entry $\mathbf{s}[\ell]$ corresponds to a
unique unordered pair $\{q,\, b{+}\ell\}$ with $b + \ell < q$.%
% \tobias{TODO: please confirm the mask direction against the kernel source --- with the panel sweep $p < \lceil (q_0+M_q)/L \rceil$, masking $b+\ell \ge q$ retains the lower triangle $b+\ell < q$.}

\myparagraph{Symmetric scatter}
We exploit the symmetry $d(q, p) = d(p, q)$ by updating two independent
top-2 trackers with the same score vector.
The \emph{query-side tracker} $T_q$ stores scores in registers $\mathbf{s}_0, \mathbf{s}_1 \in \mathbb{R}^L$ and their database-point indices in $\mathbf{i}_0, \mathbf{i}_1 \in \mathbb{N}^L$; it tracks the two smallest scores for query $q$ and resides entirely in SIMD registers throughout the panel loop, incurring no memory traffic.
The \emph{database-side tracker} $T_d[b..b{+}L{-}1]$ maintains the analogous state for the $L$ database points in panel $p$.
As it is shared across query batches, it is loaded from memory once per panel, updated in-register by all $M_q$ queries in the current batch, and stored back once the batch completes.
The unrolling optimization for GEMM benefits the memory traffic here as well.
Both trackers are updated via the branchless two-comparison cascade described next.
Note that $T_d$ uses one lane per database point, updated with the same query id, whereas $T_q$ uses $L$ lanes for one query point updated with $L$ database points.

\myparagraph{Lane-parallel top-2 insertion}
Given the tracker registers $\mathbf{s}_0, \mathbf{s}_1$ and $\mathbf{i}_0, \mathbf{i}_1$ for the running top-2, and an incoming score vector $\mathbf{s}$ and index vector $\mathbf{idx}$, the insertion proceeds independently per lane.
We first compute two Boolean masks:
$\mathbf{mask}_0 \gets \mathbf{s} < \mathbf{s}_0$ (indicating new global bests) and $\mathbf{mask}_1 \gets \neg \mathbf{mask}_0 \wedge (\mathbf{s} < \mathbf{s}_1)$ (indicating new second-bests).
The second-best registers are updated first: if the incoming score beats the current best ($\mathbf{mask}_0$), the old best is demoted to second; if it beats only the second-best ($\mathbf{mask}_1$), it replaces the second directly:
\begin{align}
  \mathbf{s}_1 &\gets \mathrm{IfThenElse}(\mathbf{mask}_0,\; \mathbf{s}_0,\; \mathrm{IfThenElse}(\mathbf{mask}_1,\; \mathbf{s},\; \mathbf{s}_1)),
  \label{eq:s2} \\
  \mathbf{i}_1 &\gets \mathrm{IfThenElse}(\mathbf{mask}_0,\; \mathbf{i}_0,\; \mathrm{IfThenElse}(\mathbf{mask}_1,\; \mathbf{idx},\; \mathbf{i}_1)),
  \notag
\end{align}
followed by the first-best registers:
\begin{align}
  \mathbf{s}_0 &\gets \mathrm{IfThenElse}(\mathbf{mask}_0,\; \mathbf{s},\; \mathbf{s}_0),
  \label{eq:s1} \\
  \mathbf{i}_0 &\gets \mathrm{IfThenElse}(\mathbf{mask}_0,\; \mathbf{idx},\; \mathbf{i}_0).
  \notag
\end{align}
Note that \cref{eq:s2} must precede \cref{eq:s1}: when $\mathbf{mask}_0$ is true, the old $\mathbf{s}_0$ must be demoted into $\mathbf{s}_1$ before being overwritten.
All four assignments are branchless $\mathrm{IfThenElse}$ blend operations executing in parallel across all lanes without cross-lane communication.
Note that this uses six blend instructions: two less than the generic top-2.

\myparagraph{Cross-lane reduction}
After the final panel, the $L$~lanes of~$T_q$ hold $L$~independent
top-2 pairs.
We extract all $2L$ score--index candidates and move them into the database-side tracker using scalar top-2 insertion.
The database-side tracker serves as the output.

\section{Accelerating \ouroldmethod{} on GPUs}
\label{sec:gpu-kernels}

\myparagraph{Overview}
GPUs are highly effective on tasks where many threads can operate on dense data with regular, contiguous memory access patterns. They provide high throughput for tiled matrix products, bulk copies, reductions, and global sorts. On the other hand, they do poorly on workloads of widely varying size, on recursive calls, and on lock-heavy data structures.

As on the CPU, the dominant cost for our GPU implementation of \ourmethod{} ought to be top-$K$ computation. GPUs can perform bulk top-$K$ very efficiently, but the rest of the work performed in \ourmethod{} --- recursive partitioning, \HashPrune{} reservoirs with locks, and the final \RobustPrune{} --- are difficult to implement performantly. %\richard{This sounds a bit contradictory. We say that the dominant cost is top-K but then that it's everything else that's hard to make efficient.}

\myparagraph{GPU background}
GPUs are built from many \emph{streaming multiprocessors} (SMs), the general-purpose cores that run threads. 
Each SM has a register file ($64$K $32$-bit registers on the H100 we use).
%, the fastest storage on the chip\lars{Isn't this just SRAM?}\tobias{Kind of, it is made out of SRAM but differs from shared memory.}.
%, in which a thread keeps its operands and running top-$K$ state.
It also has a small \emph{shared-memory} scratchpad, shared by a group of cooperating threads, that we use to stage data that is read frequently.
For matrix multiplication, the SM provides \emph{tensor cores}, units that multiply and accumulate a small tile in a single instruction.
Two \emph{asynchronous copy engines} move data into shared memory without occupying the SM's compute units so that copies overlap with computation: \texttt{cp.async} performs ordinary per-thread copies from global memory, whereas the Tensor Memory Accelerator (\texttt{TMA}) copies a whole tile characterized by a copy descriptor containing the base address, shape, stride, and swizzle to be performed.
%\lars{more specific. is it a single pointer? can you do disaggregated storage.}
Software groups threads at three levels: a \emph{warp} is $32$ threads that execute in lockstep, the GPU analogue of a SIMD register; a \emph{warpgroup} is four warps ($128$ threads) that together issue one tensor-core instruction, \texttt{wgmma}; and a \emph{thread block} is one or more warps scheduled onto a single SM, sharing its registers and shared memory. We call threads operating within a warp \emph{lanes}.

With this in mind, we change a number of \ouroldmethod{}'s algorithmic and implementation details to map it onto the GPU.
% SIGMOD submission: use the acmart list geometry and spacing defaults.
% \begin{enumerate}[label=(\arabic*),topsep=0pt,itemsep=0pt,parsep=0pt,leftmargin=15pt]
\begin{enumerate}
    \item We flatten the recursion into a fixed number of assignment-and-sort rounds (\Cref{sec:gpu-kernels:flatten}).
    \item We regularize bucket sizes during partitioning and leaf building into fixed-size work units called \emph{score tiles} (\Cref{sec:gpu-kernels:uneven}). Each scores a group of queries against a block of database points and maps one-to-one onto a thread block.
    \item We compute every distance through a single top-$K$ primitive that scores directly over the vectors without ever materializing a score matrix (\Cref{sec:gpu-kernels:skeleton}). This is \emph{segmented}, in the sense that it is one kernel that computes a top-$K$ over many independent subproblems of varying size at once.
    \item We replace the lock-based \HashPrune{} with sort-and-merge accumulation (\Cref{sec:gpu-kernels:hashprune}). This requires a transient buffer bounded by the per-batch slab cap rather than by the full candidate set (\Cref{app:gpu-prune-detail}).
    \item We replace \RobustPrune{} with a relaxation, \DirectEdgePrune{}, that reuses already-witnessed distances rather than recomputing them (\Cref{sec:gpu-kernels:direct-prune}). While \DirectEdgePrune{} produces denser graphs than \RobustPrune{} (\Cref{tab:prune_ablation}), we find that this does not affect index quality.
\end{enumerate}
%
%\tobias{Need to add a forward pointer to supplemental result here (not sure if we actually have this in paper)}.

% \ourgpumethod{} keeps the candidate source---near-neighbor witnesses inside
% overlapping local regions---and changes the representation. It flattens
% recursive buckets into sorted segment arrays, turns uneven buckets into
% bounded leaf batches, routes dense comparisons through one segmented top-$K$
% primitive, replaces synchronized \HashPrune{} updates with sort-and-merge,
% and replaces candidate-distance recomputation in the final prune with
% direct-edge tests. \Cref{sec:gpu_ann_build,sec:gpu_throughput} evaluate the
% resulting build times, query quality, and kernel behavior. The supplement
% contains the lower-level layout and kernel details.

\subsection{Flattening Recursion}
\label{sec:gpu-kernels:flatten}
\label{sec:gpu-kernels:partitioning}

% \ouroldmethod{} creates overlap by recursively assigning each point to
% one or more leader points. 
% Because recursion is a poor fit for GPUs, we replace the recursive tree with a fixed number of assignment passes separated by sorts.
\ouroldmethod{} partitions points by recursively assigning each point to one or more leader points. Because recursion is a poor fit for GPUs, we replace the explicit recursion with a fixed number of assignment passes. In each pass, we assign points to their nearest leaders within a given subproblem, then sort by the induced keys. This gives us the same partitioning structure without explicit recursion.

More formally, we first sample top-level leaders, assign every point to its $f_0$ nearest leaders, then sort the memberships by leader so that each top-level bucket is a contiguous run.
Second, within each top-level bucket, we sample bucket-local sub-leaders, assign each member to its $f_1$ nearest sub-leaders, and sort again to expose the second-level buckets as contiguous runs.
We set the \ourgpumethod{} fanout to $f_0 = 8$ at the top level and $f_1 = 4$ at the second, so each point appears in up to $f_0 \cdot f_1 = 32$ leaf memberships, comparable to the roughly $30$ in \ourmethod{} (\Cref{sec:cpu-kernels:topk}).
Each round costs a device-wide sort. Additional assign-then-sort rounds are possible, but two were adequate to achieve high recall.
% \richard{(8, 4) doesn't really strike me as higher per-level fanout than (10, 3).}
\Cref{fig:max-depth-ablation} reports the effect of recursion depth on index quality.

% After these passes we no longer manipulate a recursion tree but flat arrays of \emph{segment} descriptors.
% A segment is one independent scoring problem: a set of \emph{query} rows (the points we are scoring) against a contiguous window of \emph{candidate} rows (the points they may match), together with the output range its winners occupy.
% Every stage after partitioning---assignment, leaf scoring, accumulation, and pruning---runs in a single \emph{permuted row space}: we reorder the points so that points in the same subproblem are stored close together, and we keep a map back to the input ids for export.
% Working in one id space lets a later KNN round add witnesses to the same accumulator without renaming rows.
% We describe the routing-key layout in \Cref{app:composite-keys} and ablate the reorder in \Cref{app:reorder-ablation}.

\subsection{Regularizing Bucket Sizes}
\label{sec:gpu-kernels:uneven}

Flattening eliminates recursion from \ourgpumethod{} but does not address skew in subproblem sizes: some top-level buckets are large, some second-level buckets exceed the maximum leaf size, and some leaves are nearly empty. 
Because a GPU runs thousands of thread blocks concurrently and a launch completes only when its \emph{slowest} block finishes, a kernel whose per-block work tracks this long tail would let a few oversized buckets dominate the launch while many near-empty leaves leave their SMs idle, so utilization collapses toward the worst-case bucket rather than the average.
We reshape the uneven buckets into nearly-uniformly-sized leaves, which we process within fixed-size \emph{score tiles} (\Cref{sec:gpu-kernels:skeleton}).

\myparagraph{Reshaping buckets}
We reshape both the partitioning buckets and the final leaves the same way.
We radix-sort points by bucket id, run-length-encode to recover each bucket's size, then split oversized buckets and merge undersized ones to a target size ($\approx 1024$), mirroring the leaf-balancing step of \ouroldmethod{} \cite{rubel2026pipnn}.
Because the sizes are normalized, one fixed-tile kernel handles every leaf, so we never specialize or relaunch per bucket size.
We score each leaf as one segment and launch the leaf kernel on batches of these segments, which keeps the tensor cores occupied and amortizes launch overhead.

\subsection{A Single Fused Segmented Top-$K$ Primitive}
\label{sec:gpu-kernels:skeleton}

\myparagraph{Overview}
We serve every batched top-$K$ with one primitive that scores over the decoded \texttt{i8} codes and keeps the top-$K$ entirely in registers.
RSQ8 codes are already \texttt{i8} and used as-is, while TQ4's packed $4$-bit codes are expanded once through a $16$-entry codebook into \texttt{i8}, so the kernel sees a uniform \texttt{i8} representation for both (\Cref{app:tq4-leaf}).
We describe that kernel here.

% \lars{try to reduce comparison to the CPU version.}

\myparagraph{The tensor-core instruction}
We use the warpgroup matrix-multiply instruction \texttt{wgmma} on Hopper tensor cores.
Recall that a warpgroup ($128$ threads) issues one \texttt{wgmma} cooperatively.
On the \texttt{i8} codes of a quantized dataset, a single \texttt{wgmma} instruction operates on a tile of $M = 64$ query rows, $N = 128$ database-point columns, and contraction width $W = 32$ bytes, computing
\[
  C[m,n] \mathrel{+}= \sum_{r=0}^{W-1} A[m,r]\cdot B[r,n], \qquad 0 \le m < M,\; 0 \le n < N,
\]
i.e.\ multiplying an $M \times W$ tile of query bytes by a $W \times N$ tile of data bytes and accumulating an $M \times N$ block of \texttt{i32} results. The instruction fixes $M = 64$, while the $N = 128$ width is our tiling choice.
%\lars{What is the size of N and M?}
The accumulator lives in registers distributed across the warpgroup. Each lane owns a fixed set of $(m,n)$ cells, and the four lanes that own one output row sit at consecutive lanes of a warp.

\myparagraph{Launch interface}
At each launch we pass to the segmented top-$K$ kernel the decoded base vectors of the dataset, the per-vector scales $\sfac$ and squared norms, a single dense \emph{slab} holding the database-point rows of every segment in the launch, and the list of segment descriptors.
A descriptor is the tuple $(\mathcal{Q}_i,\, r_i,\, c_i,\, o_i)$ consisting of the global ids $\mathcal{Q}_i$ of its query rows, the first row $r_i$ and count $c_i$ of its database-point window inside the slab, and the output offset $o_i$ at which we write its top-$K$ results.
Packing all database-point rows into one slab allows faster TMA tile copies from global to shared memory.
Because that descriptor covers the whole slab, each segment accesses its rows by offset, and we never rebuild a descriptor per leaf.

\myparagraph{Score tiles}
We expand the segment sizes on the host into a list of \emph{score tiles}.
A tile $T = (i, m_0)$ pairs segment $i$ with a block of $M = 64$ query rows starting at $m_0$. A multi-warpgroup variant doubles this to $M = 128$ (\Cref{app:gpu-pipeline-modes}).
Each thread block executes one tile: a single warpgroup of $128$ threads issues the \texttt{wgmma}s, each computing $64 \times 128 \times 32$ multiply-accumulates over one $32$-byte stripe.
Because the tile size is fixed, work is assigned to thread blocks statically and evenly, no matter how segment sizes are distributed.
A segment is one independent scoring subproblem: a set of queries scored against a contiguous window of database points.
A single launch scores many segments at once---one per leaf---for a batch of leaves sized to fill the slab, which is thousands of leaves per launch.

\myparagraph{Fused scoring loop}
Each thread block sweeps its segment's database-point window in tiles of $N = 128$ rows.
Within a tile we reduce over the vector dimension $d$ in $W = 32$-byte \emph{stripes}: we stage the next stripe of database-point bytes (and, at high dimension, query bytes) into shared memory while the tensor cores multiply the current one, keeping a single \texttt{wgmma} group in flight so copies and math overlap.
We move bytes into shared memory with asynchronous copies (\texttt{cp.async}) or with the \texttt{TMA} engine, neither of which ties up the warps' registers; we use \texttt{TMA} for the database-point slab in our warp-specialized pipeline and \texttt{cp.async} otherwise (\Cref{app:gpu-pipeline-modes}).
When the database-point tile is exhausted, the warpgroup drains the accumulator and runs the epilogue.
We never write the $M \times N$ score block to global memory. Each finished dot product is consumed in registers by the top-$K$ reduction, as shown in \Cref{alg:gpu-segmented-topk}.

% \myparagraph{Distance epilogue}
% After the matrix products, the accumulator holds raw inner products, which we turn into scores.
% For a query $q$ and candidate $p$ with stored unit operands $\hat q, \hat p$ and per-vector scales $\sfac_q, \sfac_d$, we form the similarity and the score
% \[
%   s = \sfac_q\,\sfac_d\,\langle \hat q, \hat p\rangle, \qquad
%   \mathrm{score} =
%   \begin{cases}
%     \lVert q\rVert^2 + \lVert p\rVert^2 - 2s & \text{(L2),}\\[2pt]
%     -s & \text{(inner product),}
%   \end{cases}
% \]
% negating the similarity so that a most-similar search becomes the argmin our reduction expects, exactly as in the CPU epilogue.
% We map the score to an order-preserving $32$-bit key so the top-$K$ comparisons are integer comparisons, and we drop self-matches ($p = q$) at the point of insertion rather than by masking the whole block.

\myparagraph{Register top-$K$ and merge}
Our register top-$K$ mirrors the CPU displacement sort (\Cref{sec:cpu-kernels:topk}): each lane runs a private top-$K$ tournament in registers, and we defer the cross-lane merge to the very end.
We keep each lane's running top-$K$ for a query row in registers, sorted so the best candidate sits in slot $0$ and the $K$-th in slot $K-1$ (pseudocode in \Cref{alg:gpu-register-topk}).
To insert a candidate we first compare it against slot $K-1$, which rejects anything worse than the current $K$-th best. Otherwise an unrolled compare-swap chain shifts it into place, carrying the displaced entry off the end.
Registers cannot be indexed by a value computed at run time, so the top-$K$ array must live at compile-time-constant slots. If any access were data-dependent, the compiler would spill the array to \emph{local memory}---slower per-thread storage backed by global memory. In our measurements that spill moved the bottleneck off the matrix product and onto the epilogue, so we fully unroll the insertion to keep every slot access constant.
% \lars{sentence too long. revise "measured to shift" since I misunderstood a few times while reading.}
Once the database-point window is consumed, the four lanes that own one output row hold four independent top-$K$ lists, and we merge them with a two-step butterfly: a warp shuffle exchanges each lane's $K$ entries with its partner across lane masks $1$ and $2$, and we insert the partner's entries into the local list.
The base lane then writes the $K$ results in sorted order.
A leaf is scored as a self-join, so its score block is symmetric. We nonetheless compute it in full, because the tensor cores need dense rectangular tiles and both edge directions are emitted for accumulation regardless.

\myparagraph{Modes and precision}
A single kernel family---not a separate kernel per precision---serves every configuration. We run it in two interchangeable pipeline modes---a single-warpgroup mode and a warp-specialized mode that splits loads from compute---and the same kernel serves the quantized \texttt{i8} builds (TQ4 and RSQ8) and the unquantized fp16/tf32 configurations we evaluate as a supplement (\Cref{app:tq4-leaf}).
We provide more details in \Cref{app:gpu-pipeline-modes,app:tq4-leaf}, and report kernel measurements in \Cref{tab:phase8_roofline,tab:leaf_roofline_4mode}.

\begin{algorithm}[t]
  \caption{Segmented fused top-$K$ scoring}
  \label{alg:gpu-segmented-topk}
  \DontPrintSemicolon
  \KwIn{Tile $T=(i,m_0)$; segments/slab; scales/norms; $K$}
  \KwOut{Per-query sorted top-$K$ (distance, id)}
  \KwData{$M \in \{64,128\}$; $N=128$; $W=32$ bytes}

  $\mathrm{seg} \gets$ segment $i$;\quad $Q \gets \mathrm{seg}.\mathcal{Q}\,[m_0 : m_0 + M)$\;
  initialize a register top-$K$ list for each owned query row\;
  stage all $W$-byte stripes of $Q$ into shared memory\;
  \For{$n_0 = 0$ to $\mathrm{seg.count}$ in steps of $N$}{
    $N_{\mathrm{valid}} \gets \min(N,\, \mathrm{seg.count} - n_0)$\;
    stage database-point ids, scales, and squared norms for rows $[n_0, n_0 + N_{\mathrm{valid}})$; pad the remaining tile entries\;
    $\mathbf{C} \gets \mathbf{0}$ \tcp*[r]{$M \times N$ accumulators}
    \For(\tcp*[f]{$W$-byte stripes}){each stripe $r$}{
      stage $S[n_0 : n_0 + N_{\mathrm{valid}},\, r]$ into an $N$-row shared-memory tile and pad the remainder\;
      $\mathbf{C} \mathrel{+}= \texttt{wgmma}(Q[:, r],\; S_{\mathrm{tile}}[:, r])$ \tcp*[r]{one in flight}
    }
    \For{each accumulator cell $(m, n)$ owned by this lane with $n < N_{\mathrm{valid}}$}{
      $s \gets \sfac_q[m]\,\sfac_d[n]\,\mathbf{C}[m, n]$\;
      \eIf{the metric is squared L2}{
        $\mathit{score} \gets \lVert q_m\rVert^2 + \lVert p_n\rVert^2 - 2s$\;
      }{
        $\mathit{score} \gets -s$ \tcp*[r]{MIPS as argmin}
      }
      \lIf{$\mathrm{id}(n) \neq \mathrm{id}(m)$}{insert $(\mathit{score}, \mathrm{id}(n))$ into row $m$'s top-$K$}
    }
  }
  merge the four owner lanes of each row with a shuffle butterfly\;
  write each row's $K$ results in sorted order\;
\end{algorithm}

% \subsection{Leaf Scoring and Candidate Emission}
% \label{sec:gpu-kernels:leaf}

% The leaves are the only stage that creates graph candidates. In an ANN-index
% build, \ourgpumethod{} uses a small local $K$, because each point appears in
% many overlapping leaves. In direct approximate $k$-NN graph construction, the
% same leaf stage uses the requested output $k$.

% A leaf result is a witness, not a final edge. It records that two points were
% near each other in one local region, along with the source-to-target distance
% computed by the segmented top-$K$ primitive. \ourgpumethod{} emits both
% directions of each witnessed relationship. This lets one local comparison
% feed the candidate rows of both endpoints.

% The leaf stage produces a stream of candidate records. Those records are the
% input to \HashPrune{} accumulation, not to the final ANN graph. The data flow
% from leaf slabs to bounded candidate rows is shown in \Cref{fig:leaf-stage}.

\subsection{From Online \HashPrune{} to Sort-and-Merge}
\label{sec:gpu-kernels:hashprune}

% \HashPrune{} is easy to implement as an online CPU update. As candidates arrive
% for a source point, the source keeps a bounded set of nearby candidates with
% diverse residual directions. A direct GPU version would make many threads try
% to update the same bounded row. That would require locks or atomics, or it
% would make the row depend on arrival order.

Recall that \HashPrune{} keeps, for each point, a bounded reservoir of nearby candidates with diverse residual directions.
On the CPU we implement it by simply protecting the structure using a lock. Unfortunately on GPU this leads to significantly worse contention and thus degrades performance meaningfully. 
% Many GPU threads updating the same bounded row would contend on locks or atomics, or make the row depend on the order in which leaves were scheduled.
We therefore replace the synchronized update with a sort. This sort-and-merge runs once per batch of leaves during leaf scoring, and each batch's candidates merge into a global \HashPrune{} datastructure that the final prune sparsifies into the graph.
For every candidate neighbor we emit two fixed-size records, one per direction, each carrying a short sketch of the residual direction from source to target, and we radix-sort the records by owner row and distance.
The sort operates on a transient buffer of two records per candidate edge, bounded by a memory parameter (the per-batch slab cap, $\approx 6$~GB in our runs) and reused across batches---it does not grow with the full candidate set.
After the sort, every owner's candidates are contiguous, so we update each row's bounded accumulator in a single sequential pass with no concurrent writes: we de-duplicate repeated targets. For colliding hashes, we keep the nearer candidate.
% The result is deterministic for a fixed partition and scoring result, because each row observes its candidates in a canonical order rather than in scheduling order.
We describe the sketch construction and record layout in more detail in \Cref{app:gpu-prune-detail}.

\subsection{Pruning Without Candidate Recompute}
\label{sec:prune:gpu}
\label{sec:gpu-kernels:direct-prune}

ANN builds need a final sparsification step after candidate accumulation.
Exact \RobustPrune{} tests whether an accepted neighbor $p^*$ occludes a
candidate $p'$ for source $p$. This test needs the distance $d(p^*,p')$.
Recomputing such distances for every row would add a new irregular
candidate-to-candidate scoring stage.

We instead use an approximation we call \DirectEdgePrune{}.
When we prune the neighbors of point $p$, we check whether the accepted neighbor $p^*$ already has $p'$ in its own neighbor-list.
If that edge exists then we use the stored distance as the witnessed value for $d(p^*,p')$ and apply the usual $\alpha$-prune test; if the edge was not witnessed then we proceed to the next neighbor.
All rows consult a read-only snapshot of the completed candidate accumulator.
Each row writes its accepted neighbors to separate scratch storage; after all pruning decisions finish, a second pass compacts these outputs into the final graph.

This relaxation of \RobustPrune{} has one-sided error: it can miss a domination that exact \RobustPrune{}
would have found, because the needed candidate pair may be absent from the
accumulator.
It never prunes an edge that would have been retained by \RobustPrune{}. 
The effect on degree and query quality is measured in \Cref{tab:prune_ablation}: \DirectEdgePrune{} yields denser graphs of quality comparable to those produced by a final \RobustPrune{}. 
%Whether \DirectEdgePrune{} could be used in frameworks other than \ouroldmethod{} is an open question.

%When we build a direct KNN graph rather than an ANN index, we skip this prune entirely; we describe KNN-graph mode and the build-time scaling choices in \Cref{app:gpu-knn-scaling}.

\section{Experimental Evaluation}
\label{sec:experiments}

Our experiments are organized around three goals: (1) quantifying the speedups we achieve for index building and $k$-NN graph construction (\Cref{sec:cpu_ann_build,sec:gpu_ann_build}), (2) demonstrating that the quantization and algorithmic changes employed by \ourmethod{} and \ourgpumethod{} preserve high index quality (\Cref{sec:cpu-ablations,sec:gpu_ann_build}), and (3) explaining which components in the implementations are responsible for the observed speedups (\Cref{sec:cpu-ablations,sec:gpu_throughput}). In particular, we find that:
% SIGMOD submission: use the acmart list geometry and spacing defaults.
% \begin{enumerate}[leftmargin=2em,itemsep=2pt,topsep=2pt]
\begin{enumerate}
  \item On CPU, \ourmethod{} is the fastest method for both the ANNS-index and $k$-NN tasks: on average for ANNS-index build it beats the strongest quantized Vamana baseline by \textbf{5.9$\times$} and unquantized Vamana by \textbf{40.7$\times$}. For $k$-NN building it beats quantized Vamana by \textbf{10.3$\times$} and \pynn{} by \textbf{5.9$\times$} at equivalent recall (\Cref{sec:cpu_ann_build}).
  \item On GPU, \ourgpumethod{} is also the fastest method: on the ANNS-index task it beats Jasper by \textbf{8.1$\times$} on average (\textbf{13.3$\times$} with quantization) and unquantized CAGRA by \textbf{16$\times$} (\textbf{27$\times$} with quantization). On the $k$-NN task it beats Jasper, NN-Descent, CAGRA, and IVF-PQ8 by geometric means of \textbf{7.1$\times$}, \textbf{2.9$\times$}, \textbf{9.8$\times$}, and \textbf{49.8$\times$} respectively (\Cref{sec:gpu_ann_build}).
  \item Using quantization in the build keeps query performance essentially unchanged: \ouroldmethod-TQ and \ouroldmethod-RSQ8 match \ouroldmethod-FP32's QPS at any given recall within geometric-mean ratios of \textbf{0.98} and \textbf{0.99} (\Cref{sec:cpu-ablations}).
  \item We trace the speedups to our highly-optimized kernels: we show that our leaf-$2$-NN kernel achieves a \textbf{12.4$\times$} speedup over \ouroldmethod{}'s prior approach. It runs at near-peak instruction throughput on CPU, and it is bandwidth-bound at high dimension on GPU, reaching \textbf{82--95\%} of memory throughput at $d \geq 1536$ (\Cref{sec:cpu-ablations,sec:gpu_throughput}).
\end{enumerate}

\myparagraph{Machines}
We use a Google Cloud \texttt{c4-highmem-288} (288 vCPUs, 2.2~TB DRAM, Intel Granite Rapids) for billion-point datasets, and a \texttt{c3-highmem-192} (192 vCPUs, 1.5~TB DRAM, Intel Sapphire Rapids) for smaller datasets.
%Both machines support AVX-512 VNNI instructions, used in \ourmethod's quantized inner loop.
%\lars{why do we care about same cost as GPU?}\tobias{sketchy way of justifying ``apples-to-apples'' hardware comparison.}\lars{do we make the comparison somewhere?}
For GPU measurements we use an \texttt{a3-highgpu-1g} instance: one NVIDIA H100 80~GB HBM3 SXM5 (cc~9.0, 132 SMs, 3.35~TB/s HBM3) on a 26-vCPU Sapphire Rapids host with 234~GB DDR5, CUDA~12 with driver 580.
\ourgpumethod{} is compiled for Hopper SM90a and run as a single-H100 builder.
Datasets that exceed the 80~GB device memory are reported on CPU only.

\myparagraph{``High-quality'' indexes}
We measure quality in terms of query throughput and recall. Specifically, we consider an index \emph{high-quality} when its QPS-vs-recall curve is comparable to that of an unquantized ParlayANN-Vamana graph. The first version of \ouroldmethod was shown to produce high-quality indexes, and we show in this paper that it continues to do so for quantized points.

\myparagraph{CPU Methods}
Our primary baseline is Vamana~\cite{jayaram2019diskann} as implemented in the ParlayANN library~\cite{manohar2024parlayann}.
To the best of our knowledge, this is the most performant shared-memory graph-based ANNS index publicly available. We restrict our comparison to Vamana because previous work~\cite{rubel2026pipnn} already showed that it builds indexes faster and of equivalent or better quality than HCNNG~\cite{munoz2019hierarchical}, HNSW~\cite{malkov2018efficient}, MIRAGE~\cite{voruganti2025mirage}, and FastKCNA~\cite{yang2024revisiting}.%, and SymphonyQG~\cite{gou2025symphonyqg}.%, so by transitivity we improve over all of these methods.

We also consider an unpublished version of ParlayANN's Vamana~\cite{parlayann_repo} that uses 8-bit scalar quantization (SQ8) for distance computations and reranks a smaller set with float precision.
For dimensionalities $> 512$ it further uses a filtering technique based on precomputed Johnson-Lindenstrauss (JL) sketches ~\cite{johnson1984extensions} that further speeds up beam search.
Quantization and filtering offers significantly faster builds and queries  than unquantized Vamana while producing indexes of identical quality.  
Hence, we use this version when computing QPS-vs-recall curves on CPUs. 
For $k$-NN graph construction, we additionally compare with a parallel implementation of NN-descent~\pynn~\cite{mcinnes2020pynndescent} also in the ParlayANN library~\cite{manohar2024parlayann} that uses SQ8.
The per-dataset build parameter configurations for CPU methods are detailed in \Cref{app:cpu-params} (\Cref{tab:params-ann-cpu,tab:params-knn-cpu}).

\myparagraph{GPU Methods}
We compare \ourgpumethod{} against the two leading GPU-native graph ANNS systems: CAGRA~\cite{ootomo2024cagra} and Jasper~\cite{mccoy2026jasper}. 
\hide{
\lars{This can go to a relatedw work section}
Graph-based ANNS is notoriously hard to implement on GPUs because memory access patterns on dynamic graphs are highly irregular, with data-dependent control flow, random memory accesses, and dynamic candidate sets. 
CAGRA and JASPER differ in how they reconcile this irregularity with accelerator-friendly execution.
\emph{CAGRA}~\cite{ootomo2024cagra} approximates a $k$-NN graph using NN-descent, then applies a rank-based pruning and reordering process to obtain diverse edges without computing distances. 
The resulting \emph{regular} graphs feed a novel beam-search variant that achieves high query throughput on GPU. 
}
The original CAGRA paper only considered FP16 coordinates, but the library implementation supports generating the initial $k$-NN graph via product quantization (PQ) coupled with IVF, which we evaluate here (to our knowledge) for the first time.
Unquantized CAGRA does not run on unnormalized MIPS datasets, as the NN-descent implementation is specialized to L2.
cuVS~\cite{rapids_cuvs} supports MIPS graphs for CAGRA only through its \textsc{IVF-PQ} build path, not the \textsc{NN-Descent} build path that unquantized CAGRA uses.  

\emph{Jasper}~\cite{mccoy2026jasper} ports Vamana to GPU, leverages RaBitQ quantization~\cite{gao2025rabitq} for search, and optimizes the greedy search kernel for latency hiding.
We use \emph{Jasper}'s query code for our index quality studies (\Cref{fig:gpu_anns_recall_main} and \Cref{fig:gpu_anns_recall_grid}). This permits us to assess the quality of the indexes without the confounding issue of query performance. It was previously shown that Jasper's query routines are at least as good as CAGRA's~\cite{mccoy2026jasper}.
The per-dataset build parameter configurations for GPU methods are detailed in \Cref{app:gpu-params} (\Cref{tab:params-ann,tab:params-knn}).

\myparagraph{Datasets}
We evaluate on a diverse suite of datasets summarized in \Cref{tab:modern_datasets_full_compact}: three billion-scale datasets (DEEP~\cite{babenko2016efficient}, MS-Turing~\cite{simhadri2022results}, Text2Image~\cite{simhadri2022results}) and two hundred-million-scale text datasets (MSMARCO~\cite{simhadri2024results} and MSMARCO-Arctic~\cite{merrick2024arctic, craswell2021trec}) obtained through the BigANN Benchmarks framework~\cite{simhadri2024results, simhadri2022results}; Wiki-Qwen-500K (4096-d) embeddings~\cite{maknee2024wikipediaqwen8b}; two DBpedia datasets at 1536-d and 3072-d~\cite{qdrant2024dbpedia1536, qdrant2024dbpedia3072}; the 12.78M image--text DataCompDr embeddings~\cite{apple2024datacompdr}; and Wiki-Cohere~\cite{reimers2022cohere}, OpenAI-ArXiv~\cite{neelakantan2022text}, and wiki-all~\cite{rapids2024wikiall}.  We use a subset of these for our GPU experiments, the full collection can be seen in \Cref{tab:ann_build_full}.
For DEEP and MS-Turing we use both 10M and 100M slices; for Text2Image, DataCompDr, Wiki-Cohere, and MSMARCO we use 10M slices, and for MSMARCO-Arctic a 1M slice, as the GPU baselines (Jasper, CAGRA) do not complete at full scale on a single 80\,GB H100.

\begin{table}[t]
\centering
\caption{Summary of datasets used in our experiments.}
\label{tab:modern_datasets_full_compact}
\footnotesize
\setlength{\tabcolsep}{3pt}
\begin{tabular}{l c c c c c c c}
\toprule
\textbf{Name} & \textbf{Type} & \textbf{Dim} & \textbf{Size} & \textbf{Q} & \textbf{DType} & \textbf{Dist.} & \textbf{Src.} \\
\midrule
DBpedia1536 & Text & 1536 & 999k & 1k & f32 & L2 & \cite{qdrant2024dbpedia1536} \\
DBpedia3072 & Text & 3072 & 999k & 1k & f32 & L2 & \cite{qdrant2024dbpedia3072} \\
DataCompDr & T2I & 1536 & 12.78M & 1k & bf16 & L2 & \cite{apple2024datacompdr} \\
MSMARCO-Arctic & Text & 1024 & 113.52M & 10k & f32 & MIPS & \cite{merrick2024arctic, craswell2021trec} \\
MSMARCO & Text & 768 & 101.07M & 9376 & f32 & IP & \cite{simhadri2024results} \\
Wiki-Cohere & Text & 768 & 35M & 5k & f32 & IP & \cite{reimers2022cohere} \\
OpenAI-ArXiv & Text & 1536 & 2.32M & 20k & f32 & L2 & \cite{neelakantan2022text} \\
wiki-all & Text & 768 & 88M & 10k & f32 & MIPS & \cite{rapids2024wikiall} \\
Wiki-Qwen-500K & Text & 4096 & 500k & 100k & float & MIPS & \cite{maknee2024wikipediaqwen8b} \\
DEEP & Image & 96 & 1B & 10k & f32 & L2 & \cite{babenko2016efficient} \\
MS-Turing & Text & 100 & 1B & 100k & f32 & L2 & \cite{simhadri2022results} \\
Text2Image & T2I & 200 & 1B & 100k & f32 & MIPS & \cite{simhadri2022results} \\
\bottomrule
\end{tabular}
\end{table}

\subsection{Speedup Results on CPU}
\label{sec:cpu_ann_build}
% Methods: PiPNN, \ourmethod{}, Vamana, quantized Vamana, HNSW, MIRAGE, FastKCNA
% Datasets: All datasets
% \begin{enumerate}
%     \item Build time or slowdown bar plots
%     \item QPS vs recall (CPU querying)
% \end{enumerate}

\begin{figure}[tb]
  \centering
  \includegraphics[width=\linewidth]{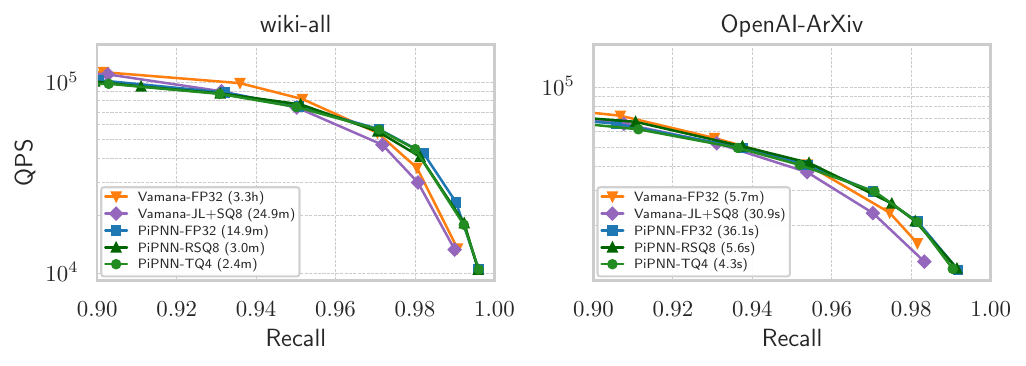}
  \caption{Index Quality (CPU). QPS vs.\ recall on two representative datasets, with per-method build times in the legends. \ouroldmethod-FP32, \ouroldmethod-TQ4, and
  \ouroldmethod-RSQ8 trace nearly identical curves; the full set of
  twelve datasets is in \Cref{fig:additional-qps}
  (\Cref{app:additional-experiments}).}
  \label{fig:cpu-index-quality-main}
\end{figure}

\myparagraph{ANNS Index Build} 
We build \ouroldmethod and Vamana indexes and subsequently query them under varying beam size to measure the throughput vs recall tradeoff. Representative QPS-recall curves on two datasets are shown in \Cref{fig:cpu-index-quality-main}; for the full set of results over all 12 datasets, refer to \Cref{fig:additional-qps} in the appendix.
%
%For each dataset, we did a parameter sweep on both algorithms; the numbers shown in the plots are those using the best build configurations for each recall. 
%as well as \ouroldmethod runs using TurboQuant vectors with 1 or 2 replicas ($\mathsf{nc}$). 
The methods have similar query performance across the datasets. 
The index build times are shown in~\cref{tab:cpu_build} and also included in the legends of each plot.
Quantization speeds up Vamana construction by 3-15$\times$ and \ouroldmethod by \textbf{2-8$\times$}, while showing no impact on index quality.
The effects on Vamana are more pronounced on high dimensional datasets due to JL. However, even with these speedups, full precision \ouroldmethod remains faster to build on most datasets.
\ourmethod is up to \textbf{8$\times$} faster than \ouroldmethod. Against quantized and unquantized Vamana, its geometric-mean speedups are \textbf{5.9$\times$} and \textbf{40.7$\times$}, respectively.

\myparagraph{$k$-NN Graph Build} To build a $k$-NN graph with Vamana we build an ANNS index and then perform a $k$-NN query for each point. While \ouroldmethod could use the same approach, we find the neighbors of a point directly from its leaves~\cite{gottesburen2024unleashing}.
We also include PyNNDescent, which is designed for direct $k$-NN graph construction.

\begin{figure}[htb]
  \centering
  \includegraphics[width=\linewidth]{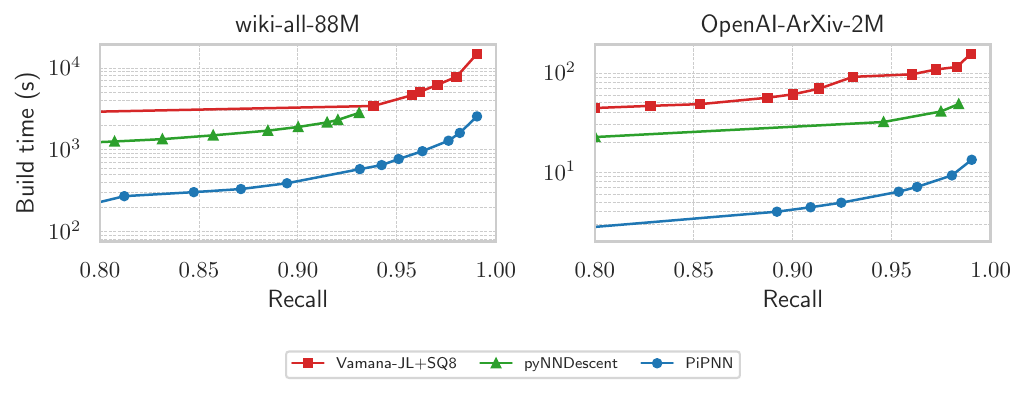}
  \caption{$k$-NN graph build time vs.\ recall on two representative
  datasets, zoomed to recall in $[0.8, 1.0]$. Results on the full set of datasets are shown in
  \Cref{fig:additional-knn-recall} (\Cref{app:additional-experiments}).
  }
  \label{fig:cpu-knn-recall-main}
\end{figure}

From a large parameter sweep, we select the fastest builds spanning different recall levels, forming a Pareto frontier for each method. \Cref{fig:cpu-knn-recall-main} shows two representative datasets; the full set is in \Cref{fig:additional-knn-recall} (\Cref{app:additional-experiments}).
All three methods are able to produce $k$-NN graphs of reasonably-high recall.
However, thanks to \ourmethod's low construction time combined with its ability to avoid materializing a complete index, it is able to build \textbf{4-10$\times$} faster than PyNNDescent and \textbf{5-20$\times$} faster than Vamana for graphs of equivalent recall (geometric means \textbf{5.9$\times$} over PyNNDescent, and \textbf{10.3$\times$} over quantized Vamana). \Cref{tab:cpu_build} reports absolute build times for both tasks across our CPU datasets, alongside the unquantized baselines.
PyNNDescent is omitted from the billion-scale datasets and MSMARCO because it exceeded the available memory.

\begin{table}[htb]
  \centering
  \label{tab:cpu_build}
  \setlength{\tabcolsep}{3pt}
  \footnotesize
    \caption{CPU build times (seconds). \emph{Left:} ANNS index build, best high-recall configuration of each method (QPS-vs-recall curves appear in \Cref{fig:additional-qps}). \emph{Right:} $k$-NN graph build, fastest swept configuration achieving recall\,$\geq 0.9$. Each half is colored independently: per-row fastest method bolded and cells shaded by log-slowdown vs the row min (green = fastest, red = slowest); ``--'' indicates method did not achieve recall cutoff. \textbf{VF32} is unquantized Vamana; \textbf{VQ} is Vamana with quantization and optional filtering; \textbf{NNDQ} is PyNNDescent with SQ8.}
  \begin{tabular}{l@{\hspace{4pt}}rrrr@{\hspace{3pt}}|@{\hspace{3pt}}rrr}
    \toprule
        & \multicolumn{4}{c}{ANNS index} & \multicolumn{3}{c}{$k$-NN @ R$\geq 0.9$} \\
    \cmidrule(lr){2-5} \cmidrule(lr){6-8}
    Dataset & \ourmethod{} & \ouroldmethod{} & VF32 & VQ & \ourmethod{} & VQ & NNDQ \\
    \midrule
    Wiki-Qwen-500K & \cellcolor[rgb]{0.780,0.920,0.660}\textbf{3.6} & \cellcolor[rgb]{0.985,0.948,0.651}18 & \cellcolor[rgb]{0.960,0.720,0.660}117 & \cellcolor[rgb]{0.882,0.934,0.655}8.0 & \cellcolor[rgb]{0.780,0.920,0.660}\textbf{2.4} & \cellcolor[rgb]{0.980,0.835,0.655}32 & \cellcolor[rgb]{0.985,0.948,0.651}12 \\
    DBP1536 & \cellcolor[rgb]{0.780,0.920,0.660}\textbf{1.9} & \cellcolor[rgb]{0.992,0.905,0.652}15 & \cellcolor[rgb]{0.960,0.720,0.660}122 & \cellcolor[rgb]{0.982,0.948,0.651}9.3 & \cellcolor[rgb]{0.780,0.920,0.660}\textbf{1.7} & \cellcolor[rgb]{0.969,0.770,0.658}37 & \cellcolor[rgb]{0.993,0.909,0.652}13 \\
    DBP3072 & \cellcolor[rgb]{0.780,0.920,0.660}\textbf{4.7} & \cellcolor[rgb]{0.999,0.942,0.650}28 & \cellcolor[rgb]{0.960,0.720,0.660}228 & \cellcolor[rgb]{0.928,0.940,0.653}15 & \cellcolor[rgb]{0.780,0.920,0.660}\textbf{4.2} & \cellcolor[rgb]{0.980,0.833,0.655}57 & \cellcolor[rgb]{0.972,0.946,0.651}19 \\
    OpenAI & \cellcolor[rgb]{0.780,0.920,0.660}\textbf{4.4} & \cellcolor[rgb]{0.991,0.900,0.652}36 & \cellcolor[rgb]{0.960,0.720,0.660}342 & \cellcolor[rgb]{0.995,0.920,0.651}31 & \cellcolor[rgb]{0.780,0.920,0.660}\textbf{4.2} & \cellcolor[rgb]{0.978,0.824,0.655}61 & \cellcolor[rgb]{0.993,0.910,0.652}32 \\
    DataComp & \cellcolor[rgb]{0.780,0.920,0.660}\textbf{35} & \cellcolor[rgb]{0.995,0.921,0.651}245 & \cellcolor[rgb]{0.960,0.720,0.660}1918 & \cellcolor[rgb]{0.999,0.943,0.650}208 & \cellcolor[rgb]{0.780,0.920,0.660}\textbf{58} & \cellcolor[rgb]{0.990,0.892,0.653}506 & \cellcolor[rgb]{0.994,0.916,0.651}421 \\
    WikiCohere & \cellcolor[rgb]{0.780,0.920,0.660}\textbf{56} & \cellcolor[rgb]{0.997,0.935,0.651}352 & \cellcolor[rgb]{0.960,0.720,0.660}4317 & \cellcolor[rgb]{0.990,0.891,0.653}491 & \cellcolor[rgb]{0.780,0.920,0.660}\textbf{212} & \cellcolor[rgb]{0.965,0.751,0.659}5303 & \cellcolor[rgb]{0.996,0.925,0.651}1442 \\
    WikiAll & \cellcolor[rgb]{0.780,0.920,0.660}\textbf{146} & \cellcolor[rgb]{0.998,0.939,0.650}892 & \cellcolor[rgb]{0.960,0.720,0.660}11780 & \cellcolor[rgb]{0.986,0.870,0.653}1492 & \cellcolor[rgb]{0.780,0.920,0.660}\textbf{576} & \cellcolor[rgb]{0.996,0.949,0.650}3137 & \cellcolor[rgb]{0.932,0.941,0.653}1903 \\
    MSMARCO & \cellcolor[rgb]{0.780,0.920,0.660}\textbf{175} & \cellcolor[rgb]{0.997,0.934,0.651}1107 & \cellcolor[rgb]{0.960,0.720,0.660}13510 & \cellcolor[rgb]{0.987,0.874,0.653}1743 & \cellcolor[rgb]{0.780,0.920,0.660}\textbf{1015} & \cellcolor[rgb]{0.988,0.879,0.653}9751 & \cellcolor[rgb]{0.92,0.92,0.92}{--} \\
    MS-Arctic & \cellcolor[rgb]{0.780,0.920,0.660}\textbf{249} & \cellcolor[rgb]{0.990,0.949,0.650}1298 & \cellcolor[rgb]{0.960,0.720,0.660}14440 & \cellcolor[rgb]{0.986,0.869,0.654}2568 & \cellcolor[rgb]{0.780,0.920,0.660}\textbf{230} & \cellcolor[rgb]{0.988,0.882,0.653}2157 & \cellcolor[rgb]{0.995,0.922,0.651}1599 \\
    DEEP & \cellcolor[rgb]{0.780,0.920,0.660}\textbf{564} & \cellcolor[rgb]{0.903,0.937,0.654}1484 & \cellcolor[rgb]{0.978,0.824,0.655}8178 & \cellcolor[rgb]{0.999,0.950,0.650}3142 & \cellcolor[rgb]{0.780,0.920,0.660}\textbf{1605} & \cellcolor[rgb]{0.953,0.944,0.652}6251 & \cellcolor[rgb]{0.92,0.92,0.92}{--} \\
    MS-Turing & \cellcolor[rgb]{0.780,0.920,0.660}\textbf{567} & \cellcolor[rgb]{0.902,0.937,0.654}1480 & \cellcolor[rgb]{0.986,0.872,0.653}5719 & \cellcolor[rgb]{0.982,0.948,0.651}2772 & \cellcolor[rgb]{0.780,0.920,0.660}\textbf{1051} & \cellcolor[rgb]{0.969,0.946,0.651}4630 & \cellcolor[rgb]{0.92,0.92,0.92}{--} \\
    Text2Image & \cellcolor[rgb]{0.780,0.920,0.660}\textbf{730} & \cellcolor[rgb]{0.910,0.938,0.654}2027 & \cellcolor[rgb]{0.983,0.851,0.654}8605 & \cellcolor[rgb]{0.990,0.949,0.650}3784 & \cellcolor[rgb]{0.780,0.920,0.660}\textbf{935} & \cellcolor[rgb]{0.981,0.839,0.655}12108 & \cellcolor[rgb]{0.92,0.92,0.92}{--} \\
    \bottomrule
  \end{tabular}
\end{table}

\subsection{Speedup Results on GPU}
\label{sec:gpu_ann_build}

\myparagraph{ANNS Index Build}
In~\cref{tab:gpu_build} we report ANNS index build time for \ourgpumethod-RSQ8 and FP16, as well as Jasper and CAGRA.

\ourgpumethod{} is fastest on every dataset.
Taking the geometric mean, \ourgpumethod-RSQ8 builds \textbf{13.3$\times$} faster than \textsc{Jasper} and \textbf{27$\times$} faster than \textsc{CAGRA}. The largest single-dataset gaps are $43.7\times$ over \textsc{Jasper} (DBpedia-3072) and $50\times$ over \textsc{CAGRA} (DEEP-10M). Without quantization, \ourgpumethod-FP16 is on average $8.1\times$ faster than \textsc{JASPER} and $16.3\times$ faster than \textsc{CAGRA}.
 Full per-dataset build times are reported in \Cref{tab:ann_build_full}.
On the unnormalized MIPS datasets, we omit results for unquantized CAGRA, since it does not support MIPS as mentioned before.
In the appendix, we show additional results with a variant that builds the index through \textsc{IVF-PQ} that does support it. 

% Auto-generated by scripts/finals_paper_plots.py (write_merged_build_table)
% Requires: \usepackage{colortbl, xcolor, booktabs}
\begin{table}[t]
  \centering
  \caption{GPU build times (s): ANNS index (\emph{left}) and fastest
  $k$-NN graph at recall\,$\geq$\,0.9 (\emph{right}); build-then-search
  includes self-search. \textbf{RSQ8}/\textbf{FP16}: quantized/unquantized
  \ourgpumethod{} modes; \textbf{NN-Desc}: NN-Descent. Coloring follows
  \Cref{tab:cpu_build}; full results in
  \Cref{tab:ann_build_full,tab:knn_build_full}.}
  \label{tab:gpu_build}
  \setlength{\tabcolsep}{2pt}
  \footnotesize
  \begin{tabular}{l@{\hspace{4pt}}cccc@{\hspace{3pt}}|@{\hspace{3pt}}cccc}
    \toprule
        & \multicolumn{4}{c}{ANNS index} & \multicolumn{4}{c}{$k$-NN @ R$\geq 0.9$} \\
    \cmidrule(lr){2-5} \cmidrule(lr){6-9}
    Dataset & RSQ8 & FP16 & Jasper & CAGRA & RSQ8 & FP16 & NN-Desc & Jasper \\
    \midrule
    DEEP-10M & \cellcolor[rgb]{0.780,0.920,0.660}\textbf{2.32} & \cellcolor[rgb]{0.793,0.922,0.659}2.57 & \cellcolor[rgb]{0.949,0.943,0.652}8.78 & \cellcolor[rgb]{0.960,0.720,0.660}115 & \cellcolor[rgb]{0.780,0.920,0.660}\textbf{4.43} & \cellcolor[rgb]{0.786,0.921,0.660}4.65 & \cellcolor[rgb]{0.996,0.929,0.651}29.1 & \cellcolor[rgb]{0.893,0.935,0.655}10.8 \\
    MS-Turing-10M & \cellcolor[rgb]{0.780,0.920,0.660}\textbf{2.64} & \cellcolor[rgb]{0.791,0.922,0.659}2.88 & \cellcolor[rgb]{0.995,0.924,0.651}18.1 & \cellcolor[rgb]{0.960,0.720,0.660}116 & \cellcolor[rgb]{0.780,0.920,0.660}\textbf{15.9} & \cellcolor[rgb]{0.783,0.920,0.660}16.3 & \cellcolor[rgb]{0.870,0.932,0.656}32.3 & \cellcolor[rgb]{0.803,0.923,0.659}19.1 \\
    DBP1536 & \cellcolor[rgb]{0.780,0.920,0.660}\textbf{0.57} & \cellcolor[rgb]{0.863,0.931,0.656}1.08 & \cellcolor[rgb]{0.970,0.775,0.658}11.9 & \cellcolor[rgb]{0.971,0.784,0.657}11.1 & \cellcolor[rgb]{0.780,0.920,0.660}\textbf{1.84} & \cellcolor[rgb]{0.805,0.923,0.659}2.24 & \cellcolor[rgb]{0.902,0.937,0.654}4.78 & \cellcolor[rgb]{0.998,0.938,0.651}11.3 \\
    DBP3072 & \cellcolor[rgb]{0.780,0.920,0.660}\textbf{0.90} & \cellcolor[rgb]{0.881,0.934,0.655}1.99 & \cellcolor[rgb]{0.960,0.720,0.660}39.4 & \cellcolor[rgb]{0.970,0.777,0.658}18.5 & \cellcolor[rgb]{0.780,0.920,0.660}\textbf{3.00} & \cellcolor[rgb]{0.817,0.925,0.658}4.00 & \cellcolor[rgb]{0.877,0.933,0.656}6.40 & \cellcolor[rgb]{0.994,0.916,0.651}21.8 \\
    Wiki-Qwen-500K & \cellcolor[rgb]{0.780,0.920,0.660}\textbf{0.61} & \cellcolor[rgb]{0.877,0.933,0.656}1.30 & \cellcolor[rgb]{0.981,0.840,0.655}7.81 & \cellcolor[rgb]{0.973,0.795,0.657}10.9 & \cellcolor[rgb]{0.780,0.920,0.660}\textbf{2.38} & \cellcolor[rgb]{0.814,0.925,0.658}3.11 & \cellcolor[rgb]{0.833,0.927,0.658}3.60 & \cellcolor[rgb]{0.997,0.950,0.650}13.1 \\
    OpenAI & \cellcolor[rgb]{0.780,0.920,0.660}\textbf{1.29} & \cellcolor[rgb]{0.872,0.933,0.656}2.66 & \cellcolor[rgb]{0.970,0.780,0.657}26.0 & \cellcolor[rgb]{0.964,0.742,0.659}34.6 & \cellcolor[rgb]{0.780,0.920,0.660}\textbf{2.46} & \cellcolor[rgb]{0.827,0.926,0.658}3.55 & \cellcolor[rgb]{0.989,0.949,0.650}12.7 & \cellcolor[rgb]{0.987,0.874,0.653}24.4 \\
    WikiAll-1M & \cellcolor[rgb]{0.780,0.920,0.660}\textbf{0.44} & \cellcolor[rgb]{0.827,0.926,0.658}0.64 & \cellcolor[rgb]{0.983,0.948,0.651}2.17 & \cellcolor[rgb]{0.92,0.92,0.92}{--} & \cellcolor[rgb]{0.780,0.920,0.660}\textbf{1.78} & \cellcolor[rgb]{0.797,0.922,0.659}2.03 & \cellcolor[rgb]{0.884,0.934,0.655}4.03 & \cellcolor[rgb]{0.901,0.936,0.655}4.60 \\
    WikiAll-10M & \cellcolor[rgb]{0.780,0.920,0.660}\textbf{3.53} & \cellcolor[rgb]{0.863,0.931,0.656}6.78 & \cellcolor[rgb]{0.989,0.886,0.653}32.2 & \cellcolor[rgb]{0.92,0.92,0.92}{--} & \cellcolor[rgb]{0.780,0.920,0.660}\textbf{8.50} & \cellcolor[rgb]{0.817,0.925,0.658}11.3 & \cellcolor[rgb]{0.975,0.947,0.651}39.2 & \cellcolor[rgb]{0.987,0.877,0.653}82.6 \\
    MS-MARCO-10M & \cellcolor[rgb]{0.780,0.920,0.660}\textbf{3.50} & \cellcolor[rgb]{0.860,0.931,0.656}6.58 & \cellcolor[rgb]{0.965,0.750,0.659}88.3 & \cellcolor[rgb]{0.92,0.92,0.92}{--} & \cellcolor[rgb]{0.780,0.920,0.660}\textbf{30.1} & \cellcolor[rgb]{0.806,0.924,0.659}37.0 & \cellcolor[rgb]{0.814,0.925,0.658}39.4 & \cellcolor[rgb]{0.980,0.832,0.655}410 \\
    DataComp-10M & \cellcolor[rgb]{0.780,0.920,0.660}\textbf{6.08} & \cellcolor[rgb]{0.92,0.92,0.92}{--} & \cellcolor[rgb]{0.981,0.838,0.655}79.3 & \cellcolor[rgb]{0.92,0.92,0.92}{--} & \cellcolor[rgb]{0.780,0.920,0.660}\textbf{15.7} & \cellcolor[rgb]{0.92,0.92,0.92}{--} & \cellcolor[rgb]{0.92,0.92,0.92}{--} & \cellcolor[rgb]{0.991,0.899,0.652}129 \\
    WikiCohere-10M & \cellcolor[rgb]{0.780,0.920,0.660}\textbf{3.59} & \cellcolor[rgb]{0.859,0.931,0.656}6.70 & \cellcolor[rgb]{0.973,0.795,0.657}64.5 & \cellcolor[rgb]{0.92,0.92,0.92}{--} & \cellcolor[rgb]{0.780,0.920,0.660}\textbf{31.1} & \cellcolor[rgb]{0.808,0.924,0.659}38.7 & \cellcolor[rgb]{0.839,0.928,0.657}49.4 & \cellcolor[rgb]{0.995,0.921,0.651}217 \\
    Text2Image-10M & \cellcolor[rgb]{0.780,0.920,0.660}\textbf{2.39} & \cellcolor[rgb]{0.807,0.924,0.659}2.96 & \cellcolor[rgb]{0.960,0.720,0.660}90.2 & \cellcolor[rgb]{0.92,0.92,0.92}{--} & \cellcolor[rgb]{0.780,0.920,0.660}\textbf{5.00} & \cellcolor[rgb]{0.791,0.922,0.659}5.47 & \cellcolor[rgb]{0.998,0.936,0.651}31.1 & \cellcolor[rgb]{0.969,0.773,0.658}106 \\
    MS-Arctic-1M & \cellcolor[rgb]{0.780,0.920,0.660}\textbf{0.55} & \cellcolor[rgb]{0.831,0.927,0.658}0.81 & \cellcolor[rgb]{0.995,0.923,0.651}3.75 & \cellcolor[rgb]{0.92,0.92,0.92}{--} & \cellcolor[rgb]{0.780,0.920,0.660}\textbf{0.96} & \cellcolor[rgb]{0.807,0.924,0.659}1.19 & \cellcolor[rgb]{0.976,0.947,0.651}4.45 & \cellcolor[rgb]{0.960,0.720,0.660}39.0 \\
    \bottomrule
  \end{tabular}
\end{table}

\begin{figure}[tb]
  \centering
  \includegraphics[width=0.80\linewidth]{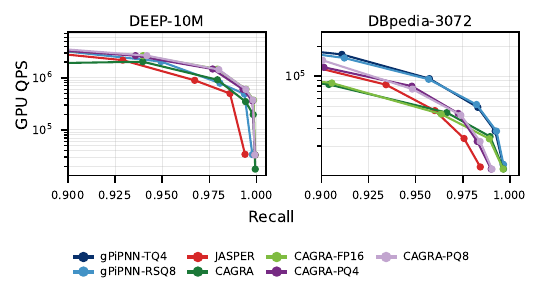}
  \caption{GPU QPS vs.\ recall for DEEP-10M ($d{=}96$) and
  DBpedia-3072 ($d{=}3072$); all datasets in
  \Cref{fig:gpu_anns_recall_grid}.}
  \label{fig:gpu_anns_recall_main}
\end{figure}

Build speed does not come at the cost of query quality.
\Cref{fig:gpu_anns_recall_main} plots \ourgpumethod's query throughput against recall for a low-dimensional ($d{=}96$, DEEP-10M) and a high-dimensional ($d{=}3072$, DBpedia-3072) dataset.
Across the recall range, \ourgpumethod's QPS is equal or higher than that of \textsc{Jasper} and the \textsc{CAGRA} variants.
The same conclusion holds on the full benchmark set, shown in \Cref{fig:gpu_anns_recall_grid}.

\myparagraph{$k$-NN Graph Build}
Next, we evaluate $k$-NN graph building.
We consider additional methods from cuVS, namely \textsc{NN-Descent} and \textsc{IVF-PQ}.
For each method and dataset, we sweep a variety of build-time parameters and retain the \emph{fastest} configuration whose resulting graph achieves recall\,$\geq$\,0.9 (\Cref{tab:gpu_build}); the full per-method, per-dataset build parameters and software versions are given in \Cref{app:gpu-params}.
Entries where no configuration meets the recall target are marked ``--''.
For \ourgpumethod-RSQ8 and cuVS \textsc{NN-Descent} we show the direct $k$-NN graph build time; for \textsc{Jasper}, \textsc{IVF-PQ}, and \textsc{CAGRA} it is the index-build time \emph{plus} the time to query all $n$ points against the built index.
% Recall is measured by sampling 5{,}000 points (seeded deterministically) and comparing the graph's stored top-10 edges against the brute-force top-10 with floats.
% \lars{Are we doing the same on CPU?}

% Richard: I will come back to this paragraph later
\ourgpumethod-RSQ8 is the fastest-building method on \textbf{all thirteen} datasets, including MS-MARCO-10M (MIPS, $d{=}768$), where it builds in 30.1~s against 39.4~s for the next-fastest \textsc{NN-Descent}.
Aggregated as a geometric mean, \ourgpumethod-RSQ8 is \textbf{7.12$\times$} faster than \textsc{Jasper}, \textbf{2.91$\times$} faster than \textsc{NN-Descent}, \textbf{9.84$\times$} faster than \textsc{CAGRA},
and \textbf{49.80$\times$} faster than \textsc{IVF-PQ8}.
\textsc{CAGRA} and \textsc{IVF-PQ8} are reported in the full comparison in
\Cref{tab:knn_build_full}. 
\textsc{CAGRA}'s ANNs construction works only on Euclidean data.

\subsection{CPU Ablations}
\label{sec:cpu-ablations}

\begin{figure}[t]
\centering
\includegraphics[width=0.85\linewidth]{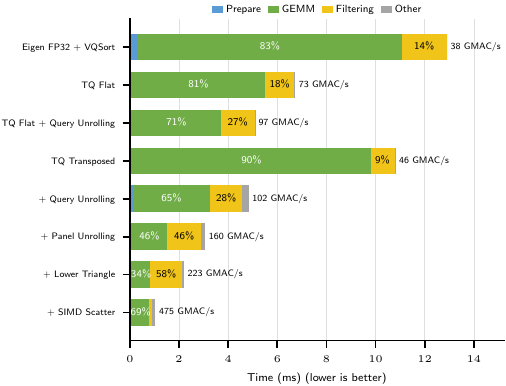}
\caption{\texttt{ProcessLeafTop2} ablation: runtime breakdown and effective
throughput across cumulative optimizations. Throughput counts $n^2d$
operations although \emph{Lower Triangle} computes $1.96\times$ fewer.}
\label{fig:leaftop2-ablation}
\end{figure}

\myparagraph{Leaf 2-NN kernel}
We evaluate the leaf 2-NN kernel described in \Cref{sec:cpu-kernels:leaf} on a single core of an Icelake CPU by scoring a leaf of $800$ points gathered from a dataset of 1M $768$-dimensional points. \Cref{fig:leaftop2-ablation} divides runtime into three components: \emph{prepare}, the dot-product \emph{GEMM}, and top-2 \emph{filtering}. \emph{Prepare} covers work done before the inner loop runs: 
decoding and packing the database vectors into the transposed panel layout, and zero-initializing the top-2 trackers. The original \ouroldmethod{}~\cite{rubel2026pipnn}, which scores the leaf with Eigen's float32 GEMM and extracts top-2 with Highway's VQSort, runs in $12.90$~ms. Our final optimized kernel runs in $1.04$~ms, a $12.4\times$ improvement.

We walk through the optimizations in \Cref{fig:leaftop2-ablation} in order. As we apply each optimization, the dominant cost increasingly shifts to the top-2 filtering step, until \emph{SIMD Scatter} addresses that bottleneck.
\emph{TQ Flat} replaces the float32 GEMM with an int8 dot product using \texttt{vpdpbusd} over the baseline's data layout (rows=points, columns=dimensions). 
%It uses AVX-512 VNNI's  quad multiply-accumulate to pack $4\times$ more elements per SIMD register. 
Processing 4 times more elements per SIMD instruction yields a $1.92\times$ speedup over the baseline.
The block-transposed panel layout (\emph{TQ Transposed}) regresses against \emph{TQ Flat} prior to query unrolling.
%because, unlike in the flat layout where each base vector has its own accumulator, all of the quads in the transposed layout accumulates into the same register, throttling instruction-level parallelism.
%\lars{the broadcast is per query. this doesnt make sense. the benefit of query unrolling is ILP, not reducing broadcast overhead. need to be more precise?}
\emph{Query Unrolling} subsequently allows the transposed layout to come out ahead of flat by enabling better instruction-level parallelism.
Transpose furthermore unlocks \emph{Panel Unrolling}, which yields another $1.58\times$ improvement by keeping the query broadcasts in registers across four database panels.

\emph{Lower Triangle} roughly halves the GEMM work ($1.96\times$), exposing filtering as the new bottleneck at $58\%$ of the total time.
Vectorizing the scatter (\emph{SIMD Scatter}, described in \Cref{sec:cpu-kernels:leaf}) is the largest single jump, $2.1\times$, and shrinks filtering back to roughly $10\%$.

In a standalone benchmark we were able to sustain $342$ GMAC/s when issuing \texttt{vpdpbusd} from registers (279 with memory loads).
The GEMM part of our kernel runs at $\frac{475}{0.69 \cdot 1.96} = 351$ GMAC/s, where $0.69$ is the fraction of time spent in GEMM.
This indicates that the GEMM runs at peak performance.
This also matches the theoretical expectation of $332.8$~GMAC/s on a $2.6$~GHz base clock core leveraging two execution ports, if we account for turboboost.

\begin{figure}
  \centering
  \includegraphics[width=0.85\linewidth]{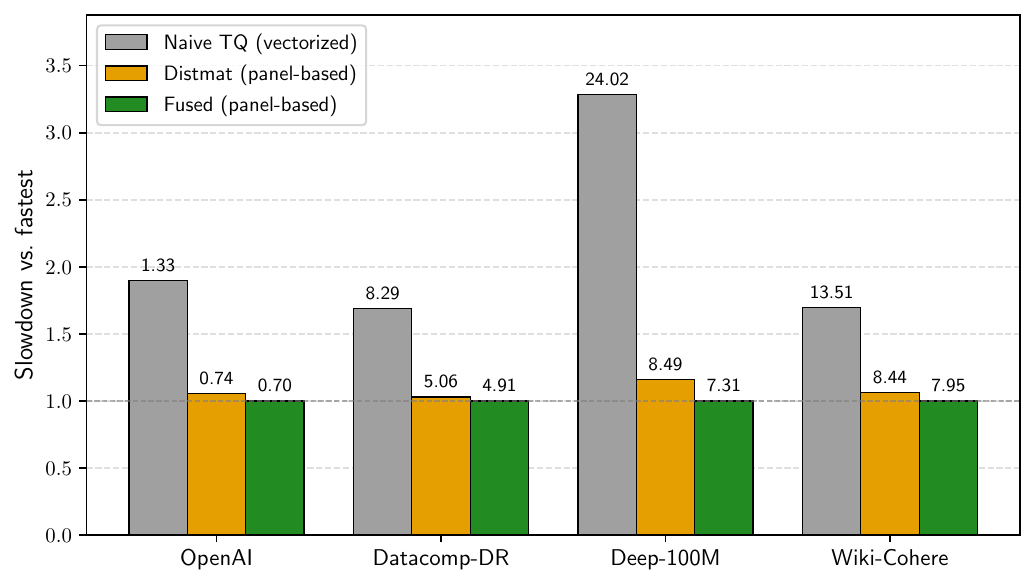}
  \caption{RobustPrune implementations absolute time in seconds (annotated above bar) and slowdown vs.\ fastest.}
  \label{fig:cpu-prune-ablation}
\end{figure}

\myparagraph{Pruning kernel} 
Recall that \ourmethod{} uses the \RobustPrune{} procedure from DiskANN \cite{jayaram2019diskann}. 
With the leaf and partitioning kernels accelerated, \RobustPrune{} becomes a non-negligible part of the runtime.
We compare three implementations which differ in how they compute distances between \RobustPrune candidates. 
The baseline ``Naive TQ (vectorized)'' computes the distances between a candidate and all prior survivors using one-to-many TurboQuant.
``Distmat (panel-based)'' precomputes the full $C \times C$
candidate--candidate distance matrix in a single GEMM pass using the techniques from \Cref{sec:cpu-kernels:overview}, then prunes using those distances.
Note that this approach is not able to exit early when a candidate gets pruned.
``Fused (panel-based)'' is a variant where we store only survivors and the next query batch (size 6) in the panel layout. 
We then compute distances from the next query batch against the survivors and itself.
After pruning, we copy over the survivors from this batch.
Compared to Distmat, this eliminates redundant work by only comparing with  pruned candidates in the same query batch.
Compared to the naive version, however, a query still does not exit early if pruned.
%\lars{We could actually do this. And make this perform better. In the first 2-3 iterations, we do a lot of wasted computation with empty lanes}

As can be seen in \Cref{fig:cpu-prune-ablation}, 
both panel-based strategies provide significant speedup over the Naive method (DistMat gets a $2.1\times$ improvement on average over Naive).
Additionally, leveraging the Fused version gives a $1.08\times$ improvement over Distmat.

\subsection{GPU Ablations}
\label{sec:gpu_throughput}

\Cref{fig:component_breakdown} in Appendix~\ref{app:perkernel-roofline} shows per-stage and per-kernel runtime breakdowns on GPU.
Leaf processing dominates the runtime on GPU as well, though less than on CPU. We perform a Speed of Light (SOL) analysis to identify the $\%$ of system utilization with respect to the theoretical hardware peak.

\myparagraph{Leaf $2$-NN kernel}
The performance bottleneck of the kernel on the H100 GPU shifts depending on the vector dimension. 
At higher dimensions ($d \geq 1536$, e.g., openai-2M and dbpedia-3072), data streaming for candidate comparisons saturates the L2 cache bandwidth.
Unified memory throughput reaches 82--95\%, indicating that L2 is the limiting tier rather than DRAM, which remains at 37--52\%. 
Meanwhile, the SM pipelines hold steady at 46--49\%, falling below Nsight Compute's (NCU) 60\% threshold for a compute-bound state.
For intermediate dimensions ($d = 768$, e.g., wiki-all-10M), neither the compute nor memory pipelines saturate: the SM sits at 50\% and the memory hierarchy at 41\%. 
Here, NCU attributes the performance gap to latency rather than throughput. 
The kernel achieves only 26--28\% occupancy, tightly constrained by its 28\% theoretical limit due to high shared-memory pressure and an allocation of 96 registers per thread.
This occupancy limitation persists at the lowest dimensions ($d = 96$, e.g., deep-10M; \Cref{tab:leaf_roofline_4mode}). The working set resides entirely within cache which drops DRAM traffic to 4\% and leaves the kernel strictly occupancy-bound.

\myparagraph{Index Quality}
\Cref{fig:gpu_anns_recall_main} shows QPS vs recall on a low- and a high-dimensional dataset. 
Across the recall range \ourgpumethod{}'s curves sit on or above those of \textsc{Jasper} and \textsc{CAGRA}, and the same holds on the full benchmark set (\Cref{fig:gpu_anns_recall_grid}). 

Using \DirectEdgePrune{} produces $30$--$58\%$ denser graphs, since the method does not prune some edges that \RobustPrune{} would. 
Still, index quality is comparable between the two pruning methods (\Cref{tab:prune_ablation}). \DirectEdgePrune{} cuts GPU build time by up to $1.4\times$ on the high-dimensional datasets. While \DirectEdgePrune{} could increase memory footprint, the maximum degree between the two pruning methods is identical, so in the Jasper query implementation there is no difference in terms of allocated memory.

\section{Discussion \& Conclusion}

\ouroldmethod{} introduced a fast, hardware-friendly route to high-quality proximity-graph construction, but its original implementation left substantial performance on the table.
%
%The bulk of its work is dense pairwise distance computation followed by per-row top-$K$ extraction, both of which were carried out by full-precision GEMM and post-hoc vectorized sorting.
%
In this paper we fully realize \ouroldmethod{}'s potential by accelerating the GEMM and top-$K$ steps, via quantization, fusing GEMM with per-query top-$K$ selection into a single in-register pass, as well as vectorizing and tailoring the top-$K$ selection to the specific $K$.
Instantiating this design on shared-memory CPUs as \ourmethod{} and on GPUs as \ourgpumethod{} yields  large speedups over the strongest prior state-of-the-art methods on both platforms while preserving equivalent query performance.

In future work we plan to scale beyond a single machine and port the design to other accelerators.
\ouroldmethod{}'s structure is well-suited to the distributed setting: partitioning creates independent subproblems, leaves can be processed independently,  and \HashPrune{} merges candidate sets without coordinating on global state.
%
%In future work, we plan to study a distributed version of \ourmethod{} on datasets with tens of billions of embeddings, a regime where single machine approaches may no longer be feasible.

%\section{Acknowledgement}
%This work is supported by NSF grants CCF-2403235 and CNS2317194 and by the National Science Foundation Graduate Research Fellowship Program under Grant No. DGE 2236417. Any opinions, findings, and conclusions or recommendations expressed in this material are those of the author(s) and do not necessarily reflect the views of the National Science Foundation.

%\clearpage

\bibliographystyle{ACM-Reference-Format}
\bibliography{references}

\clearpage
\appendix
\section{Beam Search Pseudocode}
\label{app:beam-search}

For completeness, we provide pseudocode for the standard beam-search graph-query procedure described in Section~\ref{sec:preliminaries}.
Given a navigation graph over points $\mX$, a beam search is a greedy search with intermediate storage (called the beam). Given a source node $s$ in the graph and a query point $q$, a beam search with beam width $L$ works as follows. We maintain two sets: a set $B$ containing up to $L$ encountered points that are closest to $q$ (often called the ``beam''), and the set $V$ of all points visited. Visiting a point $p \in B \setminus V$ adds all its unseen neighbors to the beam, then adds $p$ to the visited set. The algorithm begins by initializing the beam with only the source $s$ and an empty visited set. We then visit the unvisited node in the beam closest to $q$, repeating until every node in the beam has been visited. Because the distance of the points in the beam to the target decreases monotonically, this process converges, at which point we return the final state of the beam.

\begin{algorithm}
    \caption{BeamSearch}
    \label{alg:beam_search}
    \SetKwData{beam}{$B$}
    \SetKwData{beamwidth}{$L$}
    \SetKwData{visited}{$V$}
    \SetKwData{graph}{$G$}
    \SetKwData{source}{$s$}
    \SetKwData{target}{$q$}
    \SetKwData{points}{$\mX$}
    \SetKwData{curr}{$p$}
    \SetKwFunction{FNeighbors}{Neighbors}
    \KwIn{$\graph = (\points, E)$ a graph over $\points$, $\source \in \points$ an entry point, $\target$ a query point, and $\beamwidth \in \mathbb{N}$ a beam width}
    \KwOut{$\beam \subseteq \points$ containing the $\beamwidth$ closest points to $\target$ encountered during the search}
    $\beam \gets \{\source\}$\;
    $\visited \gets \varnothing$\;
    \While{$\beam \setminus \visited \neq \varnothing$}{
        $\curr \gets \argmin_{u \in \beam \setminus \visited} d(u,\target)$\;
        $\beam \gets \beam \cup \FNeighbors{\curr}$\;
        $\visited \gets \visited \cup \{\curr\}$\;
        \If{$|\beam| > \beamwidth$}{
            $\beam \gets$ the $\beamwidth$ points in $\beam$ with smallest distance to $\target$\;
        }
    }
    \Return \beam\;
\end{algorithm}

\section{\HashPrune{} Pseudocode}
\label{app:hashprune}

Here we document and give pseudocode for the \HashPrune{} online pruning procedure used by \ouroldmethod{} (\Cref{sec:pipnn}). The algorithm is reproduced verbatim from \cite{rubel2026pipnn}. 

\myparagraph{Motivation}
Pruning ensures that each vertex connects to neighbors in many different directions, so that greedy search can route quickly from any point to any other. \HashPrune{} achieves this with locality-sensitive hashing (LSH)~\cite{charikar2002similarity}: it discards candidate edges that are directionally similar to edges already kept. Working with relatively low dimensional hashes ($\approx12$) also lets the procedure operate on low-dimensional sketches rather than full vectors, so pruning avoids fetching high-dimensional data.  

\myparagraph{Residualized hashing}
\HashPrune{} draws a fixed set of random hyperplanes through the origin. For each point $p$ it derives an individualized hash $h_p$ that hashes the \emph{residual} $c - p$ of a candidate $c$: the $i$-th bit records which side of the $i$-th hyperplane the residual falls on. Two candidates collide under $h_p$ with probability $(1 - \theta/\pi)^m$, where $\theta$ is the angle between $c - p$ and $c' - p$ and $m$ is the number of hyperplanes~\cite{charikar2002similarity}. Hashing the residual thus draws probabilistic cones centered on $p$, and the number of hyperplanes controls how finely those cones partition direction space.

\myparagraph{Insertion}
To prune the candidate stream $\mathcal{V}$ for a point $p$, \HashPrune{} maintains a reservoir $M$ of size at most $\lMax$ keyed by hash value. For each candidate $c$ it computes $h_p(c)$. If the bucket is already occupied, it keeps whichever of the two candidates lies closer to $p$. Otherwise it inserts $c$, and when the reservoir is full it evicts the candidate farthest from $p$. The reservoir therefore retains the nearest point in each cone while never exceeding $\lMax$ entries, which bounds memory up front and lets it be allocated in a single contiguous block rather than grown dynamically.

\myparagraph{History independence}
The final reservoir depends only on the candidate set, not on the order candidates arrive in: dropping the size limit leaves exactly the nearest point in each bucket, and the size limit only ever evicts the farthest retained candidate. History independence lets candidates be merged in arbitrary order, partition by partition, without fragmenting the graph, and makes \ouroldmethod{}'s partitioning-based construction deterministic.

\begin{algorithm}
    \caption{\HashPrune{}}
    \label{alg:hashprune}
    \SetKwData{V}{$\mathcal{V}$}
    \SetKwData{pstar}{$p^*$}
    \SetKwData{E}{$E$}
    \SetKwData{dict}{$M$}
    \SetKwFunction{FGet}{Get}
    \SetKwFunction{FSet}{Set}
    \SetKwFunction{FHasKey}{HasKey}
    \SetKwFunction{FRemove}{Remove}
    \SetKwFunction{FValues}{Values}
    \KwIn{\V a stream of candidate points, fitted with dissimilarity
    measure $\lVert\cdot,\cdot\rVert$, point $p$, size limit $\lMax \in \mathbb{N}$}
    \KwOut{A set $C \subseteq \V$ of size at most $\min(|\V|, \lMax)$}
    $\dict \gets$ an empty key-value store mapping hashes to points\;
    \For{$c \in \V$}{
      \If{\dict.\FHasKey{$h_p(c)$}}{
            \If{$\lVert p, c \rVert < \lVert p, {}$\dict.\FGet{$h_p(c)$}$\rVert$}{
                \dict.\FSet{$h_p(c), c$}\;
              }
        }
        \Else{
          \If{$|\dict| < \lMax$}{
                \dict.\FSet{$h_p(c), c$}\;
            }
            \Else{
                $z = \argmax_{y \in \dict} \lVert p, y \rVert$\;
                \If{$\lVert p, c \rVert < \lVert p, z \rVert$}{
                  \dict.\FRemove{$h_p(z)$}\;
                    \dict.\FSet{$h_p(c), c$}\;
                }
            }
        }
    }
    \Return \dict.\FValues{}\;
\end{algorithm}

\section{Additional Experiments}
\label{app:additional-experiments}

This appendix presents the full per-dataset query-throughput and $k$-NN-graph
recall results, together with supporting ablations, for the
datasets listed in
\Cref{tab:modern_datasets_full_compact}. For datasets that we report at
multiple scales (DEEP, MS-Turing, Text2Image, MSMARCO), we show the
largest slice.

\section{Full GPU Build-Time Tables}
\label{app:gpu-build-full}
For completeness, \Cref{tab:ann_build_full,tab:knn_build_full} report the
GPU build-time comparisons of \Cref{sec:gpu_ann_build} across the full
set of \ourgpumethod{} and baseline
variants---adding \ourgpumethod-TQ4, the fp16/tf32 unquantized
modes, the four CAGRA variants, and \textsc{NN-Descent-FP16}. The table in the main part 
(\Cref{tab:gpu_build}) contains a subset
of these configurations. The build parameters behind these tables, the tuning protocol, and the
software environment are reported in \Cref{app:gpu-params}.
% Auto-generated by scripts/finals_paper_plots.py
% Requires: \usepackage{colortbl, xcolor, booktabs}
\begin{table*}[t]
  \centering
  \caption{Full GPU ANNS-index native build times (s) across all \ourgpumethod{} modes (TQ4/RSQ8, unquantized FP16/TF32) and CAGRA variants; cf.\ the headline \Cref{tab:gpu_build}. Colored as in \Cref{tab:cpu_build}. ``--'' marks a run that did not complete or exceeded the 80\,GB H100 budget; the unquantized FP16/TF32 modes carry a $2$--$4\times$ larger decoded-row footprint and so miss the largest cells (TF32 at 100M, and both on DataCompDr-10M), and plain \textsc{CAGRA} runs only on the L2 datasets (its cuVS NN-Descent build is L2-only). The 100M-scale rows (DEEP-100M, MS-Turing-100M) are shown for completeness but excluded from the main-text tables and the aggregate speedups of \Cref{tab:gpu_speedup_summary}; the \textsc{CAGRA} DEEP-100M outlier is discussed in \Cref{app:gpu-build-full}.
  }
  \label{tab:ann_build_full}
  \setlength{\tabcolsep}{4pt}
  \small
  \begin{tabular}{lccccccccc}
    \toprule
    & \multicolumn{4}{c}{gPiPNN} & \multicolumn{1}{c}{JASPER} & \multicolumn{4}{c}{CAGRA} \\
    \cmidrule(lr){2-5} \cmidrule(lr){6-6} \cmidrule(lr){7-10}
    Dataset & gPiPNN-TQ4 & gPiPNN-RSQ8 & gPiPNN-FP16 & gPiPNN-TF32 & JASPER & CAGRA & CAGRA-FP16 & CAGRA-PQ4 & CAGRA-PQ8 \\
    \midrule
    DEEP-10M & \cellcolor[rgb]{0.780,0.920,0.660}\textbf{2.31} & \cellcolor[rgb]{0.781,0.920,0.660}2.32 & \cellcolor[rgb]{0.794,0.922,0.659}2.57 & \cellcolor[rgb]{0.801,0.923,0.659}2.73 & \cellcolor[rgb]{0.950,0.943,0.652}8.78 & \cellcolor[rgb]{0.960,0.720,0.660}115 & \cellcolor[rgb]{0.960,0.720,0.660}192 & \cellcolor[rgb]{0.960,0.720,0.660}298 & \cellcolor[rgb]{0.960,0.720,0.660}266 \\
    MS-Turing-10M & \cellcolor[rgb]{0.780,0.920,0.660}\textbf{2.63} & \cellcolor[rgb]{0.780,0.920,0.660}2.64 & \cellcolor[rgb]{0.792,0.922,0.659}2.88 & \cellcolor[rgb]{0.799,0.923,0.659}3.07 & \cellcolor[rgb]{0.995,0.924,0.651}18.1 & \cellcolor[rgb]{0.960,0.720,0.660}116 & \cellcolor[rgb]{0.960,0.720,0.660}124 & \cellcolor[rgb]{0.960,0.720,0.660}176 & \cellcolor[rgb]{0.960,0.720,0.660}193 \\
    DEEP-100M & \cellcolor[rgb]{0.780,0.920,0.660}\textbf{19.4} & \cellcolor[rgb]{0.781,0.920,0.660}19.6 & \cellcolor[rgb]{0.798,0.922,0.659}22.3 & \cellcolor[rgb]{0.92,0.92,0.92}{--} & \cellcolor[rgb]{0.995,0.949,0.650}105 & \cellcolor[rgb]{0.960,0.720,0.660}3940 & \cellcolor[rgb]{0.960,0.720,0.660}5391 & \cellcolor[rgb]{0.960,0.720,0.660}3706 & \cellcolor[rgb]{0.960,0.720,0.660}3907 \\
    MS-Turing-100M & \cellcolor[rgb]{0.780,0.920,0.660}\textbf{20.1} & \cellcolor[rgb]{0.781,0.920,0.660}20.3 & \cellcolor[rgb]{0.798,0.922,0.659}23.2 & \cellcolor[rgb]{0.92,0.92,0.92}{--} & \cellcolor[rgb]{0.991,0.899,0.652}166 & \cellcolor[rgb]{0.92,0.92,0.92}{--} & \cellcolor[rgb]{0.92,0.92,0.92}{--} & \cellcolor[rgb]{0.92,0.92,0.92}{--} & \cellcolor[rgb]{0.92,0.92,0.92}{--} \\
    DBpedia-1536 & \cellcolor[rgb]{0.780,0.920,0.660}\textbf{0.54} & \cellcolor[rgb]{0.785,0.921,0.660}0.57 & \cellcolor[rgb]{0.868,0.932,0.656}1.08 & \cellcolor[rgb]{0.919,0.939,0.654}1.62 & \cellcolor[rgb]{0.969,0.770,0.658}11.9 & \cellcolor[rgb]{0.970,0.779,0.657}11.1 & \cellcolor[rgb]{0.970,0.776,0.658}11.3 & \cellcolor[rgb]{0.960,0.720,0.660}45.9 & \cellcolor[rgb]{0.960,0.720,0.660}51.8 \\
    DBpedia-3072 & \cellcolor[rgb]{0.780,0.920,0.660}\textbf{0.86} & \cellcolor[rgb]{0.786,0.921,0.660}0.90 & \cellcolor[rgb]{0.888,0.935,0.655}1.99 & \cellcolor[rgb]{0.941,0.942,0.653}3.03 & \cellcolor[rgb]{0.960,0.720,0.660}39.4 & \cellcolor[rgb]{0.969,0.771,0.658}18.5 & \cellcolor[rgb]{0.969,0.773,0.658}18.2 & \cellcolor[rgb]{0.960,0.720,0.660}97.9 & \cellcolor[rgb]{0.960,0.720,0.660}124 \\
    Wiki-Qwen-500K & \cellcolor[rgb]{0.780,0.920,0.660}\textbf{0.57} & \cellcolor[rgb]{0.789,0.921,0.660}0.61 & \cellcolor[rgb]{0.886,0.934,0.655}1.30 & \cellcolor[rgb]{0.938,0.942,0.653}1.96 & \cellcolor[rgb]{0.979,0.831,0.655}7.81 & \cellcolor[rgb]{0.972,0.786,0.657}10.9 & \cellcolor[rgb]{0.972,0.789,0.657}10.6 & \cellcolor[rgb]{0.960,0.720,0.660}64.9 & \cellcolor[rgb]{0.960,0.720,0.660}73.4 \\
    OpenAI-2M & \cellcolor[rgb]{0.780,0.920,0.660}\textbf{1.22} & \cellcolor[rgb]{0.786,0.921,0.660}1.29 & \cellcolor[rgb]{0.879,0.933,0.656}2.66 & \cellcolor[rgb]{0.935,0.941,0.653}4.12 & \cellcolor[rgb]{0.969,0.773,0.658}26.0 & \cellcolor[rgb]{0.963,0.735,0.659}34.6 & \cellcolor[rgb]{0.963,0.735,0.659}34.6 & \cellcolor[rgb]{0.960,0.720,0.660}133 & \cellcolor[rgb]{0.960,0.720,0.660}149 \\
    Wiki-all-1M & \cellcolor[rgb]{0.780,0.920,0.660}\textbf{0.39} & \cellcolor[rgb]{0.795,0.922,0.659}0.44 & \cellcolor[rgb]{0.843,0.929,0.657}0.64 & \cellcolor[rgb]{0.887,0.935,0.655}0.91 & \cellcolor[rgb]{0.998,0.950,0.650}2.17 & \cellcolor[rgb]{0.92,0.92,0.92}{--} & \cellcolor[rgb]{0.92,0.92,0.92}{--} & \cellcolor[rgb]{0.960,0.720,0.660}26.9 & \cellcolor[rgb]{0.960,0.720,0.660}29.4 \\
    Wiki-all-10M & \cellcolor[rgb]{0.780,0.920,0.660}\textbf{3.41} & \cellcolor[rgb]{0.784,0.921,0.660}3.53 & \cellcolor[rgb]{0.868,0.932,0.656}6.78 & \cellcolor[rgb]{0.929,0.940,0.653}11.0 & \cellcolor[rgb]{0.988,0.881,0.653}32.2 & \cellcolor[rgb]{0.92,0.92,0.92}{--} & \cellcolor[rgb]{0.92,0.92,0.92}{--} & \cellcolor[rgb]{0.960,0.720,0.660}381 & \cellcolor[rgb]{0.960,0.720,0.660}456 \\
    MS-MARCO-10M & \cellcolor[rgb]{0.780,0.920,0.660}\textbf{3.38} & \cellcolor[rgb]{0.784,0.921,0.660}3.50 & \cellcolor[rgb]{0.865,0.932,0.656}6.58 & \cellcolor[rgb]{0.924,0.940,0.653}10.5 & \cellcolor[rgb]{0.964,0.746,0.659}88.3 & \cellcolor[rgb]{0.92,0.92,0.92}{--} & \cellcolor[rgb]{0.92,0.92,0.92}{--} & \cellcolor[rgb]{0.960,0.720,0.660}370 & \cellcolor[rgb]{0.960,0.720,0.660}446 \\
    DataCompDr-10M & \cellcolor[rgb]{0.780,0.920,0.660}\textbf{5.81} & \cellcolor[rgb]{0.786,0.921,0.660}6.08 & \cellcolor[rgb]{0.92,0.92,0.92}{--} & \cellcolor[rgb]{0.92,0.92,0.92}{--} & \cellcolor[rgb]{0.979,0.832,0.655}79.3 & \cellcolor[rgb]{0.92,0.92,0.92}{--} & \cellcolor[rgb]{0.92,0.92,0.92}{--} & \cellcolor[rgb]{0.960,0.720,0.660}625 & \cellcolor[rgb]{0.960,0.720,0.660}790 \\
    Wiki-Cohere-10M & \cellcolor[rgb]{0.780,0.920,0.660}\textbf{3.41} & \cellcolor[rgb]{0.787,0.921,0.660}3.59 & \cellcolor[rgb]{0.866,0.932,0.656}6.70 & \cellcolor[rgb]{0.926,0.940,0.653}10.8 & \cellcolor[rgb]{0.972,0.788,0.657}64.5 & \cellcolor[rgb]{0.92,0.92,0.92}{--} & \cellcolor[rgb]{0.92,0.92,0.92}{--} & \cellcolor[rgb]{0.960,0.720,0.660}345 & \cellcolor[rgb]{0.960,0.720,0.660}415 \\
    Text2Image-10M & \cellcolor[rgb]{0.780,0.920,0.660}\textbf{2.35} & \cellcolor[rgb]{0.782,0.920,0.660}2.39 & \cellcolor[rgb]{0.809,0.924,0.659}2.96 & \cellcolor[rgb]{0.830,0.927,0.658}3.47 & \cellcolor[rgb]{0.960,0.720,0.660}90.2 & \cellcolor[rgb]{0.92,0.92,0.92}{--} & \cellcolor[rgb]{0.92,0.92,0.92}{--} & \cellcolor[rgb]{0.960,0.720,0.660}231 & \cellcolor[rgb]{0.960,0.720,0.660}250 \\
    MS-Arctic-1M & \cellcolor[rgb]{0.780,0.920,0.660}\textbf{0.46} & \cellcolor[rgb]{0.802,0.923,0.659}0.55 & \cellcolor[rgb]{0.853,0.930,0.657}0.81 & \cellcolor[rgb]{0.900,0.936,0.655}1.18 & \cellcolor[rgb]{0.991,0.900,0.652}3.75 & \cellcolor[rgb]{0.92,0.92,0.92}{--} & \cellcolor[rgb]{0.92,0.92,0.92}{--} & \cellcolor[rgb]{0.960,0.720,0.660}34.2 & \cellcolor[rgb]{0.960,0.720,0.660}37.6 \\
    \bottomrule
  \end{tabular}
\end{table*}

% Auto-generated by scripts/finals_paper_plots.py
% Requires: \usepackage{colortbl, xcolor, booktabs}
\begin{table*}[t]
  \centering
  \caption{Full $k$-NN-graph native build times (s), fastest configuration of each method reaching graph-fidelity recall $\geq$ 0.9; cf.\ the headline \Cref{tab:gpu_build}. \ourgpumethod{} is shown in TQ4/RSQ8 plus unquantized FP16/TF32 supplement modes; IVF-PQ4 is omitted (it met the floor on no dataset). Colored as in \Cref{tab:cpu_build}. ``--'' marks cells where no configuration met the floor, the build ran out of memory, or the metric is unsupported. The 4-bit TQ4 mode is too lossy to reach 0.9 on the harder cells, while the FP16/TF32 modes clear the floor wherever they fit; cuVS \textsc{CAGRA} (NN-Descent build) is L2-only, so it is absent on the MIPS datasets.
  }
  \label{tab:knn_build_full}
  \setlength{\tabcolsep}{4pt}
  \small
  \begin{tabular}{lccccccccc}
    \toprule
    & \multicolumn{4}{c}{gPiPNN} & \multicolumn{1}{c}{JASPER} & \multicolumn{2}{c}{NN-Descent} & \multicolumn{1}{c}{IVF-PQ} & \multicolumn{1}{c}{CAGRA} \\
    \cmidrule(lr){2-5} \cmidrule(lr){6-6} \cmidrule(lr){7-8} \cmidrule(lr){9-9} \cmidrule(lr){10-10}
    Dataset & gPiPNN-TQ4 & gPiPNN-RSQ8 & gPiPNN-FP16 & gPiPNN-TF32 & JASPER & NN-Descent & NN-Descent-FP16 & IVF-PQ8 & CAGRA \\
    \midrule
    DEEP-10M & \cellcolor[rgb]{0.92,0.92,0.92}{--} & \cellcolor[rgb]{0.780,0.920,0.660}\textbf{4.43} & \cellcolor[rgb]{0.786,0.921,0.660}4.65 & \cellcolor[rgb]{0.791,0.922,0.659}4.84 & \cellcolor[rgb]{0.893,0.935,0.655}10.8 & \cellcolor[rgb]{0.996,0.929,0.651}29.1 & \cellcolor[rgb]{0.996,0.925,0.651}30.0 & \cellcolor[rgb]{0.984,0.857,0.654}49.9 & \cellcolor[rgb]{0.967,0.758,0.658}106 \\
    MS-Turing-10M & \cellcolor[rgb]{0.92,0.92,0.92}{--} & \cellcolor[rgb]{0.780,0.920,0.660}\textbf{15.9} & \cellcolor[rgb]{0.783,0.920,0.660}16.3 & \cellcolor[rgb]{0.791,0.922,0.659}17.4 & \cellcolor[rgb]{0.803,0.923,0.659}19.1 & \cellcolor[rgb]{0.870,0.932,0.656}32.3 & \cellcolor[rgb]{0.862,0.931,0.656}30.3 & \cellcolor[rgb]{0.92,0.92,0.92}{--} & \cellcolor[rgb]{0.996,0.928,0.651}106 \\
    DBpedia-1536 & \cellcolor[rgb]{0.780,0.920,0.660}\textbf{1.81} & \cellcolor[rgb]{0.782,0.920,0.660}1.84 & \cellcolor[rgb]{0.807,0.924,0.659}2.24 & \cellcolor[rgb]{0.841,0.928,0.657}2.92 & \cellcolor[rgb]{0.998,0.936,0.651}11.3 & \cellcolor[rgb]{0.903,0.937,0.654}4.78 & \cellcolor[rgb]{0.903,0.937,0.654}4.78 & \cellcolor[rgb]{0.960,0.720,0.660}64.1 & \cellcolor[rgb]{0.991,0.899,0.652}14.9 \\
    DBpedia-3072 & \cellcolor[rgb]{0.780,0.920,0.660}\textbf{2.94} & \cellcolor[rgb]{0.782,0.920,0.660}3.00 & \cellcolor[rgb]{0.819,0.925,0.658}4.00 & \cellcolor[rgb]{0.860,0.931,0.656}5.52 & \cellcolor[rgb]{0.994,0.913,0.652}21.8 & \cellcolor[rgb]{0.879,0.933,0.656}6.40 & \cellcolor[rgb]{0.879,0.934,0.655}6.41 & \cellcolor[rgb]{0.960,0.720,0.660}143 & \cellcolor[rgb]{0.992,0.905,0.652}23.3 \\
    Wiki-Qwen-500K & \cellcolor[rgb]{0.780,0.920,0.660}\textbf{2.35} & \cellcolor[rgb]{0.782,0.920,0.660}2.38 & \cellcolor[rgb]{0.816,0.925,0.658}3.11 & \cellcolor[rgb]{0.857,0.931,0.656}4.29 & \cellcolor[rgb]{0.999,0.950,0.650}13.1 & \cellcolor[rgb]{0.835,0.927,0.658}3.60 & \cellcolor[rgb]{0.835,0.927,0.658}3.60 & \cellcolor[rgb]{0.960,0.720,0.660}93.5 & \cellcolor[rgb]{0.998,0.936,0.651}14.6 \\
    OpenAI-2M & \cellcolor[rgb]{0.897,0.936,0.655}6.15 & \cellcolor[rgb]{0.780,0.920,0.660}\textbf{2.46} & \cellcolor[rgb]{0.827,0.926,0.658}3.55 & \cellcolor[rgb]{0.868,0.932,0.656}4.89 & \cellcolor[rgb]{0.987,0.874,0.653}24.4 & \cellcolor[rgb]{0.989,0.949,0.650}12.7 & \cellcolor[rgb]{0.990,0.949,0.650}12.7 & \cellcolor[rgb]{0.960,0.720,0.660}186 & \cellcolor[rgb]{0.977,0.820,0.656}36.6 \\
    Wiki-all-1M & \cellcolor[rgb]{0.780,0.920,0.660}\textbf{1.77} & \cellcolor[rgb]{0.781,0.920,0.660}1.78 & \cellcolor[rgb]{0.797,0.922,0.659}2.03 & \cellcolor[rgb]{0.822,0.926,0.658}2.46 & \cellcolor[rgb]{0.902,0.937,0.654}4.60 & \cellcolor[rgb]{0.885,0.934,0.655}4.03 & \cellcolor[rgb]{0.884,0.934,0.655}4.02 & \cellcolor[rgb]{0.968,0.764,0.658}40.1 & \cellcolor[rgb]{0.92,0.92,0.92}{--} \\
    Wiki-all-10M & \cellcolor[rgb]{0.928,0.940,0.653}27.0 & \cellcolor[rgb]{0.780,0.920,0.660}\textbf{8.50} & \cellcolor[rgb]{0.817,0.925,0.658}11.3 & \cellcolor[rgb]{0.858,0.931,0.656}15.7 & \cellcolor[rgb]{0.987,0.877,0.653}82.6 & \cellcolor[rgb]{0.975,0.947,0.651}39.2 & \cellcolor[rgb]{0.976,0.947,0.651}39.5 & \cellcolor[rgb]{0.960,0.720,0.660}1721 & \cellcolor[rgb]{0.92,0.92,0.92}{--} \\
    MS-MARCO-10M & \cellcolor[rgb]{0.92,0.92,0.92}{--} & \cellcolor[rgb]{0.780,0.920,0.660}\textbf{30.1} & \cellcolor[rgb]{0.806,0.924,0.659}37.0 & \cellcolor[rgb]{0.849,0.929,0.657}51.9 & \cellcolor[rgb]{0.980,0.832,0.655}410 & \cellcolor[rgb]{0.814,0.925,0.658}39.4 & \cellcolor[rgb]{0.814,0.925,0.658}39.5 & \cellcolor[rgb]{0.92,0.92,0.92}{--} & \cellcolor[rgb]{0.92,0.92,0.92}{--} \\
    DataCompDr-10M & \cellcolor[rgb]{0.929,0.940,0.653}50.6 & \cellcolor[rgb]{0.780,0.920,0.660}\textbf{15.7} & \cellcolor[rgb]{0.92,0.92,0.92}{--} & \cellcolor[rgb]{0.92,0.92,0.92}{--} & \cellcolor[rgb]{0.991,0.899,0.652}129 & \cellcolor[rgb]{0.92,0.92,0.92}{--} & \cellcolor[rgb]{0.92,0.92,0.92}{--} & \cellcolor[rgb]{0.92,0.92,0.92}{--} & \cellcolor[rgb]{0.92,0.92,0.92}{--} \\
    Wiki-Cohere-10M & \cellcolor[rgb]{0.92,0.92,0.92}{--} & \cellcolor[rgb]{0.780,0.920,0.660}\textbf{31.1} & \cellcolor[rgb]{0.808,0.924,0.659}38.7 & \cellcolor[rgb]{0.852,0.930,0.657}54.8 & \cellcolor[rgb]{0.995,0.921,0.651}217 & \cellcolor[rgb]{0.839,0.928,0.657}49.4 & \cellcolor[rgb]{0.839,0.928,0.657}49.5 & \cellcolor[rgb]{0.960,0.720,0.660}3127 & \cellcolor[rgb]{0.92,0.92,0.92}{--} \\
    Text2Image-10M & \cellcolor[rgb]{0.92,0.92,0.92}{--} & \cellcolor[rgb]{0.780,0.920,0.660}\textbf{5.00} & \cellcolor[rgb]{0.791,0.922,0.659}5.47 & \cellcolor[rgb]{0.802,0.923,0.659}5.95 & \cellcolor[rgb]{0.969,0.773,0.658}106 & \cellcolor[rgb]{0.998,0.936,0.651}31.1 & \cellcolor[rgb]{0.998,0.937,0.651}31.1 & \cellcolor[rgb]{0.92,0.92,0.92}{--} & \cellcolor[rgb]{0.92,0.92,0.92}{--} \\
    MS-Arctic-1M & \cellcolor[rgb]{0.780,0.920,0.660}\textbf{0.90} & \cellcolor[rgb]{0.788,0.921,0.660}0.96 & \cellcolor[rgb]{0.815,0.925,0.658}1.19 & \cellcolor[rgb]{0.846,0.929,0.657}1.51 & \cellcolor[rgb]{0.960,0.720,0.660}39.0 & \cellcolor[rgb]{0.983,0.948,0.651}4.45 & \cellcolor[rgb]{0.984,0.948,0.651}4.45 & \cellcolor[rgb]{0.960,0.720,0.660}70.9 & \cellcolor[rgb]{0.92,0.92,0.92}{--} \\
    \bottomrule
  \end{tabular}
\end{table*}

\myparagraph{Aggregate speedups}
\Cref{tab:gpu_speedup_summary} summarizes the per-dataset build times of
\Cref{tab:gpu_build} as the geometric-mean and
maximum build-time speedup of \ourgpumethod-RSQ8 over each GPU baseline,
taken over the datasets that baseline completes. On proximity-graph
(ANNS index) construction \ourgpumethod-RSQ8 is the fastest builder on
every dataset, a geometric-mean $13.3\times$ over \textsc{Jasper} and
$27\times$ over unquantized \textsc{CAGRA}, peaking at $43.7\times$
(DBpedia-3072) and $50\times$ (DEEP-10M). On $k$-NN-graph construction
it is a geometric-mean $7.12\times$, $2.91\times$, and $9.84\times$
faster than \textsc{Jasper}, \textsc{NN-Descent}, and \textsc{CAGRA}
respectively, and is the fastest builder on every cell.
These ANNS-index aggregates are taken over the $\leq$10M-scale datasets;
the two 100M-scale rows of \Cref{tab:ann_build_full} are excluded, for the
reason discussed next.
\begin{table}[t]
  \centering
  \caption{Build-time speedup of \ourgpumethod-RSQ8 over each GPU baseline,
  reported as the geometric mean and maximum over the datasets that
  baseline completes (computed from \Cref{tab:gpu_build} and the full
  \Cref{tab:knn_build_full}); $n$ is the number of such datasets. The
  ANNS-index rows exclude the two 100M-scale datasets, whose \textsc{CAGRA}
  build is an outlier (see \Cref{app:gpu-build-full}).}
  \label{tab:gpu_speedup_summary}
  \setlength{\tabcolsep}{6pt}
  \footnotesize
  \begin{tabular}{llccc}
    \toprule
    Build & Baseline & $n$ & Geo.\ mean & Max \\
    \midrule
    ANNS index   & \textsc{Jasper}     & 13 & $13.3\times$  & $43.7\times$ \\
    ANNS index   & \textsc{CAGRA}      & 6 & $27\times$    & $50\times$  \\
    \midrule
    $k$-NN graph & \textsc{Jasper}     & 13 & $7.12\times$  & $40.7\times$ \\
    $k$-NN graph & \textsc{NN-Descent} & 12 & $2.91\times$  & $6.56\times$ \\
    $k$-NN graph & \textsc{CAGRA}      & 6 & $9.84\times$ & $23.8\times$ \\
    \bottomrule
  \end{tabular}
\end{table}

\myparagraph{The 100M-scale \textsc{CAGRA} build is an outlier}
\Cref{tab:ann_build_full} also reports the two 100M-scale datasets
(DEEP-100M, MS-Turing-100M). We exclude these from the aggregate speedups
above and from the main-text tables because the \textsc{CAGRA} DEEP-100M
build is anomalous. At $3940$~s it is $\approx\!4.7\times$ slower than the
$825.6$~s that Jasper report for \textsc{CAGRA} on the \emph{same} dataset
and graph degree~\cite{mccoy2026jasper}---measured there on a slower
A100. We reproduced the
figure ($3868$~s on our H100) and traced it to the cuVS~25.04 large-scale
CAGRA graph-build path rather than to an intrinsic property of the
dataset: it is insensitive to memory-pool, device-residency, and
batching settings. \ourgpumethod{} itself builds both 100M graphs in
$\approx\!20$~s ($5.4\times$ and $8.2\times$ faster than Jasper on
DEEP-100M and MS-Turing-100M, respectively), so its scaling is
unaffected; we omit the 100M cells from the headline averages only to
avoid reporting speedups against an unreliable \textsc{CAGRA} baseline.

%\clearpage

\begin{table}[htbp]
  \centering
  \caption{GPU final prune: direct-edge reuse vs.\ candidate-distance recompute
  (\ourgpumethod{}-TQ4, \Cref{sec:gpu-kernels:direct-prune}). Average
  out-degree of the built graph and the area under the GPU
  QPS-vs-recall curve above recall $0.9$
  (AUC$_{>0.9}$, in $10^3$ QPS$\cdot$recall; higher is better). Both
  modes use the identical build recipe and \HashPrune{} pass and differ
  only in the final pruning rule. The direct-edge rule yields $30$--$58\%$ denser graphs of comparable quality in the high recall regime.}
  \label{tab:prune_ablation}
  \begin{tabular}{lcccc}
    \toprule
    & \multicolumn{2}{c}{avg.\ out-degree} & \multicolumn{2}{c}{AUC$_{>0.9}$ ($\times 10^3$)} \\
    \cmidrule(lr){2-3}\cmidrule(lr){4-5}
    Dataset & recomp. & direct & recomp. & direct \\
    \midrule
    DEEP-10M ({\footnotesize L2, $d{=}96$})        & 28.7 & 37.3 & 167.8 & 182.5 \\
    Wiki-all-10M ({\footnotesize IP, $d{=}768$}) & 17.6 & 27.7 &  20.4 &  20.1 \\
    OpenAI-2M ({\footnotesize L2, $d{=}1536$})     & 21.4 & 30.6 &   9.0 &   9.1 \\
    DBpedia ({\footnotesize L2, $d{=}3072$})  & 18.6 & 27.6 &   9.0 &  10.4 \\
    \bottomrule
  \end{tabular}
\end{table}

\section{\ourgpumethod{} Kernel Details}
\label{app:gpu-kernel-details}
This appendix collects implementation detail deferred from
\Cref{sec:gpu-kernels}.

\subsection{Flattened routing keys}
\label{app:composite-keys}
\Cref{sec:gpu-kernels:flatten} flattens \ouroldmethod{}'s recursion into
two scoring assignment passes joined by radix sorts. The current
implementation has no third assignment or refinement pass: after the
second-level assignments, oversized second-level buckets are split by the
bin-packing walk described below.

\myparagraph{Pass 1: top-level leaders}
Every point scores against the sampled top-level leader set and keeps its
$f_0$ best leaders (default $f_0=8$). The kernel emits $f_0n$ pairs
$(L,p)$ with key $L$, and a radix sort makes each top-level bucket a
contiguous run. The point array is also permuted by each point's best
top-level leader; all later graph work uses this permuted row space until
the final export remap.

\myparagraph{Pass 2: sub-leaders}
Inside each top-level bucket, the host samples a bucket-local sub-leader
candidate set. Each bucket member scores against only that bucket's
sub-leaders and keeps its $f_1$ best (default $f_1=4$). The pass emits
$(s,p)$ pairs keyed by the group-local sub-leader label $s$; sorting and
run-length encoding these keys exposes the second-level buckets. With the
default fanout, a point can therefore appear in up to $f_0f_1=32$ leaf
memberships.

\myparagraph{Bin packing}
The second-level buckets are then packed into leaf-sized work units. This
step does not introduce new leaders or new geometric assignments. It only
splits oversized buckets and merges small adjacent buckets within the same
parent range so the leaf scorer sees regular batch sizes. The implementation
uses a count/assign/broadcast walk over the sorted bucket runs: first count
the bins each parent range will produce, then assign compact leaf ids to
runs, then broadcast those ids back to the membership array. The broadcast
keys are monotone, so no second post-pack sort is needed. Sub-leader sampling
is keyed by the global top-level bucket id, so changing the number of streamed
leader groups changes memory use but not the sampled partition.

\subsection{Partitioning-pass operand staging}
\label{app:partition-staging}
As is discussed in \Cref{sec:gpu-kernels:partitioning}, the assignment passes
are expressed as segmented top-$K$ scoring. Here we provide more details on how the slabs of segments work.

\myparagraph{Top-level leader slab}
The top-level pass samples $L_0$ leaders (auto-tuned up to 4096) and gathers
those rows once into a dense leader slab. The single segment then scores all
points against that slab and emits $f_0$ synthetic leader labels per point.
The synthetic labels avoid self-skip behavior in the shared scoring
primitive, since a sampled leader must be allowed to assign to itself.

\myparagraph{Sub-leader slabs}
For each streamed range of top-level buckets, the host samples sub-leaders
inside each bucket and resolves them to permuted row ids. All sub-leaders
for the group are gathered into one dense slab, and the launch contains one
score segment per nonempty top-level bucket. Each segment points to its
bucket's contiguous sub-leader slice with an offset and count. This lets one
scoring launch handle buckets with different candidate counts while keeping
the operand layout dense.

\subsection{Decoded rows across precisions}
\label{app:tq4-leaf}
The host-side TurboQuant encoder stores TQ4 and RSQ8 in their packed code
formats, but the GPU pipeline expands them before the assignment and leaf
scoring stages. TQ4 is decoded through its $16$-entry codebook into signed
INT8 rows, RSQ8 is copied into the same signed-byte decoded layout, and
padding beyond the true dimension is zeroed in the decoded domain. The
segmented top-$K$ primitive therefore sees the same decoded INT8 row
contract for both TQ modes.

The unquantized supplement uses the same decoded-base interface. The
loader reads float32 rows, normalizes each row to unit length, records the
original norm, and the device decode converts the normalized base to
fp16 or tf32 rows. The scoring primitive then switches operand traits while
the surrounding gather, accumulation, canonicalization, pruning, and export
stages remain the same. The TQ paths are validated by bitwise-equality
tests against a reference decode; the unquantized fp16 and tf32 supplement
paths, which are not bit-exact, are validated by recall-threshold tests.

\subsection{Scoring pipeline modes}
\label{app:gpu-pipeline-modes}
The segmented top-$K$ primitive of \Cref{sec:gpu-kernels:skeleton} runs in two
pipeline modes behind one interface. In the \emph{unified} mode a single
warpgroup performs both the copies and the matrix products, synchronizing on
shared memory once per stripe; we use it for small segments, where it is the
simplest and fastest path. In the \emph{warp-specialized} mode a dedicated
producer warp issues the \texttt{TMA} and gather copies while the consumer
warpgroups compute, so loads and matrix products run on separate warps. The two
modes produce identical results. In either mode two consumer warpgroups can
share a thread block to score a $128$-row query tile ($M = 128$) against one
shared-memory ring of database-point rows, so each row is loaded once and reused
across both $64$-row halves. We use this variant for the small-$K$ builds
($K \le 4$). We stage the query operands into shared memory
once and reuse them when a $64$-row query tile fits---up to $768$ bytes of
decoded operand per row---and stream them alongside the database points above that
width. The same kernel serves the quantized \texttt{i8} builds and the
unquantized fp16/tf32 modes, which differ only in the operand type and
accumulator (\Cref{app:tq4-leaf}).

\subsection{Register top-$K$ epilogue}
\label{app:gpu-register-topk-detail}
The scoring epilogue of \Cref{sec:gpu-kernels:skeleton} keeps each query row's
top-$K$ in registers and, once the candidate sweep finishes, merges the four
lanes that own a row with a butterfly shuffle (\Cref{alg:gpu-register-topk}).

\begin{algorithm}[h]
  \caption{Register top-$K$ insertion and four-lane butterfly merge}
  \label{alg:gpu-register-topk}
  \DontPrintSemicolon
  \KwData{%
    Per-lane registers $\mathbf{d}[0 {:} K], \mathbf{i}[0 {:} K]$, sorted ascending in canonical order ($\mathbf{d}[0]$ best, ties broken by smaller id)\;
  }

  \tcp{Insert candidate $(c, j)$ into the sorted top-$K$}
  \lIf{$(c, j) \not< (\mathbf{d}[K{-}1], \mathbf{i}[K{-}1])$}{\textbf{return}} \tcp*[r]{worse than $K$-th best: one compare}
  \For{$s = 0$ \KwTo $K - 1$}{
    \lIf{$(c, j) < (\mathbf{d}[s], \mathbf{i}[s])$}{$(c, j) \leftrightarrow (\mathbf{d}[s], \mathbf{i}[s])$}
  }
  \BlankLine
  \tcp{Merge the four lanes that own one output row}
  \For{$\mathit{mask} \in \{1, 2\}$}{
    $(\mathbf{d}', \mathbf{i}') \gets \mathrm{shfl\_xor}(\mathbf{d}, \mathbf{i}, \mathit{mask})$ \tcp*[r]{snapshot the partner first}
    \lForEach{$(c, j)$ of $(\mathbf{d}', \mathbf{i}')$}{insert it as above}
  }
\end{algorithm}

\subsection{Leaf kernel across precision and dimension}
\label{app:leaf-roofline-4mode}
\Cref{tab:phase8_roofline} reports the RSQ8 leaf kernel on the
high-dimensional datasets; \Cref{tab:leaf_roofline_4mode} gives the full
sweep across the four build precisions (RSQ8, TQ4, and the unquantized
fp16/tf32 supplement) and all four datasets.
Wider operands push the kernel into the memory hierarchy: at a fixed
dimension the unquantized fp16/tf32 paths load $2$--$4\times$ the bytes per
coordinate, so they saturate L2 ($\approx 98\%$ of unified memory throughput)
even at $d{=}768$, where the INT8 paths are still occupancy-bound.
Dimension has the same effect: the INT8 leaf is L2-bound at $d{\geq}1536$ but
occupancy-bound at $d{\leq}768$. Only low-dimensional deep-10M ($d{=}96$)
stays occupancy-bound in every precision: its working set is small enough
that neither pipe saturates regardless of operand width. Achieved occupancy
is $22$--$28\%$ throughout, set by the kernel's $96$ registers per thread and
shared-memory footprint.

\begin{table}[tb]
\centering
\scriptsize
\caption{Leaf $2$-NN kernel SOL analysis for
\ourgpumethod-RSQ8 (full precision sweep in
\Cref{tab:leaf_roofline_4mode}). Columns: ``$2$-NN\%'' is the leaf
kernel's share of the build time. ``SM\%''/``Mem\%'' are NCU compute- and memory-pipe
throughput; ``Occ\%'' is achieved occupancy
($28\%$ ceiling, set by $96$ registers/thread and SMEM); ``regime''
names the binding resource.
% \lars{please revise this caption. a lot of the text can go in the main body.}
% \lars{if we use the terminology speed-of-light, let's introduce it. unfamiliar readers will think we're trolling. same for other GPU perf tuning terminology}
}
\setlength{\tabcolsep}{3pt}
\begin{tabular}{lrrrrrrl}
\toprule
Dataset ($d$) & $2$-NN\% & SM\% & Mem\% & DRAM\% & L1\% & Occ\% & regime \\
\midrule
WikiAll-10M ($768$)  & $24.1\%$ & $50.1$ & $40.6$ & $13.9$ & $40.1$ & $27.7$ & occupancy \\
OpenAI ($1536$)      & $20.6\%$ & $48.7$ & $81.8$ & $36.6$ & $43.6$ & $27.6$ & L2-bound \\
DBP3072 ($3072$)     & $25.5\%$ & $45.6$ & $94.8$ & $51.9$ & $51.0$ & $27.7$ & L2-bound \\
\bottomrule
\end{tabular}
\label{tab:phase8_roofline}
\end{table}

\begin{table}[t]
\centering
\scriptsize
\caption{Leaf $2$-NN kernel speed-of-light utilization across build precision
and dimension (single H100, \ourgpumethod{}; columns as in
\Cref{tab:phase8_roofline}). The fp16/tf32 rows are the unquantized
supplement; ``occupancy'' marks the latency/occupancy-bound regime where
neither the SM nor the memory pipe reaches $60\%$ of peak.}
\setlength{\tabcolsep}{3pt}
\begin{tabular}{llrrrrrl}
\toprule
Dataset ($d$) & mode & SM\% & Mem\% & DRAM\% & L1\% & Occ\% & regime \\
\midrule
WikiAll-10M ($768$) & RSQ8 & $50.1$ & $40.6$ & $13.9$ & $40.1$ & $27.7$ & occupancy \\
WikiAll-10M ($768$) & TQ4 & $50.1$ & $40.8$ & $14.0$ & $40.1$ & $27.6$ & occupancy \\
WikiAll-10M ($768$) & FP16 & $37.3$ & $98.0$ & $30.3$ & $52.8$ & $22.4$ & L2-bound \\
WikiAll-10M ($768$) & TF32 & $30.2$ & $97.9$ & $45.8$ & $53.6$ & $22.5$ & L2-bound \\
\addlinespace
OpenAI ($1536$) & RSQ8 & $48.7$ & $81.8$ & $36.6$ & $43.6$ & $27.6$ & L2-bound \\
OpenAI ($1536$) & TQ4 & $49.0$ & $78.3$ & $35.0$ & $43.7$ & $27.6$ & L2-bound \\
OpenAI ($1536$) & FP16 & $31.2$ & $97.6$ & $43.9$ & $53.8$ & $22.5$ & L2-bound \\
OpenAI ($1536$) & TF32 & $27.9$ & $97.8$ & $55.8$ & $56.5$ & $22.6$ & L2-bound \\
\addlinespace
DBP3072 ($3072$) & RSQ8 & $45.6$ & $94.8$ & $51.9$ & $51.0$ & $27.7$ & L2-bound \\
DBP3072 ($3072$) & TQ4 & $46.0$ & $90.4$ & $50.9$ & $51.1$ & $27.7$ & L2-bound \\
DBP3072 ($3072$) & FP16 & $27.7$ & $97.8$ & $54.1$ & $56.4$ & $22.4$ & L2-bound \\
DBP3072 ($3072$) & TF32 & $26.2$ & $97.2$ & $64.0$ & $54.4$ & $22.5$ & L2-bound \\
\addlinespace
DEEP ($96$) & RSQ8 & $49.4$ & $23.1$ & $4.4$ & $23.1$ & $26.3$ & occupancy \\
DEEP ($96$) & TQ4 & $49.4$ & $23.1$ & $4.4$ & $23.1$ & $26.3$ & occupancy \\
DEEP ($96$) & FP16 & $46.5$ & $26.1$ & $7.0$ & $26.1$ & $24.9$ & occupancy \\
DEEP ($96$) & TF32 & $46.8$ & $42.0$ & $12.4$ & $31.4$ & $24.9$ & occupancy \\
\bottomrule
\end{tabular}
\label{tab:leaf_roofline_4mode}
\end{table}

\subsection{All major build kernels across the speed-of-light}
\label{app:perkernel-roofline}
\Cref{tab:phase8_roofline,tab:leaf_roofline_4mode} characterize the leaf $2$-NN
kernel; here we apply the same speed-of-light methodology to every major build
kernel. \Cref{tab:perkernel_roofline} gives the RSQ8 picture across dimension, and
\Cref{tab:perkernel_roofline_scorefamily,tab:perkernel_roofline_support} sweep all
four build precisions. The comparable-cost kernels
sit in different speed-of-light regimes, so no single kernel dominates the build:
\begin{itemize}
  \item The \textbf{leaf $2$-NN} and \textbf{sub-leader} scans share the
  \texttt{score\_topk} primitive and the same regime shift---L2-bound at
  $d\,{\geq}\,1536$, latency/occupancy-bound at $d\,{\leq}\,768$---at
  ${\approx}\,28\%$ achieved occupancy.
  \item The \textbf{top-leader} scan ($K{=}8$) is capped at ${\approx}\,19\%$
  occupancy by register pressure: latency-bound at low dimension, reaching the L2
  ceiling only at $d\,{\geq}\,1536$. It is individually small and folds into the
  \texttt{cluster} stage (hence no \%build entry).
  \item The \textbf{edge-hash} kernel is firmly \textbf{SM-bound} ($87$--$89\%$ of
  SM throughput at ${\approx}\,95\%$ occupancy): each warp folds a $32$-lane
  SimHash off the operand load it already performs, with no operand re-streaming.
  \item The \textbf{direct-edge prune} is \textbf{SM-bound} ($68$--$76\%$) and
  \emph{precision- and dimension-invariant}: because it reuses stored
  direct-edge witnesses instead of recomputing candidate distances, its utilization
  is identical across RSQ8/TQ4/FP16/TF32 (\Cref{tab:perkernel_roofline_support}) and
  its share of the build shrinks as $d$ grows ($23.7\%$ at $d{=}96$ down to $3.9\%$
  at $d{=}3072$, as the $d$-dependent score and hash kernels grow around it).
\end{itemize}
The unquantized fp16/tf32 paths push the score-family kernels deeper into the L2
(wider operands, ${\approx}\,98\%$ unified memory throughput) but leave the SM-bound
edge-hash and prune essentially unchanged.

\begin{table}[t]
\centering
\scriptsize
\caption{Per-kernel speed-of-light utilization on a single H100
(\ourgpumethod-RSQ8; \texttt{R=64}, \texttt{max\_cand=64}, leaf target $1024$,
fanout $8{,}4$; columns as in \Cref{tab:phase8_roofline}). ``\%build'' is the
kernel's share of the native GPU build from \texttt{--diag}; the top-leader scan is
not separable from the \texttt{cluster} stage (``--''). The leaf and sub-leader
sub-timers run on two overlapped streams, so their shares are taken against the
native build and need not sum within a stage.}
\setlength{\tabcolsep}{2pt}
% AUTO-GENERATED by summarize_gpipnn2_sol.py -- per-kernel SOL, TQ8 (all major build kernels)
\begin{tabular}{llrrrrrrl}
\toprule
Kernel & Dataset ($d$) & $\%$build & SM\% & Mem\% & DRAM\% & L1\% & Occ\% & regime \\
\midrule
top-leader scan & WikiAll-10M (768) & -- & 50.3 & 29.5 & 2.5 & 20.6 & 18.7 & occupancy \\
top-leader scan & OpenAI (1536) & -- & 46.2 & 63.5 & 10.6 & 34.2 & 18.7 & L2-bound \\
top-leader scan & DBP3072 (3072) & -- & 42.0 & 83.3 & 32.6 & 49.5 & 18.6 & L2-bound \\
top-leader scan & DEEP (96) & -- & 63.1 & 9.6 & 0.6 & 9.6 & 18.7 & SM-bound \\
\addlinespace
sub-leader scan & WikiAll-10M (768) & 19.0\% & 49.0 & 55.0 & 2.5 & 51.7 & 28.0 & occupancy \\
sub-leader scan & OpenAI (1536) & 15.6\% & 44.8 & 78.4 & 37.1 & 54.4 & 28.0 & L2-bound \\
sub-leader scan & DBP3072 (3072) & 7.2\% & 42.8 & 88.3 & 44.7 & 62.1 & 27.9 & L2-bound \\
sub-leader scan & DEEP (96) & 20.2\% & 49.1 & 28.6 & 0.9 & 28.5 & 27.7 & occupancy \\
\addlinespace
leaf 2-NN & WikiAll-10M (768) & 24.1\% & 50.1 & 40.6 & 13.9 & 40.1 & 27.7 & occupancy \\
leaf 2-NN & OpenAI (1536) & 20.6\% & 48.7 & 81.8 & 36.6 & 43.6 & 27.6 & L2-bound \\
leaf 2-NN & DBP3072 (3072) & 25.5\% & 45.6 & 94.8 & 51.9 & 51.0 & 27.7 & L2-bound \\
leaf 2-NN & DEEP (96) & 15.6\% & 49.4 & 23.1 & 4.4 & 23.1 & 26.3 & occupancy \\
\addlinespace
edge-hash & WikiAll-10M (768) & 14.7\% & 87.1 & 31.1 & 6.0 & 31.1 & 93.9 & SM-bound \\
edge-hash & OpenAI (1536) & 20.9\% & 88.5 & 29.3 & 5.4 & 29.3 & 94.9 & SM-bound \\
edge-hash & DBP3072 (3072) & 24.7\% & 89.2 & 28.4 & 4.4 & 28.3 & 95.6 & SM-bound \\
edge-hash & DEEP (96) & 5.1\% & 78.9 & 41.8 & 3.6 & 41.8 & 86.9 & SM-bound \\
\addlinespace
direct-edge prune & WikiAll-10M (768) & 12.7\% & 67.9 & 21.1 & 3.2 & 21.1 & 53.6 & SM-bound \\
direct-edge prune & OpenAI (1536) & 7.5\% & 69.9 & 22.1 & 3.0 & 22.1 & 55.5 & SM-bound \\
direct-edge prune & DBP3072 (3072) & 3.9\% & 73.9 & 24.1 & 3.9 & 24.2 & 60.5 & SM-bound \\
direct-edge prune & DEEP (96) & 23.7\% & 76.1 & 22.4 & 5.3 & 22.4 & 65.4 & SM-bound \\
\addlinespace
\bottomrule
\end{tabular}

\label{tab:perkernel_roofline}
\end{table}

\begin{table}[t]
\centering
\scriptsize
\caption{Score-family kernels (top-leader $K{=}8$, sub-leader $K{=}4$)
speed-of-light across the four build precisions (single H100, \ourgpumethod{};
columns as in \Cref{tab:leaf_roofline_4mode}). Both track the leaf: L2-bound at
$d\,{\geq}\,1536$ and under the wider fp16/tf32 operands, latency/occupancy-bound
otherwise.}
\setlength{\tabcolsep}{3pt}
% AUTO-GENERATED: gpipnn2 4-mode SOL grouped s1_score_k8+s3_score_k4
\begin{tabular}{llrrrrrl}
\toprule
Dataset ($d$) & mode & SM\% & Mem\% & DRAM\% & L1\% & Occ\% & regime \\
\midrule
\multicolumn{8}{l}{\textit{top-leader scan}} \\
\midrule
WikiAll-10M (768) & RSQ8 & 50.3 & 29.5 & 2.5 & 20.6 & 18.7 & occupancy \\
WikiAll-10M (768) & TQ4 & 50.5 & 29.1 & 2.5 & 20.7 & 18.7 & occupancy \\
WikiAll-10M (768) & FP16 & 42.7 & 60.5 & 10.2 & 33.5 & 18.7 & L2-bound \\
WikiAll-10M (768) & TF32 & 40.0 & 83.5 & 32.4 & 47.5 & 18.7 & L2-bound \\
\addlinespace
OpenAI (1536) & RSQ8 & 46.2 & 63.5 & 10.6 & 34.2 & 18.7 & L2-bound \\
OpenAI (1536) & TQ4 & 46.6 & 60.9 & 10.2 & 34.1 & 18.7 & L2-bound \\
OpenAI (1536) & FP16 & 40.5 & 80.5 & 31.2 & 47.9 & 18.7 & L2-bound \\
OpenAI (1536) & TF32 & 37.4 & 88.3 & 34.9 & 54.6 & 18.7 & L2-bound \\
\addlinespace
DBP3072 (3072) & RSQ8 & 42.0 & 83.3 & 32.6 & 49.5 & 18.6 & L2-bound \\
DBP3072 (3072) & TQ4 & 42.0 & 83.2 & 32.5 & 49.6 & 18.6 & L2-bound \\
DBP3072 (3072) & FP16 & 37.5 & 90.0 & 35.5 & 54.6 & 18.6 & L2-bound \\
DBP3072 (3072) & TF32 & 35.5 & 92.7 & 37.4 & 56.5 & 18.6 & L2-bound \\
\addlinespace
DEEP (96) & RSQ8 & 63.1 & 9.6 & 0.6 & 9.6 & 18.7 & SM-bound \\
DEEP (96) & TQ4 & 63.2 & 9.6 & 0.6 & 9.6 & 18.7 & SM-bound \\
DEEP (96) & FP16 & 62.7 & 11.9 & 1.0 & 11.9 & 18.7 & SM-bound \\
DEEP (96) & TF32 & 61.2 & 15.9 & 1.6 & 16.0 & 18.7 & SM-bound \\
\addlinespace
\midrule
\multicolumn{8}{l}{\textit{sub-leader scan}} \\
\midrule
WikiAll-10M (768) & RSQ8 & 49.0 & 55.0 & 2.5 & 51.7 & 28.0 & occupancy \\
WikiAll-10M (768) & TQ4 & 49.1 & 55.0 & 2.5 & 51.7 & 28.0 & occupancy \\
WikiAll-10M (768) & FP16 & 39.1 & 97.7 & 20.6 & 52.5 & 23.3 & L2-bound \\
WikiAll-10M (768) & TF32 & 30.2 & 98.7 & 39.3 & 52.4 & 23.4 & L2-bound \\
\addlinespace
OpenAI (1536) & RSQ8 & 44.8 & 78.4 & 37.1 & 54.4 & 28.0 & L2-bound \\
OpenAI (1536) & TQ4 & 44.8 & 78.6 & 37.2 & 54.4 & 28.0 & L2-bound \\
OpenAI (1536) & FP16 & 31.2 & 98.2 & 38.9 & 52.5 & 23.3 & L2-bound \\
OpenAI (1536) & TF32 & 27.0 & 98.3 & 40.5 & 56.6 & 23.3 & L2-bound \\
\addlinespace
DBP3072 (3072) & RSQ8 & 42.8 & 88.3 & 44.7 & 62.1 & 27.9 & L2-bound \\
DBP3072 (3072) & TQ4 & 43.0 & 86.8 & 44.0 & 62.0 & 27.9 & L2-bound \\
DBP3072 (3072) & FP16 & 28.6 & 96.1 & 38.2 & 55.8 & 23.2 & L2-bound \\
DBP3072 (3072) & TF32 & 25.9 & 97.2 & 40.0 & 56.4 & 23.3 & L2-bound \\
\addlinespace
DEEP (96) & RSQ8 & 49.1 & 28.6 & 0.9 & 28.5 & 27.7 & occupancy \\
DEEP (96) & TQ4 & 49.1 & 28.6 & 0.9 & 28.6 & 27.7 & occupancy \\
DEEP (96) & FP16 & 39.2 & 17.9 & 1.0 & 17.9 & 23.1 & occupancy \\
DEEP (96) & TF32 & 40.2 & 28.4 & 1.6 & 22.9 & 23.2 & occupancy \\
\addlinespace
\bottomrule
\end{tabular}

\label{tab:perkernel_roofline_scorefamily}
\end{table}

\begin{table}[t]
\centering
\scriptsize
\caption{Support kernels (edge-hash, direct-edge prune) speed-of-light across the
four build precisions. Both are SM-bound; the prune is precision-invariant because
it reuses stored witnesses rather than reloading operands.}
\setlength{\tabcolsep}{3pt}
% AUTO-GENERATED: gpipnn2 4-mode SOL grouped s5_hash+s7_prune
\begin{tabular}{llrrrrrl}
\toprule
Dataset ($d$) & mode & SM\% & Mem\% & DRAM\% & L1\% & Occ\% & regime \\
\midrule
\multicolumn{8}{l}{\textit{edge-hash}} \\
\midrule
WikiAll-10M (768) & RSQ8 & 87.1 & 31.1 & 6.0 & 31.1 & 93.9 & SM-bound \\
WikiAll-10M (768) & TQ4 & 87.1 & 31.1 & 6.2 & 31.1 & 93.8 & SM-bound \\
WikiAll-10M (768) & FP16 & 75.4 & 33.6 & 15.6 & 33.6 & 93.0 & SM-bound \\
WikiAll-10M (768) & TF32 & 78.4 & 50.6 & 37.3 & 35.0 & 92.4 & SM-bound \\
\addlinespace
OpenAI (1536) & RSQ8 & 88.5 & 29.3 & 5.4 & 29.3 & 94.9 & SM-bound \\
OpenAI (1536) & TQ4 & 88.5 & 29.3 & 5.4 & 29.3 & 95.0 & SM-bound \\
OpenAI (1536) & FP16 & 78.8 & 36.2 & 15.3 & 32.0 & 94.5 & SM-bound \\
OpenAI (1536) & TF32 & 82.4 & 57.9 & 39.0 & 33.5 & 94.1 & SM-bound \\
\addlinespace
DBP3072 (3072) & RSQ8 & 89.2 & 28.4 & 4.4 & 28.3 & 95.6 & SM-bound \\
DBP3072 (3072) & TQ4 & 89.2 & 28.2 & 4.4 & 28.3 & 95.7 & SM-bound \\
DBP3072 (3072) & FP16 & 80.8 & 37.2 & 14.8 & 31.2 & 95.3 & SM-bound \\
DBP3072 (3072) & TF32 & 84.2 & 62.0 & 41.0 & 32.5 & 95.2 & SM-bound \\
\addlinespace
DEEP (96) & RSQ8 & 78.9 & 41.8 & 3.6 & 41.8 & 86.9 & SM-bound \\
DEEP (96) & TQ4 & 78.8 & 41.8 & 3.6 & 41.8 & 87.0 & SM-bound \\
DEEP (96) & FP16 & 71.8 & 41.6 & 5.4 & 41.6 & 85.6 & SM-bound \\
DEEP (96) & TF32 & 70.7 & 41.3 & 11.0 & 41.3 & 85.3 & SM-bound \\
\addlinespace
\midrule
\multicolumn{8}{l}{\textit{direct-edge prune}} \\
\midrule
WikiAll-10M (768) & RSQ8 & 67.9 & 21.1 & 3.2 & 21.1 & 53.6 & SM-bound \\
WikiAll-10M (768) & TQ4 & 67.9 & 21.1 & 3.2 & 21.1 & 53.7 & SM-bound \\
WikiAll-10M (768) & FP16 & 67.9 & 21.1 & 3.2 & 21.1 & 53.6 & SM-bound \\
WikiAll-10M (768) & TF32 & 67.8 & 21.1 & 3.2 & 21.1 & 53.6 & SM-bound \\
\addlinespace
OpenAI (1536) & RSQ8 & 69.9 & 22.1 & 3.0 & 22.1 & 55.5 & SM-bound \\
OpenAI (1536) & TQ4 & 70.2 & 22.1 & 3.0 & 22.1 & 55.7 & SM-bound \\
OpenAI (1536) & FP16 & 70.0 & 22.1 & 3.0 & 22.1 & 55.5 & SM-bound \\
OpenAI (1536) & TF32 & 70.0 & 22.1 & 3.0 & 22.1 & 55.5 & SM-bound \\
\addlinespace
DBP3072 (3072) & RSQ8 & 73.9 & 24.1 & 3.9 & 24.2 & 60.5 & SM-bound \\
DBP3072 (3072) & TQ4 & 74.0 & 24.1 & 3.9 & 24.2 & 60.5 & SM-bound \\
DBP3072 (3072) & FP16 & 73.9 & 24.1 & 3.9 & 24.2 & 60.5 & SM-bound \\
DBP3072 (3072) & TF32 & 73.9 & 24.1 & 3.9 & 24.2 & 60.5 & SM-bound \\
\addlinespace
DEEP (96) & RSQ8 & 76.1 & 22.4 & 5.3 & 22.4 & 65.4 & SM-bound \\
DEEP (96) & TQ4 & 76.3 & 22.4 & 5.4 & 22.5 & 65.8 & SM-bound \\
DEEP (96) & FP16 & 76.2 & 22.4 & 5.3 & 22.4 & 65.4 & SM-bound \\
DEEP (96) & TF32 & 76.2 & 22.4 & 5.2 & 22.4 & 65.4 & SM-bound \\
\addlinespace
\bottomrule
\end{tabular}

\label{tab:perkernel_roofline_support}
\end{table}

\subsection{Build component breakdown}
\label{app:component-breakdown}
\begin{figure*}[t]
  \centering
  \includegraphics[width=0.84\linewidth]{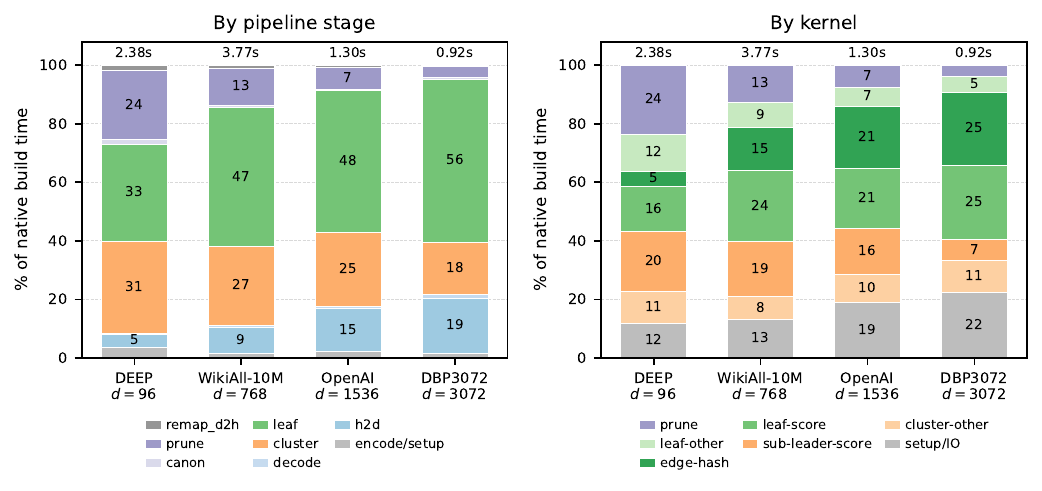}
  \caption{Build component breakdown of \ourgpumethod-RSQ8 by pipeline stage
  (left) and by kernel (right), as a percentage of the native build time (absolute
  build time above each bar). Datasets are ordered by dimension. No single kernel
  dominates, and the bottleneck shifts from the prune and sub-leader scan at low
  dimension to the leaf score and edge-hash at high dimension.}
  \label{fig:component_breakdown}
\end{figure*}

\Cref{fig:component_breakdown} breaks the native GPU build time down by component on
the four ablation datasets, both by pipeline stage and by kernel (RSQ8,
\texttt{R=64}---the same configuration as \Cref{tab:phase8_roofline}). The dominant component
\emph{shifts with dimension}: at $d{=}96$ (deep) the direct-edge prune ($24\%$) and
the sub-leader scan ($20\%$) are the largest kernels; at $d{=}3072$ (dbpedia) the
leaf score ($25\%$) and edge-hash ($25\%$) take over while the prune falls to $4\%$.
The \texttt{leaf-other} and \texttt{cluster-other} slices collect the per-stage
remainder---gather, sort, merge, emit, and the top-leader assignment---after
subtracting the named sub-timers; because the leaf runs two overlapped streams these
are stage wall time minus the measured sub-timers, not separately profiled kernels.

\subsection{\HashPrune{} sketch construction}
\label{app:gpu-prune-detail}
\HashPrune{} on \ourgpumethod{} (\Cref{sec:prune:gpu}) sketches each edge
$(s, t)$ by folding the per-coordinate sign of the residual into a $12$-bit
SimHash,
\[
  h(s \!\to\! t) \;=\;
    \mathrm{fold}_{12}\!\bigl(\mathrm{sign}(t[j] - s[j])\bigr)_{j=0}^{d-1};
\]
the reverse sketch rides the same operand load, packed into the other half
of a $32$-bit word.\footnote{The reverse is not the bitwise complement of
the forward: equal decoded coordinates contribute $0$ to both
directions.} One warp takes one edge and its $32$ lanes coalesce the
contraction-axis loads. The emit kernel writes a fixed forward and reverse
candidate record for each witnessed edge. The sort key is
$(\texttt{owner} \mathbin{\texttt{<<}} 16) \mid \texttt{dist}_{16}$, and the payload stores
the target id together with the $12$-bit sketch. Thus each owner row sees
candidates in distance order before the deterministic merge applies
duplicate-target and same-sketch rules. Because the emit buffer holds only the
current batch's records, its size is bounded by the per-batch slab cap and does
not scale with the graph.

\subsection{Direct KNN-graph mode and build-time scaling}
\label{app:gpu-knn-scaling}
We also use the same pipeline to build direct KNN graphs. In KNN mode we set the
leaf top-$K$ and accumulator capacity to the requested $k$, disable same-sketch
\HashPrune{} filtering, and deduplicate only repeated target ids, keeping the
nearer copy. We then skip the final prune and export the canonicalized
accumulator as a directed graph; this avoids
the build-then-search pattern that graph-index baselines use for KNN
construction (\Cref{tab:gpu_build,tab:knn_build_full}). For harder KNN cells we
run several rounds. Round $0$ creates the permuted row space; later rounds
resample the top-level partition with new seeds but keep the row order fixed, so
their witnesses merge into the same accumulator and export through the same id
map, changing the candidate set without creating incompatible row spaces. We
plan the remaining scaling choices before allocation: a memory planner chooses
how many top-level bucket ranges to stream and bounds the leaf and accumulation
buffers, keeping output-neutral choices (such as the number of streamed ranges)
separate from quality-coupled ones (such as candidate capacity). The full
build-time tables are in \Cref{tab:ann_build_full,tab:gpu_speedup_summary}.

\section{GPU Build Parameter Configurations}
\label{app:gpu-params}
This section reports the exact per-(method, dataset) build parameters behind the GPU build-time
results (\Cref{tab:gpu_build,tab:ann_build_full,tab:knn_build_full}), how they were selected, and
the software and hardware environment. All runs were driven by a single comparison harness that
builds each method from source and records the invocation. \ourgpumethod{} here is the H100 port;
its four modes---TQ4, RSQ8, FP16, and TF32 (CLI \texttt{-{}-tq 4}, \texttt{-{}-tq 8}, and
\texttt{-{}-unquant fp16|tf32})---change only the score-kernel precision, so the graph-topology
parameters in \Cref{tab:params-ann,tab:params-knn} are shared across all four modes.

\subsection{Tuning protocol}
\label{app:params-tuning}
For each (method, dataset) pair we ran a parameter sweep and kept a single configuration. For the
ANNS index we kept the \emph{fastest build whose index is high-quality}---its QPS-versus-recall
curve comparable to an unquantized ParlayANN-Vamana graph (\Cref{sec:experiments})---scored by the
area under that curve. For the $k$-NN graph we kept the \emph{fastest build reaching graph-fidelity
recall\,$\geq$\,0.9} (sampling 5{,}000 seeded points against the brute-force top-10); a ``--'' in
\Cref{tab:knn_build_full} marks a method$\times$dataset cell where no swept configuration cleared the
floor. The swept ranges were: \ourgpumethod{} $\alpha\in\{0.90,\dots,1.20\}$, leaf size
$\in\{512,1024,2048,4096\}$, prune degree $R$ and prune-candidate cap $\in\{32,64\}$, and, for
$k$-NN, leaders and fanout; \textsc{Jasper} $\alpha$ (same grid), \texttt{use\_rabitq},
\texttt{use\_reorder}; CAGRA graph degree $\in\{32,48,64\}$ and intermediate degree
$\in\{64,128,192\}$. Methods otherwise run at library defaults. The $\alpha$ default is
distance-aware ($0.9$ for MIPS, $1.1$ for L2) for both \ourgpumethod{} and \textsc{Jasper}; RaBitQ
was not used during queries for any method.

\subsection{ANNS-index build parameters}
\label{app:params-ann}
\ourgpumethod{} always builds with GPU \HashPrune{} and the direct-edge final prune,
both enabled by default. \Cref{tab:params-ann} lists the
tuned parameters per dataset:
\ourgpumethod{} angular slack $\alpha$, prune degree $R$, and leaf size $C$ (prune-candidate cap
$\mathit{pc}{=}64$, reduced to $32$ at 100M to fit the
accumulators); \textsc{Jasper} degree $R$, $\alpha$, and reorder flag ($L_{\mathrm{cap}}{=}64$,
$\mathit{mbr}{=}0.02$ throughout, $0.004$ at 100M to avoid an illegal access at $2^{19}$ nodes); and
CAGRA graph / intermediate degree. Unquantized CAGRA and CAGRA-FP16 build the initial graph with
\textsc{NN-Descent} (L2 only) and therefore do not run on the MIPS sets (Wiki-all, MS-MARCO,
MSMARCO-Arctic, Wiki-Cohere, Text2Image) except for Wiki-Qwen-500K, which is L2-normalized; only CAGRA-PQ4/PQ8 (\textsc{IVF-PQ} path) do, sharing the listed degrees
and differing only in PQ bit-width. On the L2 DataCompDr-10M set the unquantized CAGRA build did not
complete. \ourgpumethod-TF32 exceeds the 80\,GB budget at 100M, and both \ourgpumethod-FP16 and TF32
exceed it on DataCompDr-10M ($10^7{\times}1536$).

\begin{table*}[t]
  \centering
  \caption{Per-dataset ANNS-index build parameters (cf.\ \Cref{tab:ann_build_full}). $R$: degree;
  $C$: leaf size; reord.: \textsc{Jasper} reorder; deg./inter.: CAGRA graph / intermediate degree.}
  \label{tab:params-ann}
  \small
  \setlength{\tabcolsep}{5pt}
  \begin{tabular}{lcccccccc}
    \toprule
     & \multicolumn{3}{c}{\ourgpumethod{} (all modes)} & \multicolumn{3}{c}{Jasper} & \multicolumn{2}{c}{CAGRA} \\
    \cmidrule(lr){2-4}\cmidrule(lr){5-7}\cmidrule(lr){8-9}
    Dataset & $\alpha$ & $R$ & $C$ & $R$ & $\alpha$ & reord. & deg. & inter. \\
    \midrule
    DEEP-10M       & 1.1  & 64 & 512  & 64 & 1.05 & Y & 64 & 128 \\
    MS-Turing-10M  & 1.1  & 64 & 512  & 64 & 1.1  & Y & 64 & 128 \\
    DEEP-100M      & 1.1  & 32 & 512  & 64 & 1.05 & Y & 64 & 128 \\
    MS-Turing-100M & 1.1  & 32 & 512  & 64 & 1.1  & Y & 64 & 128 \\
    DBpedia-1536   & 1.0  & 32 & 512  & 32 & 1.05 & Y & 32 & 128 \\
    DBpedia-3072   & 1.0  & 32 & 512  & 32 & 1.1  & N & 32 & 192 \\
    Wiki-Qwen-500K & 0.95 & 32 & 512  & 32 & 1.0  & Y & 32 & 192 \\
    OpenAI-2M      & 1.05 & 64 & 1024 & 64 & 1.1  & Y & 32 & 192 \\
    Wiki-all-1M    & 0.95 & 32 & 512  & 32 & 1.0  & Y & 32 & 128 \\
    Wiki-all-10M   & 0.95 & 32 & 512  & 32 & 0.95 & Y & 32 & 128 \\
    MS-MARCO-10M   & 0.9  & 32 & 512  & 32 & 0.9  & Y & 32 & 128 \\
    DataCompDr-10M  & 1.05 & 32 & 1024 & 32 & 1.1  & Y & 32 & 192 \\
    Wiki-Cohere-10M & 0.95 & 32 & 512  & 32 & 0.95 & Y & 32 & 128 \\
    Text2Image-10M  & 0.9  & 64 & 512  & 64 & 0.9  & Y & 64 & 128 \\
    MS-Arctic-1M    & 0.9  & 32 & 512  & 32 & 0.9  & Y & 32 & 128 \\
    \bottomrule
  \end{tabular}
\end{table*}

\subsection{$k$-NN-graph build parameters}
\label{app:params-knn}
\Cref{tab:params-knn} lists the $k$-NN build parameters ($K{=}10$). \ourgpumethod{} (published as
RSQ8) is set by leaders, leaf size $C$, fanout $(f_0,f_1)$, and rounds. TQ4 is too lossy to reach the recall floor on DEEP-10M, MS-Turing-10M,
MS-MARCO-10M, Wiki-Cohere-10M, and Text2Image-10M. For some datasets it requires significantly higher fanout than RSQ8 and is thus \emph{slower}.  FP16/TF32 exceed the 80\,GB budget on
DataCompDr-10M. \textsc{Jasper} uses $R{=}64$, $L_{\mathrm{cap}}{=}64$, $\mathit{mbr}{=}0.02$, RaBitQ
and reorder off, with $\alpha$ and beam size as shown. \textsc{NN-Descent} (and its FP16 variant) is
set by intermediate degree / iterations; \textsc{IVF-PQ8} by probe count $n_p$ (list count and PQ
dimension auto); cuVS \textsc{CAGRA} uses graph / intermediate degree $32/128$ on every dataset. At
100M (DEEP, MS-Turing; not in \Cref{tab:knn_build_full}) \ourgpumethod{} scales leaders to $4096$,
with leaf $2048/4096$ and fanout $(8,8)/(8,10)$ respectively.

\begin{table*}[t]
  \centering
  \caption{Per-dataset $k$-NN-graph build parameters (cf.\ \Cref{tab:knn_build_full}). Values are
  the swept configuration retained per method; ``--'' in \Cref{tab:knn_build_full} marks cells where
  no configuration reached recall\,$\geq$\,0.9. deg./iter.: \textsc{NN-Descent} intermediate
  degree / iterations.}
  \label{tab:params-knn}
  \small
  \setlength{\tabcolsep}{5pt}
  \begin{tabular}{lcccccccc}
    \toprule
     & \multicolumn{4}{c}{\ourgpumethod{} (all modes)} & \multicolumn{2}{c}{Jasper} & NN-Descent & IVF-PQ8 \\
    \cmidrule(lr){2-5}\cmidrule(lr){6-7}\cmidrule(lr){8-8}\cmidrule(lr){9-9}
    Dataset & leaders & $C$ & $(f_0,f_1)$ & rounds & $\alpha$ & beam & deg./iter. & $n_p$ \\
    \midrule
    DEEP-10M       & 512 & 1024 & (8,4)  & 1 & 0.95 & 64  & 64/15 & 32 \\
    MS-Turing-10M  & 512 & 4096 & (8,10) & 1 & 1.05 & 64  & 48/15 & 64 \\
    DBpedia-1536   & 512 & 512  & (8,8)  & 1 & 0.95 & 128 & 48/20 & 32 \\
    DBpedia-3072   & 512 & 512  & (8,8)  & 1 & 0.95 & 128 & 48/30 & 32 \\
    Wiki-Qwen-500K & 512 & 512  & (8,10) & 1 & 0.95 & 128 & 48/30 & 32 \\
    OpenAI-2M      & 512 & 512  & (8,4)  & 1 & 0.95 & 128 & 48/15 & 32 \\
    Wiki-all-1M    & 512 & 2048 & (8,10) & 1 & 1.0  & 64  & 48/15 & 32 \\
    Wiki-all-10M   & 512 & 2048 & (8,4)  & 1 & 0.95 & 64  & 48/15 & 64 \\
    MS-MARCO-10M   & 512 & 2048 & (8,4)  & 4 & 0.9  & 128 & 48/15 & 64 \\
    DataCompDr-10M  & 512 & 4096 & (8,4)  & 1 & 1.05 & 128 & --    & --  \\
    Wiki-Cohere-10M & 512 & 2048 & (8,4)  & 4 & 0.95 & 128 & 48/20 & 128 \\
    Text2Image-10M  & 512 & 1024 & (8,4)  & 1 & 0.9  & 64  & 48/20 & --  \\
    MS-Arctic-1M    & 512 & 4096 & (8,4)  & 1 & 0.9  & 512 & 48/20 & 32  \\
    \bottomrule
  \end{tabular}
\end{table*}

\subsection{Software and hardware}
\label{app:params-env}
\Cref{tab:params-env} gives the full version manifest abbreviated in \Cref{sec:experiments}.
\ourgpumethod{} is compiled for Hopper SM90a with fused multiply-add disabled
(\texttt{-fmad=false} device, \texttt{-ffp-contract=off} host) for deterministic builds. All builds
run on a single H100, device-resident where the dataset fits the 80\,GB budget; the DEEP-100M CAGRA
outlier (\Cref{tab:ann_build_full}) stems from the cuVS~25.04 large-scale host-input build path.

\begin{table*}[t]
  \centering
  \caption{Software and hardware environment for the GPU experiments.}
  \label{tab:params-env}
  \small
  \setlength{\tabcolsep}{5pt}
  \begin{tabular}{ll}
    \toprule
    Component & Version \\
    \midrule
    GPU              & NVIDIA H100 80\,GB HBM3 SXM5 (cc~9.0) \\
    Host / instance  & GCP \texttt{a3-highgpu-1g}, 26-vCPU Sapphire Rapids, 234\,GB DDR5 \\
    GPU driver       & 580.126.09 \\
    CUDA toolkit     & 12.9 (compiled sm\_80/89/90; run as SM90a) \\
    OS               & Ubuntu 24.04.4 LTS \\
    \ourgpumethod{} libs & CuTe/CUTLASS 3.7.0, RaBitQ, Eigen 3.4.0, Abseil 20250512.0, mimalloc 2.2.2 \\
    Compiler         & gcc/g++ 13, C++17, CMake $\geq$ 3.15 \\
    Jasper           & git \texttt{1f03773} (CUDA $\geq$ 12.1, cc $>$ 8.0) \\
    cuVS / RAFT / RMM & 25.04 (cuvs-bench 25.04, Python 3.12) \\
    Harness          & Python 3.10, NumPy $\geq$ 1.24 \\
    \bottomrule
  \end{tabular}
\end{table*}

\section{CPU Build Parameter Configurations}
\label{app:cpu-params}
This section reports the per-(method, dataset) build parameters behind the CPU
results: the build times in \Cref{tab:cpu_build} and the QPS-vs-recall curves in
\Cref{fig:additional-qps}. \ouroldmethod{} appears in three modes (TQ4, RSQ8, and
FP32) that differ only in build precision. They share the build parameters below,
so its columns cover all three, but the differing precision yields different
graphs; all three are queried identically. For the ANNS index we fixed the
graph degree $R{=}64$ and build beam $L{=}128$ and tuned the slack $\alpha$ per
dataset (\Cref{app:cpu-params-ann}). For the $k$-NN graph we swept the per-method knobs of
\Cref{tab:params-knn-cpu} and kept the fastest build reaching
recall\,$\geq$\,0.9.

\subsection{ANNS-index build parameters}
\label{app:cpu-params-ann}
\Cref{tab:params-ann-cpu} lists the tuned parameters per dataset. The parameter
$\alpha$ is the only knob that was varied for both methods, as it directly controls the density of the produced graph.

\begin{figure*}[t]
  \centering
  \begin{subfigure}[t]{0.31\textwidth}
    \includegraphics[width=\linewidth]{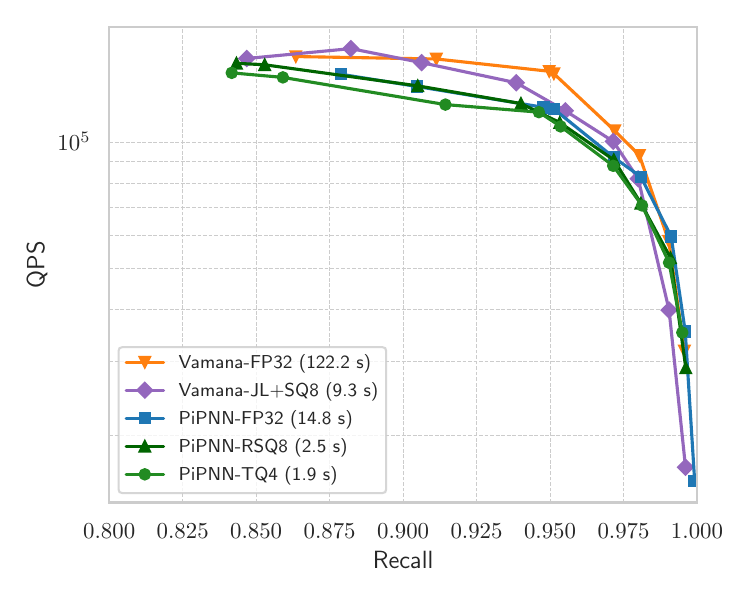}
    \caption{DBpedia1536}
  \end{subfigure}\hfill
  \begin{subfigure}[t]{0.31\textwidth}
    \includegraphics[width=\linewidth]{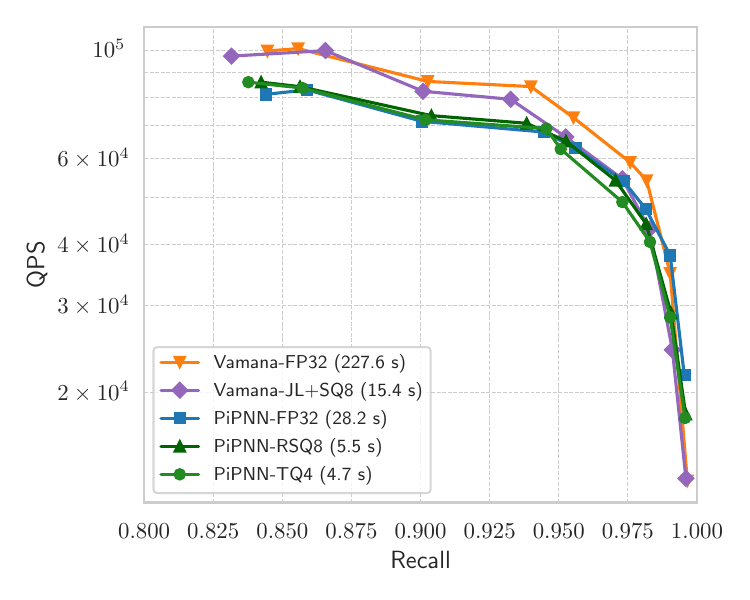}
    \caption{DBpedia3072}
  \end{subfigure}\hfill
  \begin{subfigure}[t]{0.31\textwidth}
    \includegraphics[width=\linewidth]{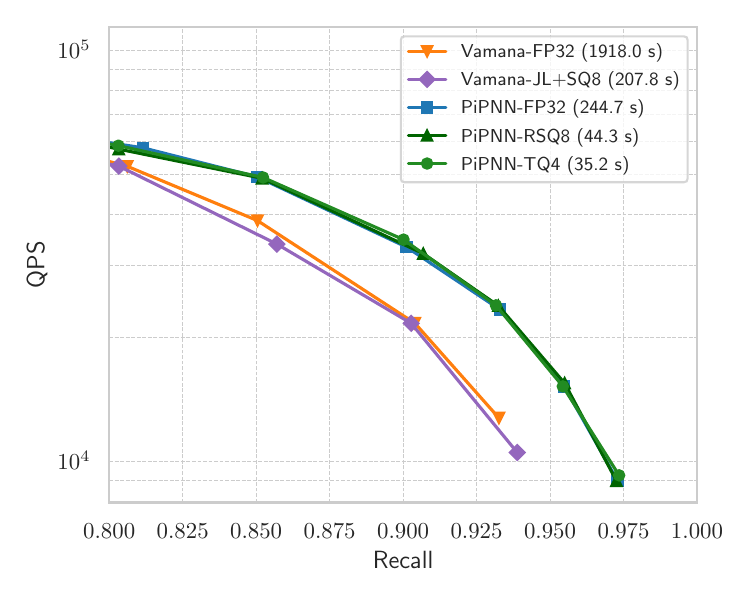}
    \caption{DataCompDr}
  \end{subfigure}\\[4pt]
  \begin{subfigure}[t]{0.31\textwidth}
    \includegraphics[width=\linewidth]{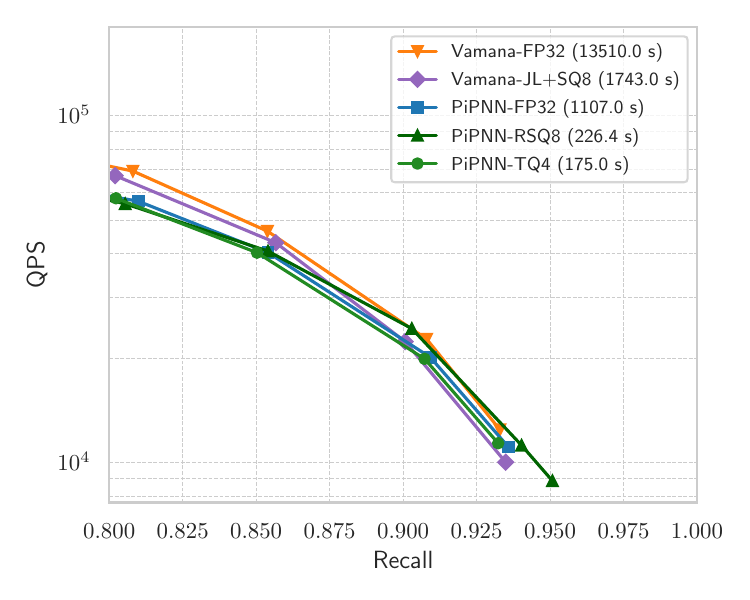}
    \caption{MSMARCO}
  \end{subfigure}\hfill
  \begin{subfigure}[t]{0.31\textwidth}
    \includegraphics[width=\linewidth]{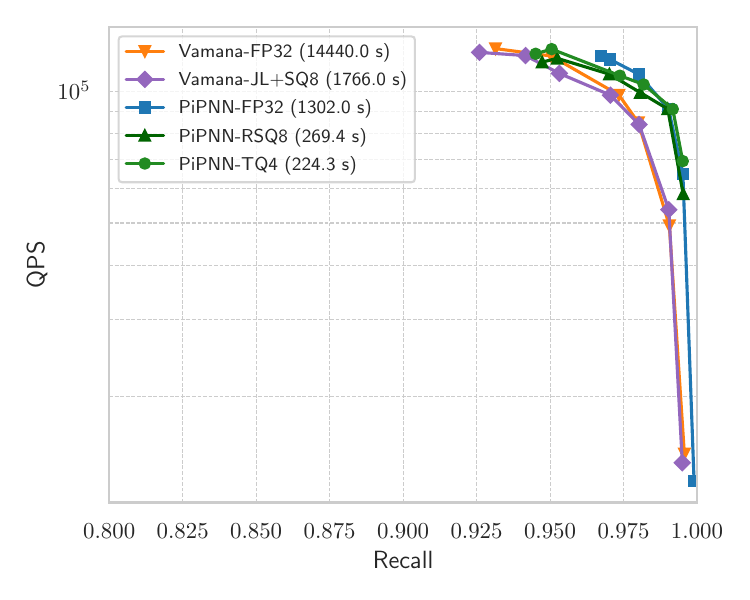}
    \caption{MSMARCO-Arctic}
  \end{subfigure}\hfill
  \begin{subfigure}[t]{0.31\textwidth}
    \includegraphics[width=\linewidth]{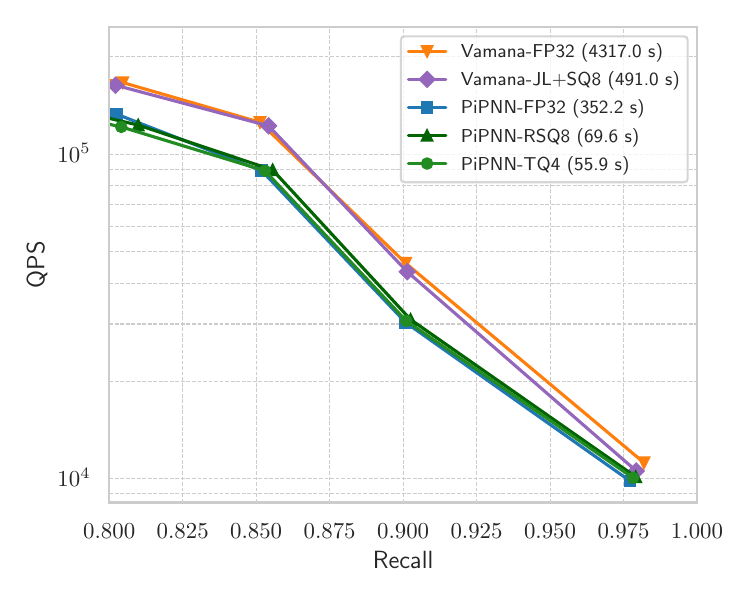}
    \caption{Wiki-Cohere}
  \end{subfigure}\\[4pt]
  \begin{subfigure}[t]{0.31\textwidth}
    \includegraphics[width=\linewidth]{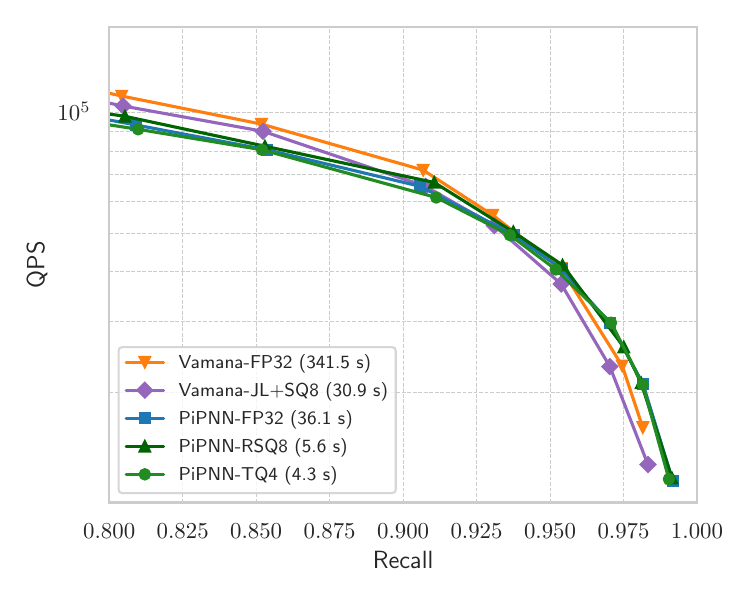}
    \caption{OpenAI-ArXiv}
  \end{subfigure}\hfill
  \begin{subfigure}[t]{0.31\textwidth}
    \includegraphics[width=\linewidth]{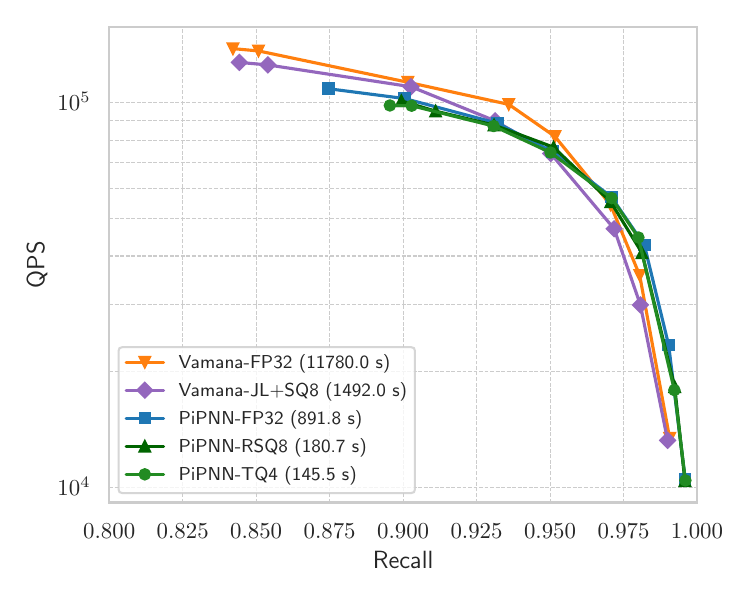}
    \caption{wiki-all}
  \end{subfigure}\hfill
  \begin{subfigure}[t]{0.31\textwidth}
    \includegraphics[width=\linewidth]{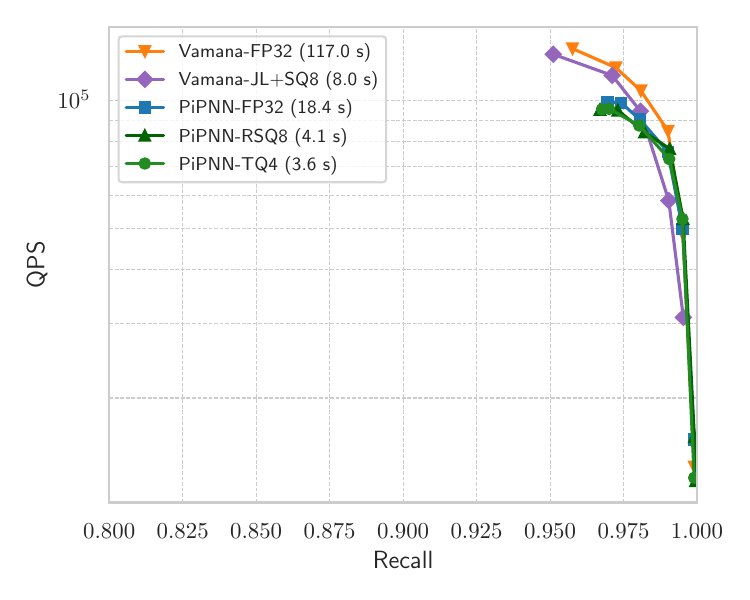}
    \caption{Wiki-Qwen-500K}
  \end{subfigure}\\[4pt]
  \begin{subfigure}[t]{0.31\textwidth}
    \includegraphics[width=\linewidth]{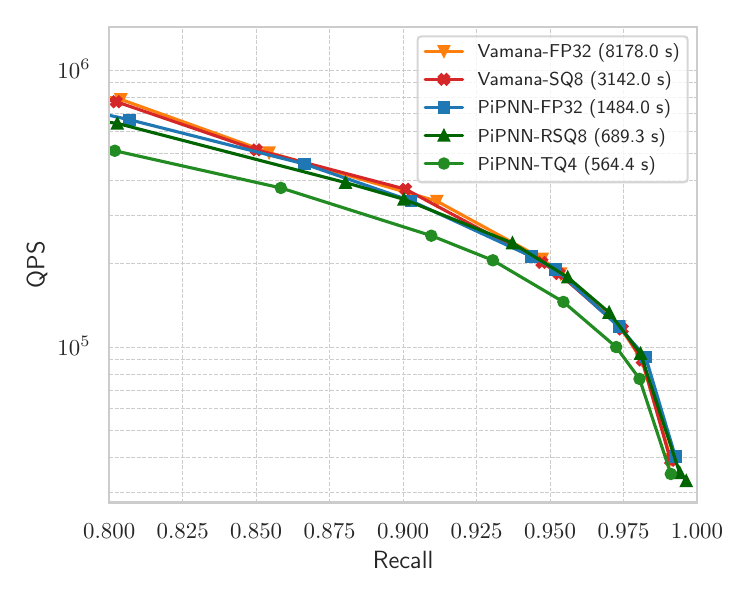}
    \caption{DEEP-1B}
  \end{subfigure}\hfill
  \begin{subfigure}[t]{0.31\textwidth}
    \includegraphics[width=\linewidth]{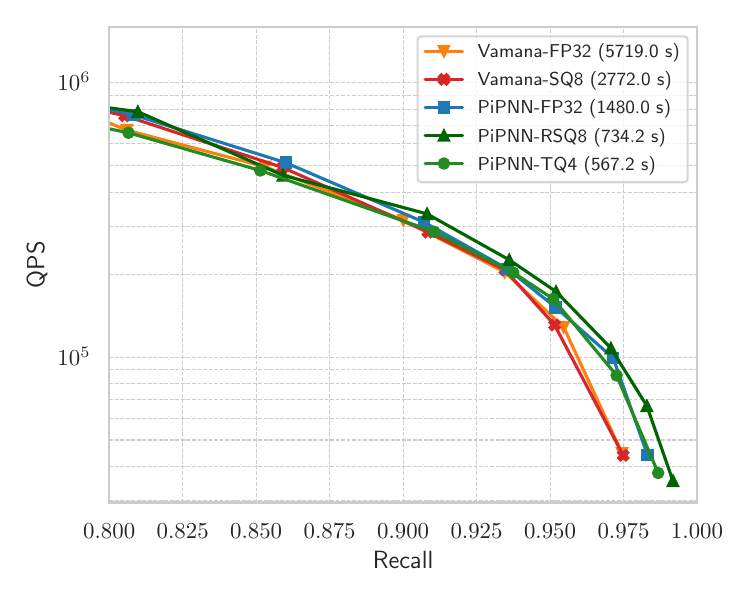}
    \caption{MS-Turing-1B}
  \end{subfigure}\hfill
  \begin{subfigure}[t]{0.31\textwidth}
    \includegraphics[width=\linewidth]{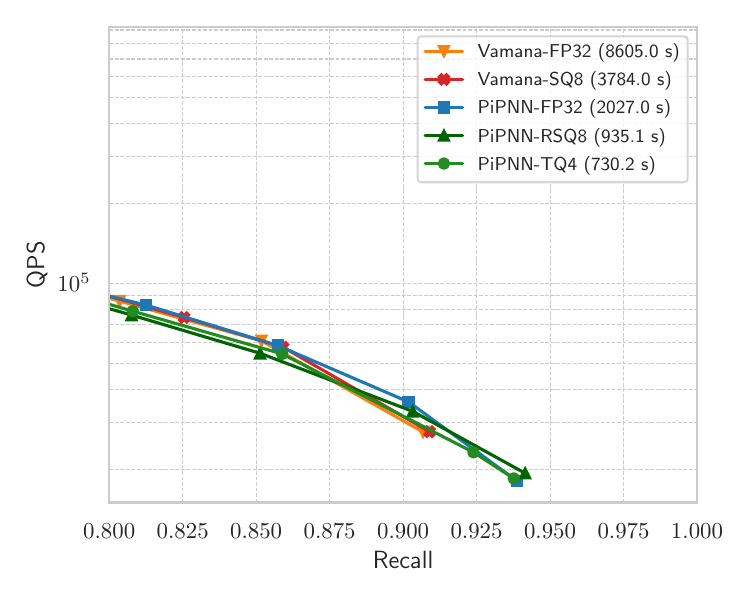}
    \caption{Text2Image-1B}
  \end{subfigure}
  \caption{Index Quality (CPU). QPS vs.\ recall on all datasets from
  \Cref{tab:modern_datasets_full_compact}, with per-method build times in
  the legends. \ouroldmethod-FP32, \ouroldmethod-TQ4, and \ouroldmethod-RSQ8
  trace nearly identical curves.
  }
  \label{fig:additional-qps}
\end{figure*}

\myparagraph{Tuning $\alpha$} For each dataset we set \ouroldmethod{}'s
$\alpha$ so that its graph's average degree falls within $\pm 2$ of the Vamana's average degree. 
We note that varying $\alpha$ does not meaningfully change the construction time for \ourmethod{}, and this tuning was done to produce an apples-to-apples comparison between the quality of the graphs produced by the two methods (\Cref{fig:additional-qps}).

\begin{table*}[t]
  \centering
  \caption{Per-dataset CPU ANNS-index build parameters (cf.\ \Cref{tab:cpu_build}
  and \Cref{fig:additional-qps}). $R$: graph degree; $C$: \ouroldmethod{}
  leaf/cluster size; $L$: Vamana build beam. \ouroldmethod{} runs a single cluster
  tree ($\mathit{nc}{=}1$) with LSH table size $T{=}96$ (dropped to $64$ on the
  billion-scale sets in FP32 mode to fit memory); its TQ4, RSQ8, and FP32 modes
  differ only in build precision and so produce different graphs from these shared
  parameters, all queried identically. Vamana quant.: SQ8 is
  8-bit scalar quantization, JL+SQ8 adds a Johnson-Lindenstrauss filter; the unquantized VF32
  baseline uses the same $R$, $L$, and $\alpha$.}
  \label{tab:params-ann-cpu}
  \setlength{\tabcolsep}{5pt}
  \small
  \begin{tabular}{lccccccc}
    \toprule
     & \multicolumn{3}{c}{\ouroldmethod{} (all modes)} & \multicolumn{4}{c}{Vamana} \\
    \cmidrule(lr){2-4}\cmidrule(lr){5-8}
    Dataset & $\alpha$ & $R$ & $C$ & $R$ & $L$ & $\alpha$ & quant. \\
    \midrule
    Wiki-Qwen-500K & 0.75  & 64 & 1024 & 64 & 128 & 1.0   & JL+SQ8 \\
    DBP1536    & 1.25  & 64 & 1024 & 64 & 128 & 1.0   & JL+SQ8 \\
    DBP3072    & 1.25  & 64 & 1024 & 64 & 128 & 1.0   & JL+SQ8 \\
    OpenAI     & 1.25  & 64 & 1024 & 64 & 128 & 1.04  & JL+SQ8 \\
    DataComp   & 1.25  & 64 & 1024 & 64 & 128 & 1.1   & JL+SQ8 \\
    WikiCohere & 0.98  & 64 & 1024 & 64 & 128 & 0.99  & JL+SQ8 \\
    WikiAll    & 0.9   & 64 & 1024 & 64 & 128 & 0.93  & JL+SQ8 \\
    MSMARCO    & 0.985 & 64 & 1024 & 64 & 128 & 1.0   & JL+SQ8 \\
    MS-Arctic  & 0.95  & 64 & 1024 & 64 & 128 & 0.95  & JL+SQ8 \\
    DEEP       & 1.15  & 64 & 1024 & 64 & 128 & 1.025 & SQ8 \\
    MS-Turing  & 1.07  & 64 & 1024 & 64 & 128 & 1.01  & SQ8 \\
    Text2Image & 0.95  & 64 & 1024 & 64 & 128 & 1.0   & SQ8 \\
    \bottomrule
  \end{tabular}
\end{table*}

\subsection{$k$-NN-graph build parameters}
\label{app:cpu-params-knn}
\Cref{tab:params-knn-cpu} lists the retained $k$-NN configuration per method: the
fastest swept build reaching recall\,$\geq$\,0.9.
PyNNDescent is omitted on MS-MARCO, DEEP, MS-Turing, and Text2Image since it runs out of memory.

\begin{table*}[t]
  \centering
  \caption{Per-dataset CPU $k$-NN-graph build parameters: the fastest swept
  configuration reaching recall\,$\geq$\,0.9 (the configurations timed in the
  right half of \Cref{tab:cpu_build}). \ouroldmethod{} is set by its leaf fanout
  scheme and cluster count $\mathit{nc}$ at fixed leaf size $C{=}1024$, per-leaf
  top-$10$, heap size $k{=}20$, and TQ8 scoring. Vamana sweeps $R$ (build beam $L{=}2R$) and is read out at query
  beam $Q$, with SQ8 or JL+SQ8 as in \Cref{tab:params-ann-cpu}. \pynn{} runs one
  fixed configuration ($R{=}40$, $\mathit{nc}{=}5$, leaf $100$, $\delta{=}0.02$,
  SQ8). ``--'' marks where \pynn{} exceeded memory (MSMARCO and the billion-scale
  sets).}
  \label{tab:params-knn-cpu}
  \setlength{\tabcolsep}{5pt}
  \small
  \begin{tabular}{lcccccc}
    \toprule
     & \multicolumn{2}{c}{\ouroldmethod{}} & \multicolumn{2}{c}{Vamana} & \multicolumn{2}{c}{\pynn{}} \\
    \cmidrule(lr){2-3}\cmidrule(lr){4-5}\cmidrule(lr){6-7}
    Dataset & fanout & $\mathit{nc}$ & $R$ & $Q$ & $R$ & $\mathit{nc}$ \\
    \midrule
    Wiki-Qwen-500K & (10,1)   & 2 & 16 & 225 & 40 & 5  \\
    DBP1536    & (10,1)   & 2 & 32 & 12  & 40 & 5  \\
    DBP3072    & (10,1)   & 3 & 16 & 500 & 40 & 5  \\
    OpenAI     & (10,1)   & 2 & 16 & 160 & 40 & 5  \\
    DataComp   & (10,2,1) & 2 & 32 & 120 & 40 & 5  \\
    WikiCohere & (10,3,1) & 3 & 64 & 500 & 40 & 5  \\
    WikiAll    & (10,3,1) & 3 & 32 & 300 & 40 & 5  \\
    MSMARCO    & (10,3,1) & 5 & 64 & 160 & -- & -- \\
    MS-Arctic  & (10,2,1) & 1 & 32 & 34  & 40 & 5  \\
    DEEP       & (10,3,1) & 3 & 64 & 30  & -- & -- \\
    MS-Turing  & (10,3,1) & 2 & 32 & 140 & -- & -- \\
    Text2Image & (10,2,1) & 2 & 32 & 200 & -- & -- \\
    \bottomrule
  \end{tabular}
\end{table*}

\begin{figure*}[t]
    \centering
    \includegraphics[width=\linewidth]{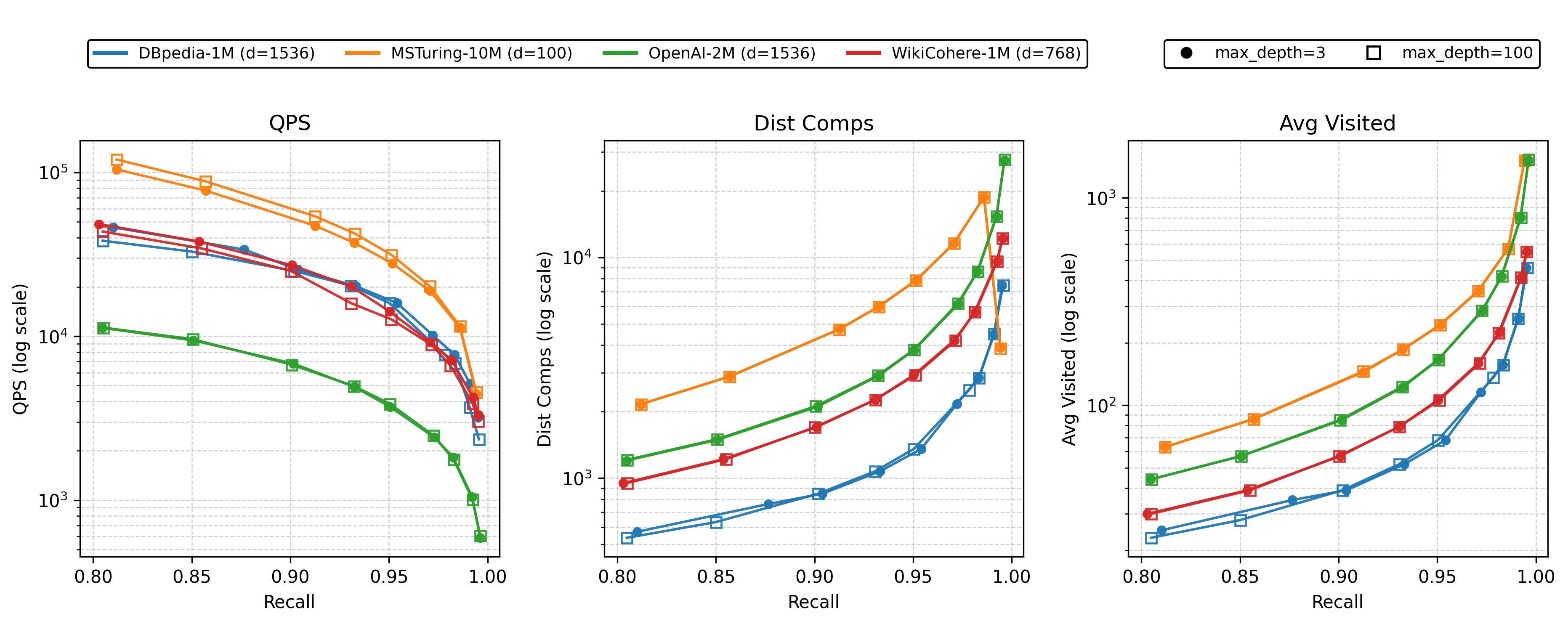}
    \caption{Effect of maximum recursion depth on PiPNN index quality. On four different datasets, increasing recursion depth cap beyond 3 has little to no impact on query performance.}
    \label{fig:max-depth-ablation}
\end{figure*}

\begin{figure*}[t]
  \centering
  \begin{subfigure}[t]{0.32\textwidth}
    \includegraphics[width=\linewidth]{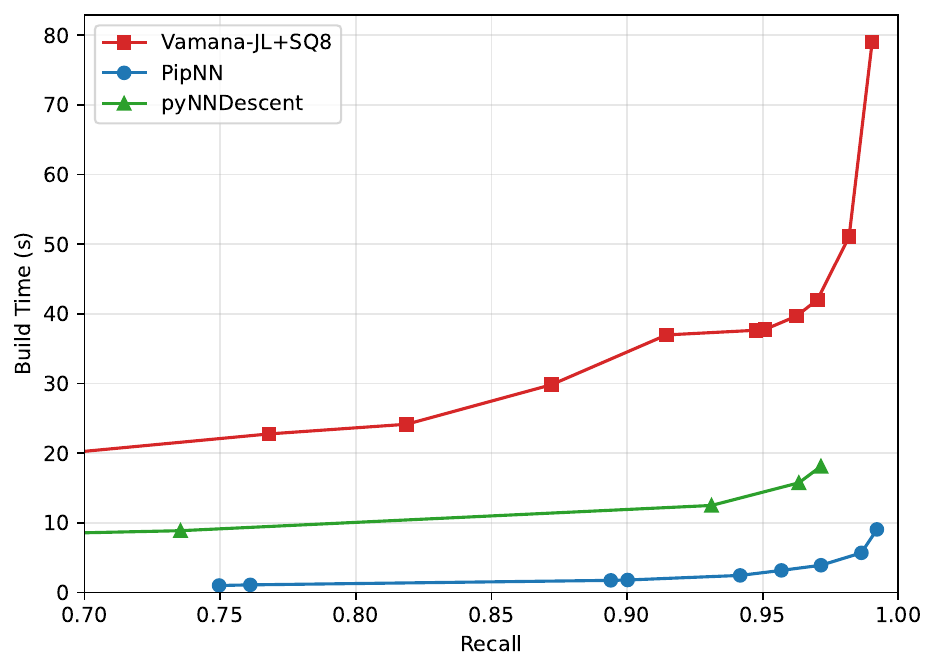}
    \caption{DBpedia-1536}
  \end{subfigure}\hfill
  \begin{subfigure}[t]{0.32\textwidth}
    \includegraphics[width=\linewidth]{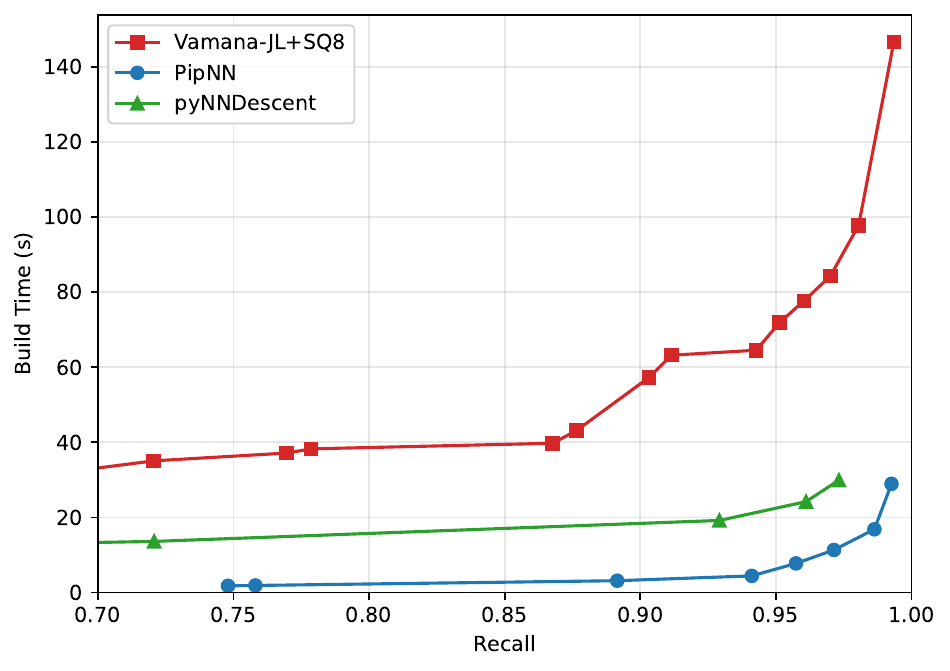}
    \caption{DBpedia-3072}
  \end{subfigure}\hfill
  \begin{subfigure}[t]{0.32\textwidth}
    \includegraphics[width=\linewidth]{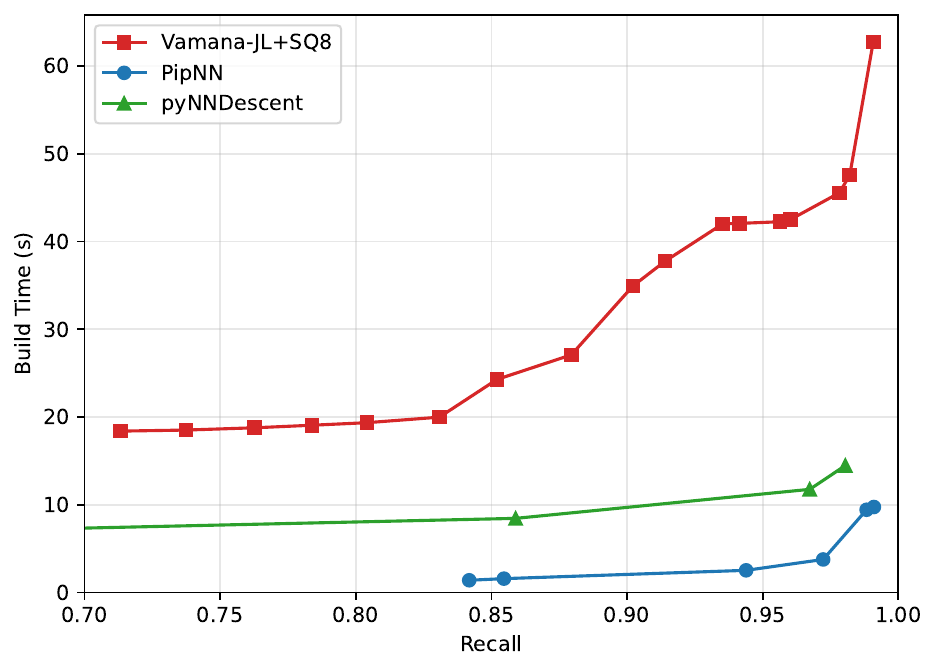}
    \caption{Wiki-Qwen-500K}
  \end{subfigure}\\[4pt]
  \begin{subfigure}[t]{0.32\textwidth}
    \includegraphics[width=\linewidth]{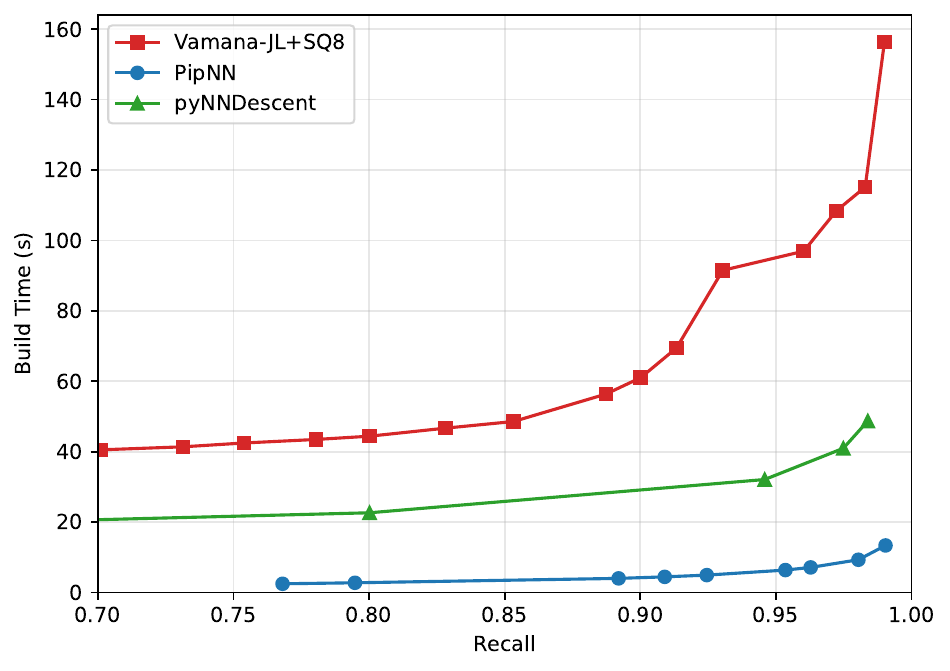}
    \caption{OpenAI-2M}
  \end{subfigure}\hfill
  \begin{subfigure}[t]{0.32\textwidth}
    \includegraphics[width=\linewidth]{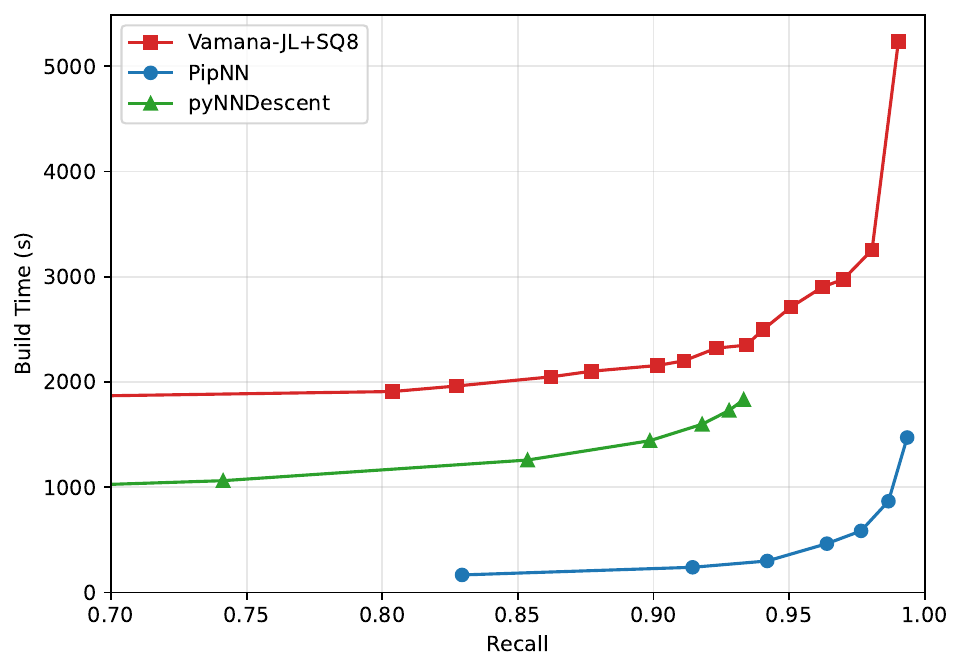}
    \caption{MSMARCO-Arctic-114M}
  \end{subfigure}\hfill
  \begin{subfigure}[t]{0.32\textwidth}
    \includegraphics[width=\linewidth]{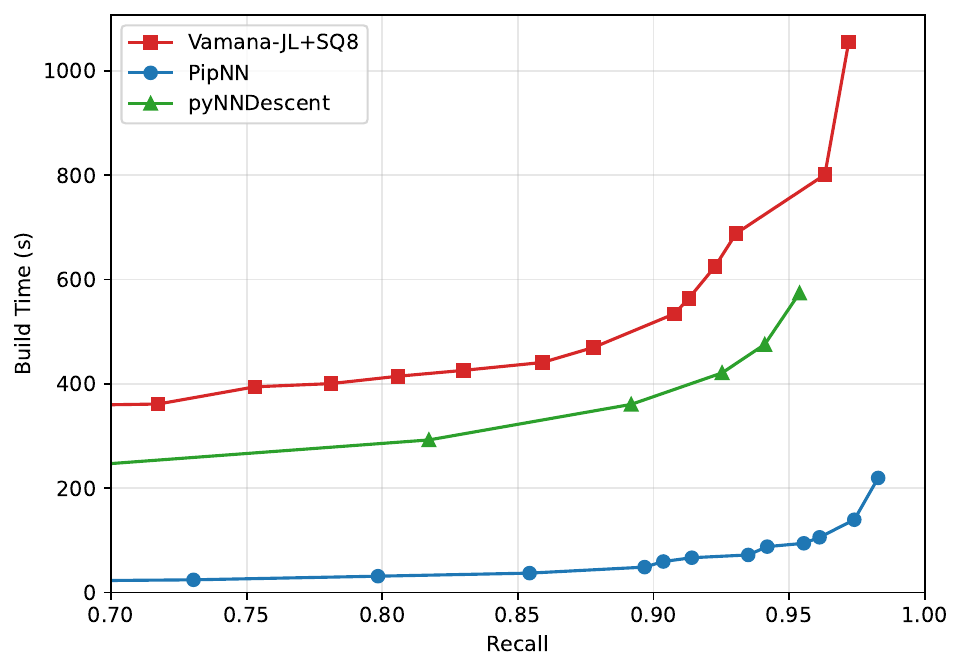}
    \caption{DataCompDR-12M}
  \end{subfigure}\\[4pt]
  \begin{subfigure}[t]{0.32\textwidth}
    \includegraphics[width=\linewidth]{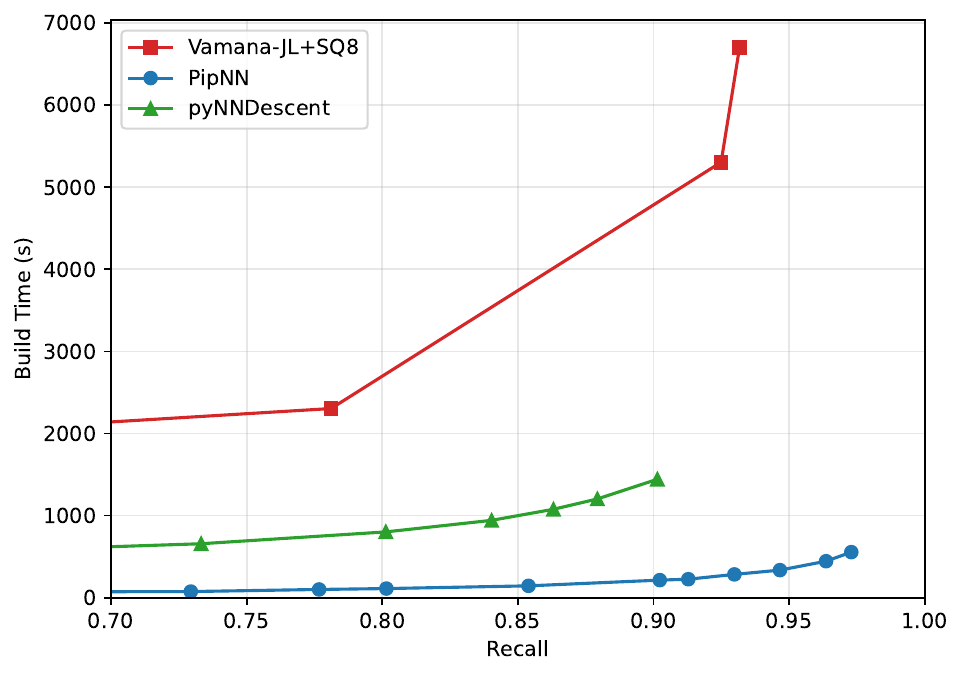}
    \caption{Wiki-Cohere-35M}
  \end{subfigure}\hfill
  \begin{subfigure}[t]{0.32\textwidth}
    \includegraphics[width=\linewidth]{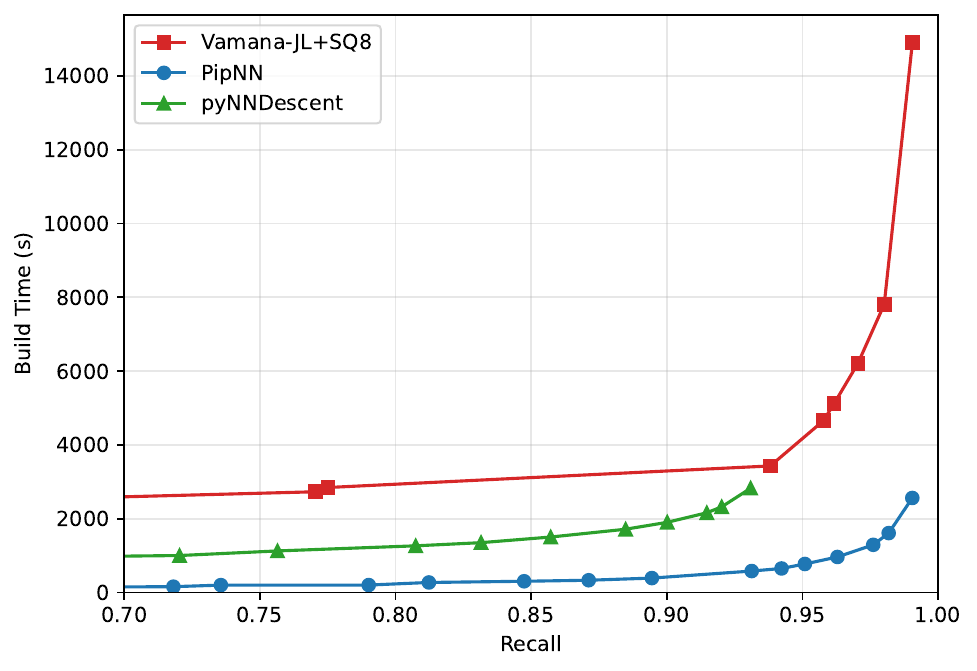}
    \caption{Wiki-all-88M}
  \end{subfigure}\hfill
  \begin{subfigure}[t]{0.32\textwidth}
    \includegraphics[width=\linewidth]{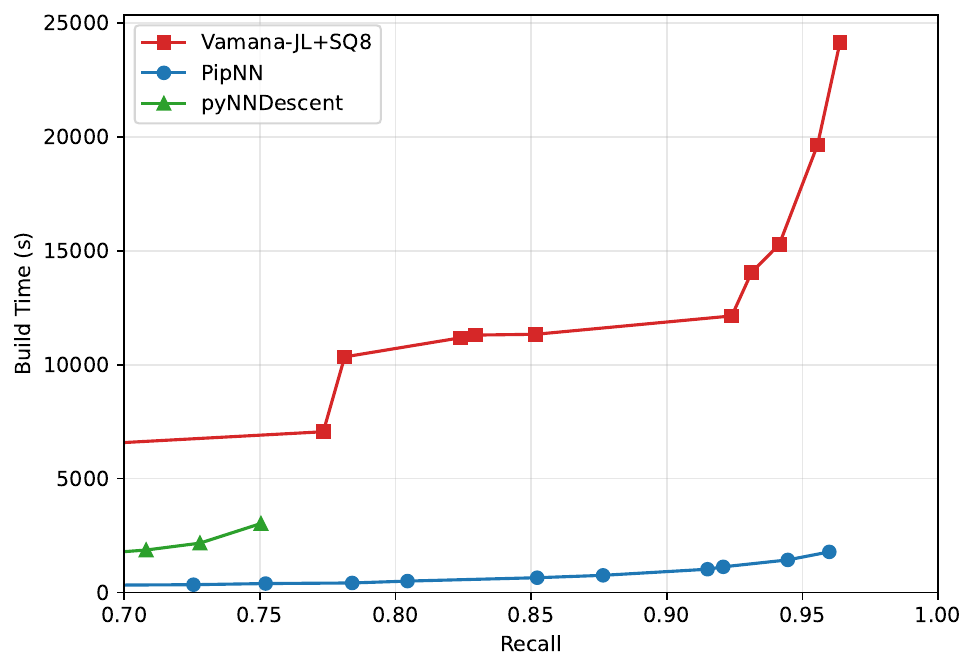}
    \caption{MSMARCO-101M}
  \end{subfigure}\\[4pt]
  \begin{subfigure}[t]{0.32\textwidth}
    \includegraphics[width=\linewidth]{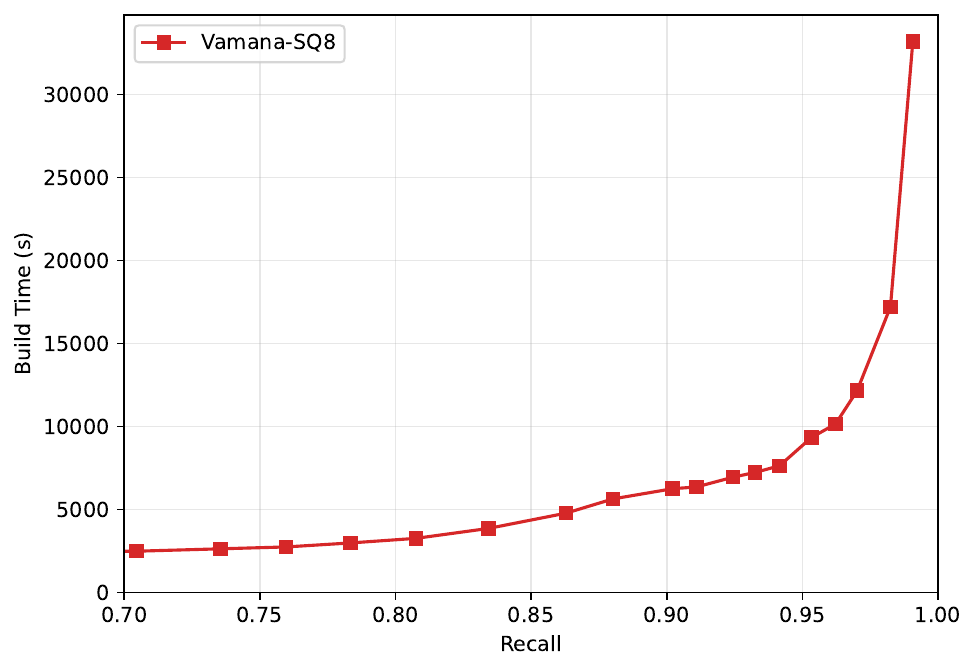}
    \caption{Deep-1B}
  \end{subfigure}\hfill
  \begin{subfigure}[t]{0.32\textwidth}
    \includegraphics[width=\linewidth]{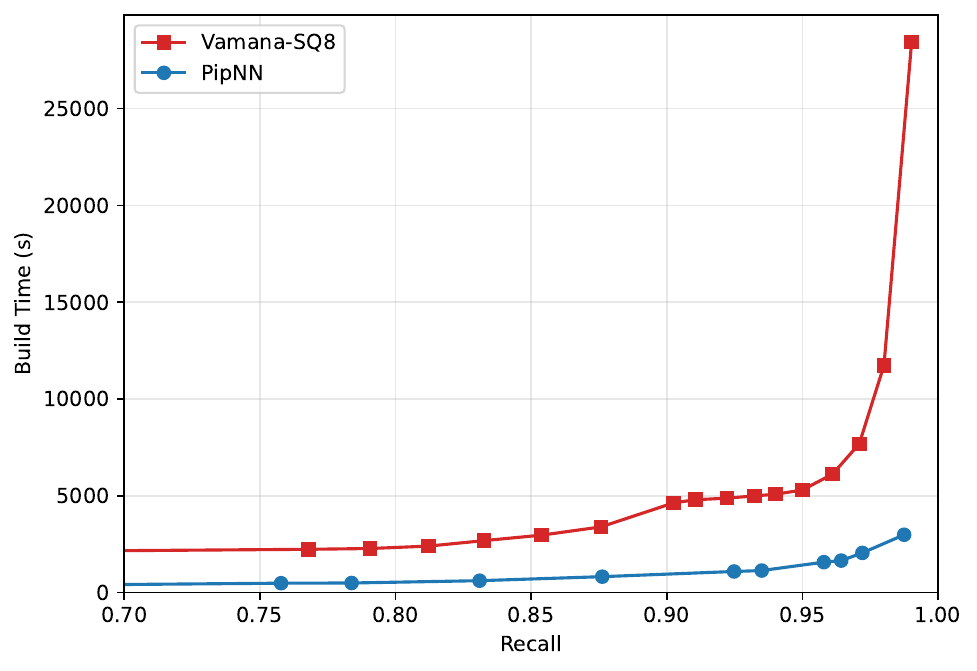}
    \caption{MSTuring-1B}
  \end{subfigure}\hfill
  \begin{subfigure}[t]{0.32\textwidth}
    \includegraphics[width=\linewidth]{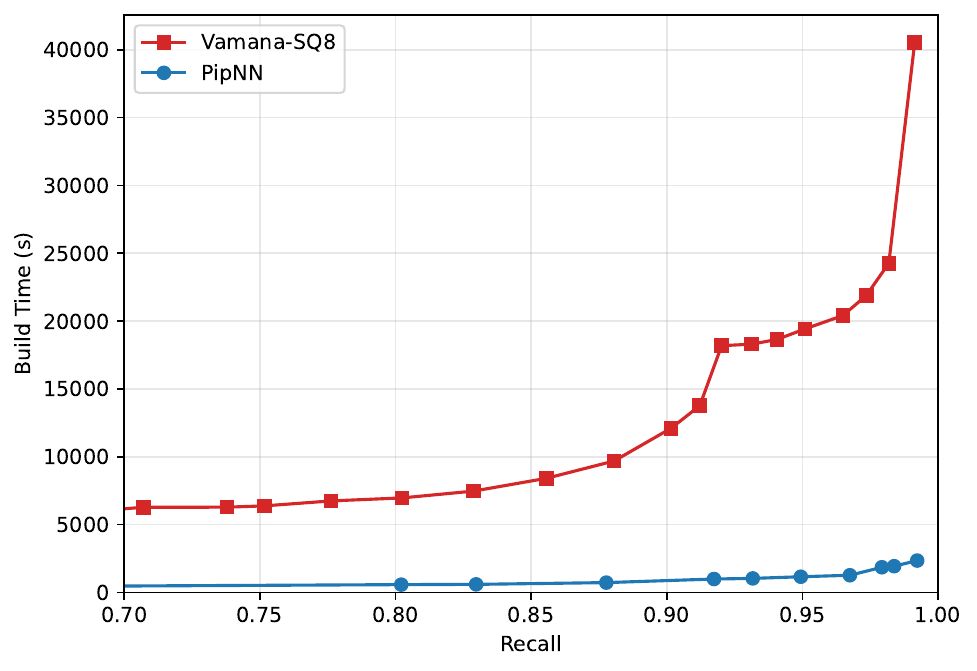}
    \caption{Text2Image-1B}
  \end{subfigure}
  \caption{Full set of $k$-NN-graph recall vs.\ build time results on CPUs.
  We note that the optimized version of pyNNDescent in ParlayANN runs out of memory on MSMARCO-101M,
  Deep-1B, MSTuring-1B, and Text2Image-1B.
  }
  \label{fig:additional-knn-recall}
\end{figure*}

\begin{figure*}
    \centering
    \includegraphics[width=\linewidth]{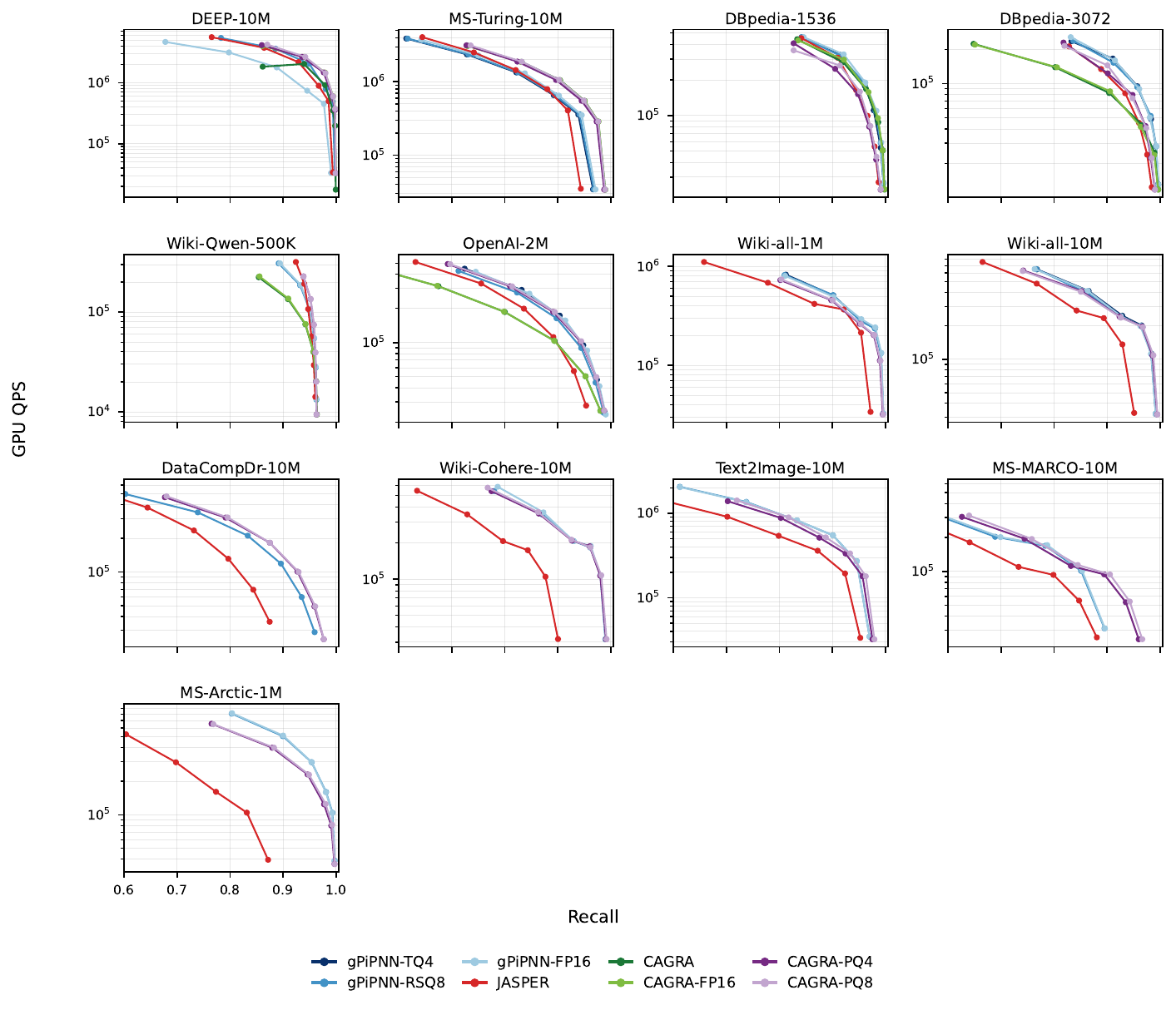}
    \caption{Full GPU query throughput vs.\ recall grid for all thirteen
    datasets (top-right is best); the body (\Cref{fig:gpu_anns_recall_main})
    shows the DEEP-10M and DBpedia-3072 panels. Queries use Jasper's GPU
    search kernel. The DataCompDr-10M, Wiki-Cohere-10M, Text2Image-10M, and
    MS-MARCO-10M panels use ground truth computed for the 10M crops, and the
    MS-Arctic-1M panel ground truth for the 1M slice; on the
    unnormalized MIPS sets only the \textsc{CAGRA}-PQ (\textsc{IVF-PQ}) variant
    is available.}
    \label{fig:gpu_anns_recall_grid}
\end{figure*}

\end{document}